\documentclass[aoas]{imsart}

\RequirePackage{amsthm,amsmath,amsfonts,amssymb}
\RequirePackage[authoryear]{natbib}
\RequirePackage[colorlinks,citecolor=blue,urlcolor=blue]{hyperref}
\RequirePackage{graphicx, makecell}
\usepackage{diagbox}
\usepackage[dvipsnames]{xcolor}
\usepackage{bbm}
\usepackage{subcaption}
\startlocaldefs
\theoremstyle{plain}

\newtheorem{theorem}{Theorem}[section]

\newtheorem{proposition}[theorem]{Proposition}
\theoremstyle{remark}

\newtheorem*{remark}{Remark}

\newcommand{\ve}[1]{{\mbox{\boldmath ${#1}$}}}
\DeclareMathOperator*{\argminB}{argmin}   % Jan Hlavacek
\endlocaldefs

\begin{document}

\begin{frontmatter}
\title{Generalized Matrix Decomposition Regression: \\ Estimation and Inference for Two-way Structured Data}
%\title{A sample article title with some additional note\thanksref{t1}}
\runtitle{GMDR}
%\thankstext{T1}{A sample additional note to the title.}

\begin{aug}
%%%%%%%%%%%%%%%%%%%%%%%%%%%%%%%%%%%%%%%%%%%%%%%
%% Only one address is permitted per author. %%
%% Only division, organization and e-mail is %%
%% included in the address.                  %%
%% Additional information can be included in %%
%% the Acknowledgments section if necessary. %%
%%%%%%%%%%%%%%%%%%%%%%%%%%%%%%%%%%%%%%%%%%%%%%%
\author[A]{\fnms{Yue} \snm{Wang}\ead[label=e1]{yue.2.wang@cuanschutz.edu}},
\author[B]{\fnms{Ali} \snm{Shojaie}\thanksref{t1}\ead[label=e2]{ashojaie@uw.edu}},
\author[C]{\fnms{Timothy} \snm{Randolph}\thanksref{t2}\ead[label=e3]{trandolp@fredhutch.org}},
\author[D]{\fnms{Parker} \snm{Knight}\ead[label=e4]{ pknight@g.harvard.edu}}
\and
\author[E]{\fnms{Jing} \snm{Ma}\thanksref{t3}\ead[label=e5]{jingma@fredhutch.org}}
%%%%%%%%%%%%%%%%%%%%%%%%%%%%%%%%%%%%%%%%%%%%%%
%% Addresses                                %%
%%%%%%%%%%%%%%%%%%%%%%%%%%%%%%%%%%%%%%%%%%%%%%
\address[A]{Department of Biostatistics and Informatics, University of Colorado Anschutz Medical Campus,
\printead{e1}}

\address[B]{Department of Biostatistics, University of Washington,
\printead{e2}}

\address[C]{Clinical Research Division, Fred Hutchinson Cancer Center,
\printead{e3}}

\address[D]{Department of Biostatistics, Harvard University,
\printead{e4}}

\address[E]{Public Health Sciences Division, Fred Hutchinson Cancer Center,
\printead{e5}}

\thankstext{t1}{Supported by National Institutes of Health R01GM133848.}
\thankstext{t2}{Supported by National Institutes of Health R01GM129512, R01HL1554178, and P50CA228944.}
\thankstext{t3}{Supported by National Institutes of Health R01GM145772.}

\end{aug}

\begin{abstract}
{Motivated by emerging applications in ecology, microbiology, and neuroscience,}
this paper studies high-dimensional regression with two-way structured data.
To estimate the high-dimensional coefficient vector, we propose the generalized matrix decomposition regression (GMDR) to efficiently leverage auxiliary information on row and column structures. GMDR extends the principal component regression (PCR) to two-way structured data, but unlike PCR, GMDR selects the components that are most predictive of the outcome, leading to more accurate prediction.
For inference on regression coefficients of individual variables, we propose the generalized matrix decomposition inference (GMDI), a general high-dimensional inferential framework for a large family of estimators that include the proposed GMDR estimator.
GMDI provides more flexibility for incorporating relevant auxiliary row and column structures. 
{As a result, GMDI  
does not require the true regression coefficients to be sparse, but constrains the coordinate system representing the regression coefficients according to the column structure.
GMDI also allows dependent and heteroscedastic observations. }
% whose dependencies are informed by the auxiliary row structure.
We study the theoretical properties of GMDI in terms of both the type-I error rate and power and demonstrate the effectiveness of GMDR and GMDI in simulation studies and an application to human microbiome data. 
%{to identify age-associated human gut microbes}.
\end{abstract}


\end{frontmatter}
%%%%%%%%%%%%%%%%%%%%%%%%%%%%%%%%%%%%%%%%%%%%%%
%% Please use \tableofcontents for articles %%
%% with 50 pages and more                   %%
%%%%%%%%%%%%%%%%%%%%%%%%%%%%%%%%%%%%%%%%%%%%%%
%\tableofcontents

\section{Introduction}
We consider the problem of regressing a scalar outcome from $n$ observations on a vector of $p$ predictors, formally, $\mathbb{E}(y)={\bf x}^\intercal\ve \beta$, in settings where it may be implausible to assume that the $p$ variables or the $n$ samples are independent. To address this problem, we account for the sample- and variable-wise dependencies to provide a framework for estimation of the coefficient vector, $\ve \beta$, and inference on the individual coefficients, $\beta_j$ ($j=1,...,p$). The proposed framework is motivated by the increasing occurrence of high-dimensional two-way structured data---that is, data with structures among the variables (columns) and samples (rows)---in ecology, microbiology, and neuroscience. Informative two-way structures can often be obtained from various auxiliary sources {\it a priori} \citep{Allen2014, li2021inference}. In many applications, the goal is to examine associations between such structured data and an outcome of interest. {One application
that motivated the current work comes from human microbiome data which record the composition and function of bacterial taxa. These data are used to investigate the role of human microbiome in health and diseases.  An interesting property of these data is that taxa are related to one another, both evolutionarily and functionally. Evolutionary relationships among taxa are typically characterized by a phylogenetic tree, or dendrogram, whose nodes represent taxonomic assignments based on genomic similarities \citep{Washburne2018}. Their functional relationships may be characterized by genomic content known to contribute to a biological process \citep{Sharifi2017}.}

\begin{figure}[t!]
    \centering
    \includegraphics[width = 0.9\textwidth]{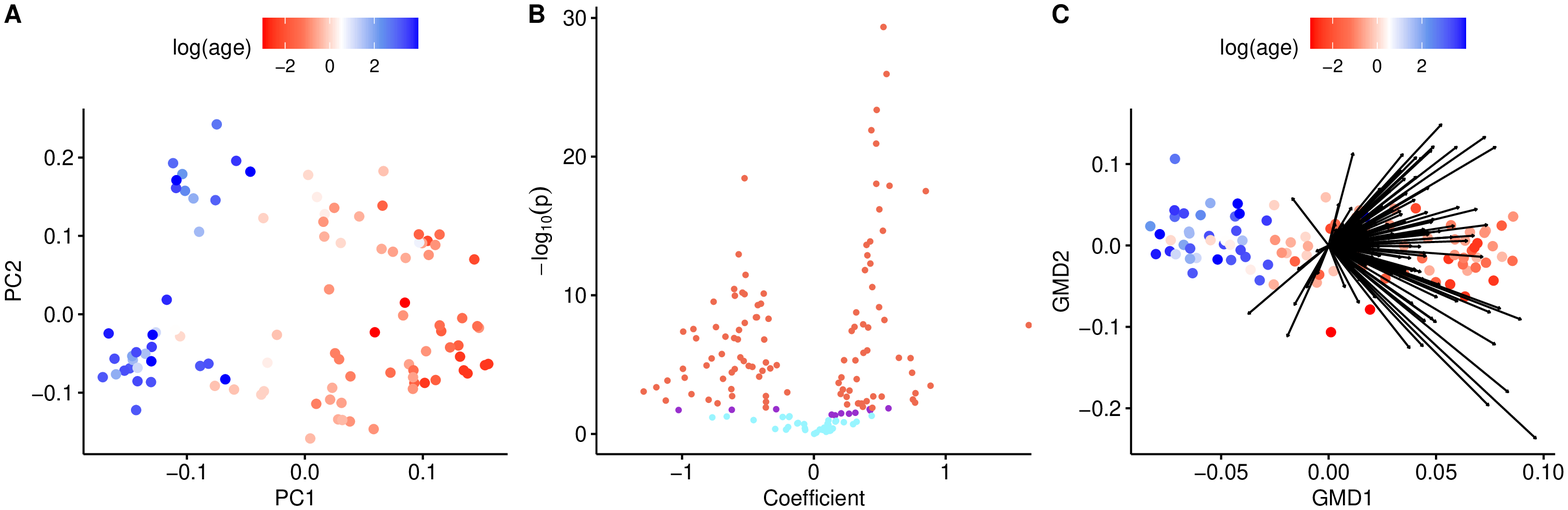}
    \caption{(A): The PC-plot of data from \cite{Yatsunenko2012}. (B): The volcano plot showing the ${\log 10}$-transformed $p$-values for the associations of the bacteria with age versus the corresponding regression coefficients.
%  %The x-axis represents the marginal spearman correlation between each bacterial and age; the y-axis represents the $p-$value associated with each correlation. 
  Cyan dots represent bacteria for which $p$-values are greater than 0.05; purple dots represent bacteria for which $p$-values are less than 0.05 but no longer show statistically significant associations after controlling the FDR at 0.1; red dots represent bacteria that still show statistically significant associations after controlling the FDR at 0.1. 
  (C): The GMD-plot of data from \cite{Yatsunenko2012}: metagenomic similarities among samples and phylogenetic similarities among taxa are considered.}
    \label{AOAS:intro:fig1}
\end{figure}

{
To motivate our regression framework, we consider data from a study investigating age-associated microbial signatures across geographic regions \citep{Yatsunenko2012}.  In this example, stool samples from $n = 100$ individuals from the Amazonas of Venezuela, rural Malawi, and US metropolitan areas were processed to identify $p = 149$ genus-level bacterial abundances.  Figure \ref{AOAS:intro:fig1}A shows a principal-component (PC) plot of the configuration of samples based on the first two PCs of the $n\times p$ microbiome data matrix; samples are colored by the logarithm of each individual's age, which range from a few months to over 50 years. This plot suggests a strong association between age and microbial composition. This is further supported by Fig.~\ref{AOAS:intro:fig1}B, a volcano plot of the log 10-transformed $p$-values versus the estimated coefficients obtained from a univariate regression of each genus on age. 
Red dots represent 
%bacteria that show \YW{statistically significant marginal associations with age} 
{bacteria that have statistically significant marginal associations with age}
after controlling the false discovery rate (FDR) at $0.1$ using the Benjamini–Yekutieli procedure \citep{benjamini2001control}; 
%\YW{purple dots represent bacteria for which $p$-values are less than $0.05$ but no longer show statistically significant associations with age after controlling the FDR;} 
{purple dots represent bacteria with $p$-values less than $0.05$ that are no longer statistically significant after controlling the FDR;} 
cyan dots represent bacteria for which $p$-values are greater than $0.05$. Figure \ref{AOAS:intro:fig1}B shows that the majority of bacteria (105 out of 149) are marginally associated with age after controlling the FDR at 0.1. 
This type of analysis, however, does not account for  the relationships between either the taxa or the individuals from which the samples were taken. As noted above, bacteria tend to be correlated via their phylogeny, and individuals also tend to be correlated in their microbial composition due to shared households, diets, and/or cultures \citep{zeevi2019structural, hullar2021associations}.
}

{
These structures are commonly acknowledged in the analysis of microbiome data. 
For example, phylogeny-aware distances between samples (e.g., UniFrac,  \citealp{lozupone2005unifrac}) are used in the principal coordinate analysis (PCoA) and in kernel-based association tests \citep{zhao2015testing}. In an extension of PCoA, \cite{Wange00504-19} used the generalized matrix decomposition (GMD, \citealp{Allen2014}) to produce dimension-reduced plots like PCoA while leveraging similarities among the taxa and among the samples. This approach is illustrated in Fig.~\ref{AOAS:intro:fig1}C, which shows a GMD-biplot of sample configurations (dots) and corresponding variable loadings (arrows) in these coordinates. Here, the coordinate system is derived by extending the singular value decomposition in a manner that accounts for both row and column structures. More specifically, the structure among taxa is characterized by a $p\times p$ similarity kernel derived from the patristic distance between each pair of tips of a phylogenetic tree. The structure among samples is derived from extrinsic data based on bacterial genes: the functional protein content produced by the bacteria in each sample is estimated by classifying genes according to Enzyme Commission (EC) numbers  \citep{Cuesta2015}; see Section \ref{sec:realdata} for more details.  Then, an $n\times n$ matrix of pairwise sample similarities based on EC numbers provides a biologically-informed auxiliary representation of sample-based structures. 
%(more information of the EC data can be found in Section \ref{sec:realdata}). 
The two axes in Fig.~\ref{AOAS:intro:fig1}C are the first two columns of the right GMD vectors. Each sample is represented by the coordinates of the projection of its microbial abundance vector onto the two axes and is colored by the logarithm of the subject’s age. An arrow is then plotted for each taxon, its coordinates coming from the first two columns of the right GMD vectors. Compared to Fig.~\ref{AOAS:intro:fig1}A, the GMD-biplot provides an alternative two-dimensional configuration of samples; it shows a strong age-dependent variation and many tightly clustered arrows (genera) contributing to this configuration. Consistent with Fig.~\ref{AOAS:intro:fig1}B, this biplot suggests that there are many correlated age-associated taxa. This analysis, however, is unsupervised and any inference made about the associations is circumstantial. It is desirable, therefore, to develop a supervised analytical framework of high-dimensional regression that leverages auxiliary row and column structures, and, importantly, provides valid inference for the associations between the taxa and a response variable.
}

%--------------------------------------------
% \subsection{Related Literature}
% Methods for incorporating auxiliary information into high-dimensional linear regression have received much recent interest. For example, network-constrained regularization approaches have been developed to incorporate prior knowledge on the structures of the variables \citep{lili2008, li2010variable}. The more recent kernel penalized regression (KPR, \citealp{randolph2018}) can incorporate pre-specified two-way structures  using a generalized least squares formulation with a kernel-guided $l_2$ type penalty. However, there is a paucity of methods that can provide inference for associations between an outcome and the high-dimensional two-way structured covariates. 
% %The problem we study here is certainly related to the high-dimensional regression problem in the conventional setting where no auxiliary information is available. Popular regularization methods in this setting include lasso \citep{tibshirani1996regression}, the ridge regression \citep{Hoerl1970}, the elastic net \citep{zou2005regularization} , SCAD \citep{fan2001variable} and group lasso \citep{yuan2006model}.

%----------------------------------------------------------------------------------
\subsection{Our Contributions}
This paper introduces the GMD regression (GMDR), a dimension reduction-based estimation procedure that efficiently leverages pre-specified two-way structures. GMDR is built upon the generalized matrix decomposition (GMD, \citealp{Allen2014, Escoufier2006}), which extends the singular value decomposition (SVD) to incorporate auxiliary two-way structures and will be reviewed in Section \ref{sec:gmdr}. 
Thus, GMDR can be viewed as an extension of principal component regression (PCR) for analyzing two-way structured data.  However, unlike PCR which uses top principal components as the predictors, our GMDR selects the GMD components that are most predictive of the outcome. This novel selection procedure ensures a more accurate prediction using GMDR. 

{We further define a broad class of estimators for high-dimensional regression on two-way structured data by leveraging the connection between dimension reduction-based regression (e.g., PCR) and penalized regression (e.g., ridge regression), which is discussed in detail in Section \ref{sec:gmdr}. 
This connection also allows us to develop the GMD inference (GMDI) framework, 
a high-dimensional inference (HDI) procedure that can assess the statistical significance of individual variables based on any arbitrary estimator in this class.  
As such, 
GMDI can be applied to not only the proposed GMDR but also many existing {estimation procedures} that lack inferential procedures for individual variables, such as PCR, {generalized ridge regression \citep{golub2013matrix}, and the kernel penalized regression (KPR, \citealp{randolph2018})}. 
%We develop the GMD inference (GMDI) framework, a high-dimensional inference (HDI) procedure that can assess the significance of individual variables based on any arbitrary estimator in this class. 
}
GMDI has three distinct features. First, unlike most existing HDI tools that assume {\it i.i.d} samples, which may not hold for two-way structured data, GMDI allows for dependent and heteroscedastic samples by efficiently leveraging auxiliary row structures. 
%Imposing structures among samples is common in %semiparametric models and kernel machine regression \citep{liu2007semiparametric}, 
Ignoring sample correlations may lead to incorrect inference even in low-dimensional settings.
Second, existing HDI tools, including \cite{buhlmann2013, zhangzhang2014, javanmard14a, javanmard2014hypothesis, vandegeer2014, belloni2015uniform, zhao2016, mitra2016benefit, ning2017, zhubradic2018}, all require at least one of the following assumptions: (i) the regression coefficient vector is sparse, (ii) the design matrix satisfies a restricted eigenvalue-type condition if a fixed design is considered, and (iii) the precision matrix of the variables in the design matrix has row sparsity if a random design is considered.
However, these conditions may fail when strong correlations exist among variables, which is common for two-way structured data. 
Third, GMDI provides flexibility for users to specify relevant auxiliary row and column structures. In particular, we provide methods
%to examine the informativeness of the pre-specified structures so as 
to avoid uninformative structures and to incorporate partially informative structures,
leading to well-controlled type-I error rates and guaranteed power.

Regarding the second property, it may be that a majority of variables are {\it marginally} associated with the outcome, as appears to be the case in Fig. \ref{AOAS:intro:fig1}B.  This has two possible explanations: (i) a large number of the variables
are also {\it conditionally} associated with the outcome; (ii) these variables are highly correlated, but only a few
of these are conditionally associated with the outcome. In the first situation, the vector of regression coefficients is not sparse; in the second situation, any restricted eigenvalue-type condition may fail \citep[see][]{van2009conditions}, and likely, the precision matrix of the variables is not sparse. As an alternative to these assumptions, GMDI assumes the pre-specified column structure informs the structure of the regression coefficients, which reduces to sparsity when no column structure is pre-specified. 

%{This paper develops the GMD inference (GMDI), an effective HDI procedure for a broad class of estimators that account for the structures in data, including the existing KPR.}

%The GMDI assumes that the pre-specified row and column structures can, respectively, inform the covariance among the samples and the structure of the regression coefficients. As a result, GMDI does not require the true regression coefficients to be sparse, allows for dependent and heteroscedastic samples, and provides more flexibility for users to specify relevant auxiliary row and column structures. 

GMDI follows the general idea of bias correction for ridge-type estimators \citep{buhlmann2013} but uses a novel initial estimator
that efficiently leverages the pre-specified two-way structures.
We derive the asymptotic distribution of the bias-corrected estimator. Based on this,
we construct asymptotically valid two-sided $p$-values and provide sufficient conditions under
which GMDI offers guaranteed power. We introduce a procedure that selects against uninformative sample structure. We also show that the GMDI results are robust to misspecification. Our numerical studies demonstrate the superior
performance of GMDI for two-way structured regression compared to existing HDI methods, even when pre-specified structures are not fully informative.
%-----------------------------------------------------------------------------------

%---------------------------------------------------------------------------------
\subsection{Organization and Notation}
The rest of the paper is organized as follows. In Section~\ref{sec:gmdr}, we first introduce the GMDR estimation/prediction framework, accompanied by the novel  procedure for the selection of  GMD components. We then link the GMDR estimator to a broad class of estimators. In Section~\ref{sec:gmdi}, we present the GMDI procedure for any arbitrary estimator in this class, explain the rationale behind the key assumptions for GMDI, and provide ways to assess the informativeness of the pre-specified structures and  incorporate partially informative structures.  
Multiple simulation studies, including one based on real data, are presented in Section~\ref{sec:simu} to examine the finite-sample performance of GMDI. In Section~\ref{sec:realdata}, we demonstrate the effectiveness of GMDR and GMDI on an application to microbiome data. Section \ref{sec:discussion} summarizes our findings and outlines potential extensions. Technical proofs are provided in the supplement \citep{wang2023gmdrsupp}.

Throughout the paper, we use normal typeface to denote scalars, bold lowercase typeface to denote vectors, and bold uppercase typeface to denote matrices.  For any vector ${\bf v} \in \mathbb{R}^p$, we use $v_j$ to denote the $j$-{th} element of $\bf v$ for $j = 1, \ldots, p$. For any matrix ${\bf M} \in \mathbb{R}^{n \times p}$, $ {\bf m}_j$ and $m_{ij}$ denote, respectively, the $j$-{th} column and $(i,j)$ entry of $\bf M$ for $i = 1, \ldots, n$ and $j = 1, \ldots, p$. 
For any index set $\mathcal{I} \subset \{1, \ldots, p\}$, ${\bf v}_\mathcal{I}$ and ${\bf M}_{\mathcal{I}}$ denote, respectively, the subvector of ${\bf v}$ whose elements are indexed by $\mathcal{I}$ and the submatrix of ${\bf M}$ whose columns are indexed by $\mathcal{I}$. 
%${\bf M}^-$ denotes the Moore-Penrose inverse of $\bf M$.
The indicator function $\mathbbm{1}(\mathcal{A})$ denotes the occurrence of the event $\mathcal{A}$; i.e., $\mathbbm{1}(\mathcal{A}) = 1$ if $\mathcal{A}$ is true, and $\mathbbm{1}(\mathcal{A}) = 0$, otherwise. 
%We write $a_n \asymp b_n$ when $\lim_{n \rightarrow \infty} a_n/b_n = C \in (0, \infty)$. 
 We denote
$ \|{\bf v}\|_0 = \sum_{j = 1}^p \mathbbm{1}(v_j \neq 0), ~~ \|{\bf v}\|_q = \left(\sum_{j = 1}^p |v_j|^q\right)^{1/q} $ for any $ 0 < q < \infty$, $\|{\bf v}\|_{\infty} = \max_j|v_j|  $, $\|{\bf v}\|_{\bf K}^2 = {{\bf v}^\intercal {\bf K} {\bf v}}$ for any positive semi-definite matrix $\bf K$,
 $\|{\bf M}\|_q = \sup_{\|{\bf v}\|_q = 1}\|{\bf Mv}\|_q$ for any $q > 0$ and $\|{\bf M}\|_F^2 = \sum_{i=1}^n \sum_{j=1}^p m_{ij}^2$. Finally, for any square matrix ${\bf S}$, we denote 
 the trace of $\bf S$ as $\mbox{tr}({\bf S})$.

%-----------------------------------------------------------
\section{The GMD regression}\label{sec:gmdr}
Consider the following linear model
\begin{equation}\label{main:model:1}
  {\bf y} = {\bf X}\ve \beta^*  + \ve \epsilon,  
\end{equation}
where ${\bf X} \in \mathbb{R}^{n \times p}$ denotes the structured design matrix, ${\bf y} \in \mathbb{R}^n$ is the response variable, and
$\ve \beta^* \in \mathbb{R}^p$ is the underlying true regression coefficient. We allow $p$ to be greater than $n$. 
In addition, we assume that $\ve \epsilon$ is a vector of random noises with $\mathbb{E}[\ve \epsilon \mid {\bf X}] = {\bf 0}_n$ and $\mbox{Cov}(\ve \epsilon \mid {\bf X}) = \ve \Psi$, where ${\bf 0}_n$ is an $n \times 1$ vector of zeros and $\ve \Psi$ is an $n \times n$ positive definite matrix.  
{By considering the non-identity matrix $\ve \Psi$},
we do not assume that entries of $\ve \epsilon$ are {\it i.i.d.}, allowing for samples to be correlated and heteroscedastic.
%Technical assumptions on $\ve \Psi$ will be made later in Section 3.1. 
%Let ${\bf X} \in \mathbb{R}^{n \times p}$ denote a data matrix with potential row and column structures.  
% Let ${\bf H} \in \mathbb{R}^{n \times n}$ and ${\bf Q} \in \mathbb{R}^{p \times p}$, respectively, inform the row and column structures of ${\bf X}$ that are obtained from auxiliary sources. Throughout the article, we assume that both $\bf H$ and $\bf Q$ are positive definite similarity matrices, whose entries, respectively, characterize similarities between samples and between variables, after adjusting for other samples or variables.  
Let ${\bf H} \in \mathbb{R}^{n \times n}$ and ${\bf Q} \in \mathbb{R}^{p \times p}$ denote two auxiliary positive definite matrices, {capturing similarities among rows and columns of $\bf X$, respectively.}
{More specifically, we assume that entries of $\bf H$ ($\bf Q$) inform the {\it conditional similarity}
between samples (variables); that is, the similarity between samples (variables) after
the effects of other samples (variables) are removed. 
This implies that, for instance, $\bf H$ provides information about $\ve \Psi$, and their connection will be made explicit in Assumption (A1).}
%We will provide additional characterizations of the informativeness of $\bf H$ and $\bf Q$ within a high-dimensional linear regression framework in Section 3.1. 
We assume that $\bf X, H$ and $\bf Q$ are deterministic quantities and refer to the triple $({\bf X, H, Q})$ as two-way structured data hereafter. Throughout the article, we assume that $\bf X$ and $\bf y$ are appropriately centered such that ${\bf 1}_n^\intercal{\bf Hy} = 0$ and ${\bf 1}_n^\intercal{\bf HX} = {\bf 0}_p^\intercal$, where ${\bf 1}_n$ is an $n \times 1$ vector of all ones.
%Suppose we observe $\{{\bf x}_i, y_i\}$ for $i = 1, \ldots, n$, where ${\bf x}_i \in \mathbb{R}^p$ is a $p \times 1$ vector of variables and $y_i \in \mathbb{R}$ is an outcome of interest. 
% Let ${\bf y} = (y_1, \ldots, y_n)^\intercal$ denote the $n \times 1$ vector of outcomes of interest. We consider a linear model $\mathbb{E}({\bf y}) = {\bf X}\ve \beta^*$, where $\ve \beta^* = (\beta_1^*, \ldots, \beta_p^*)^\intercal$ is a $p \times 1$ vector of unknown parameters of interest.
%In the high-dimensional regime, we consider $p = p(n)$ and allow $p$ to be greater than $n$. 
We will study the estimation and inference of the high-dimensional parameters $\ve \beta^*$, while leveraging the information from $\bf H$ and $\bf Q$.
%Thus (\ref{gmdr:motiv}) is essentially a high-dimensional linear model. 
%Second, by assuming that $\bf H$ and $\bf Q$, respectively, characterize the structures among samples and variables, we mean that $\mbox{Cov}({\bf x}_i) = {\bf Q}^-$ and $\mbox{Cov}(\ve \epsilon | {\bf X}) = \sigma_\epsilon^2 {\bf H}^-$.
%Let ${\bf Q} \in \mathbb{R}^{p \times p}$ be a positive definite matrix characterizing relations among variables. 
%Let ${\bf H} = {\bf L_HL_H}^\intercal$ denote the Cholesky decomposition of ${\bf H}$. Without loss of generosity, 
%It can be easily seen that ${\bf H = L_HL_H}^\intercal$ and ${\bf Q = L_QL_Q}^\intercal$.
% we assume that $\mbox{rank}({\bf X}) = \min(n,p)$, and ${\bf X}$ and ${\bf y}$ are appropriately scaled such that ${\bf 1}_n^\intercal{\bf L}_{\bf H}{\bf X} = 0$ and ${\bf 1}_n^\intercal{\bf L}_{\bf H}{\bf y} = 0$. 

Our idea is built upon the generalized matrix decomposition (GMD), which we will review next. The GMD of $\bf X$ with respect to $\bf H$ and $\bf Q$ is ${\bf X = USV}^{\intercal}$, where the components are obtained by solving the optimization problem
\begin{equation}\label{est:1}
    \text{argmin}_{\bf U,S,V}\|{\bf X} - {\bf USV}^\intercal\|_{\bf H,Q},
\end{equation}
subject to ${\bf U}^\intercal {\bf HU} = {\bf I}_K, {\bf V}^\intercal{\bf QV} = {\bf I}_K$ and ${\bf S} = \text{diag}(\sigma_1, \ldots, \sigma_K)$. Here, $K \leq \min(n,p)$ is the rank of ${\bf X}^\intercal{\bf HXQ}$ and  $\|{\bf M}\|_{\bf H, Q}^2 = \mbox{tr}({\bf M}^\intercal{\bf HMQ})$ for any matrix ${\bf M} \in \mathbb{R}^{n \times p}$. Note that unlike SVD, the GMD vectors ${\bf U}$ and ${\bf V}$ are not orthogonal in the Euclidean norm unless ${\bf H} = {\bf I}_n$ and ${\bf Q} = {\bf I}_p$. 
%It can be seen from (\ref{est:1}) that 
GMD directly extends  SVD by replacing the Frobenius norm with the $(\bf H, Q)$-norm $\|\cdot\|_{\bf H, Q}$. As such, GMD preserves appealing properties of SVD such as ordering the component vectors according to a nonincreasing set of GMD values, $\sigma_1, \ldots, \sigma_K$, indicating that the
decomposition of the total variance of $\bf X$ into each dimension is nonincreasing. 
%incorporates two-way structures from both ${\bf H}$ and ${\bf Q}$ from a matrix decomposition perspective. We utilize the generalized matrix decomposition (GMD) \citep{Allen2014}, which generalizes singular value decomposition (SVD) by incorporating external structures. 
%Recall that (\ref{est:1}) defines the GMD of $\bf X$ with respect to $\bf H$ and $\bf Q$.
% The GMD of $\bf X$ as ${\bf X} =  {\bf U SV}^\intercal$ can be obtained from the following optimization problem:
% \begin{equation}\label{est:1}
%     \text{argmin}_{\bf U,S,V}\|{\bf X} - {\bf USV}^\intercal\|_{\bf H,Q},
% \end{equation}
% subject to ${\bf U}^\intercal {\bf HU} = {\bf I}_K, {\bf V}^\intercal{\bf QV} = {\bf I}_K$ and ${\bf S} = \text{diag}(\sigma_1, \ldots, \sigma_K)$. Here, for any matrix ${\bf M} \in \mathbb{R}^{n \times p}$, $\|{\bf M}\|_{\bf H, Q}^2 = \text{tr}({\bf M}^\intercal{\bf HMQ})$ and $K = \mbox{rank}({\bf X}^\intercal{\bf HXQ})$.
% The $\sigma_1, \ldots, \sigma_K$ is a sequence of positive numbers, called the GMD values of ${\bf X}$ with respect to $\bf H$ and $\bf Q$.
% The GMD is a direct extension of the SVD by accounting for any known two-way structures and preserves nice properties of the SVD; for instance, $\sigma_1, \ldots, \sigma_K$ are non-increasing, indicating that the GMD provides a good dimension reduction of ${\bf X}$ with ${\bf H}$ and ${\bf Q}$ taken into account. 
An efficient algorithm was proposed by \cite{Allen2014} to iteratively solve for each column of $\bf U, S$ and $\bf V$ in (\ref{est:1}).
%(\ref{est:1}) provides an exact decomposition of $\bf X$ with ${\bf H}$ and ${\bf Q}$. 
Analogous to the SVD of ${\bf X}$, which is closely related to the eigen-decomposition of ${\bf X}^\intercal{\bf X}$, the GMD of ${\bf X}$ with respect to ${\bf H}$ and ${\bf Q}$ is related to the eigen-decomposition of ${\bf X}^\intercal{\bf HXQ}$. In fact, \cite{escoufier1987duality} and \cite{Allen2014} show that the squared GMD values $\sigma_1^2, \ldots, \sigma_K^2$ are non-zero eigenvalues of ${\bf X}^\intercal{\bf HXQ}$, and columns of ${\bf V}$ are the corresponding eigenvectors. Note that ${\bf X}^\intercal{\bf HXQ}$ may not be symmetric, again implying that columns of $\bf V$ may not be orthogonal in the Euclidean norm. Given ${\bf V}$ and ${\bf S}$, the $n \times K$ matrix ${\bf U}$ can be uniquely defined by ${\bf US} = {\bf XQV}$. 
%Clearly, if ${\bf H} = {\bf I}_n$ and ${\bf Q} = {\bf I}_p$, GMD will reduce to the ordinary SVD. 

Similar to PCR, the GMDR estimate of $\ve \beta^*$ in (\ref{main:model:1}) is obtained by regressing ${\bf y}$ on a reduced subset of GMD components. 
%We now introduce our two-step GMDR estimation procedure to estimate $\ve \beta^*$ in (\ref{note:1}). 
More specifically, let $\ve \nu_j = {\bf u}_j\sigma_j$ be the $j$-th GMD component for $j = 1, \ldots, K$ and set $\ve \Upsilon = \left[\ve \nu_1 \cdots \ve \nu_K\right] \in \mathbb{R}^{n \times K}$. For any fixed index set $\mathcal{I} \subset \{1, \ldots, K\}$, the GMDR estimator of $\ve \beta^*$, $\widehat{\ve \beta}_{\mbox{\tiny GMDR}}(\mathcal{I})$, can be obtained in two steps:
\begin{itemize}
\item[(i)]  Regress ${\bf y}$ on $\ve \Upsilon_\mathcal{I}$ and obtain 
   % \begin{equation}\label{est:2}
        $\widehat{\ve \gamma}(\mathcal{I}) = \text{argmin}_{\ve \gamma}\left\|{\bf y} - {\ve \Upsilon}_{\mathcal{I}}\ve \gamma\right\|^2_{\bf H}$.
   % \end{equation}
    %where ${\ve \nu}_S$ is the submatrix of $\ve \nu$ whose columns are indexed by $S$.
\item[(ii)] Calculate $\widehat{\ve \beta}_{\mbox{\tiny GMDR}}(\mathcal{I}) = 
        \left({\bf QV}\right)_{\mathcal{I}}\widehat{\ve \gamma}(\mathcal{I}).$
\end{itemize}
%We treat a few examples of $\widehat{\ve \beta}_{\mbox{\tiny GMDR}}(\mathcal{I})$ as follows.\\
%{\bf Example 1:} 
Letting $w_j = \mathbbm{1}(j \in \mathcal{I})$ for $j = 1, \ldots, K$ and ${\bf W}_{\mathcal{I}} = \text{diag}(w_1, \ldots, w_K)$,
$\widehat{\ve \beta}_{\mbox{\tiny GMDR}}(\mathcal{I})$ can be explicitly expressed as
%\textcolor{red}{AS: we haven't defined ${\ve \beta}_{GMDR}(S) yet$ and haven't discussed how this is a good estimator of $\beta^\ast$ -- need to first motivate this estiamtor before saying that it is not hard to find the explicit form...also, since we use $S$ in the SVD, we should use a different symbol for the index set, perhaps $\mathcal{I}$}
\begin{equation}\label{gmdr:est}
    \widehat{\ve \beta}_{\mbox{\tiny GMDR}}(\mathcal{I}) = {\bf QVW_{\mathcal{I}}S}^{-1}{\bf U}^\intercal{\bf Hy}, 
\end{equation}
where $\bf U, S, V$ are the GMD components of $\bf X$ with respect to $\bf H$ and $\bf Q$.

\begin{remark}
Similar to SVD, GMD is not invariant to a scale transformation of the variables unless the same scale transformation is applied to all variables. 
Thus, 
our GMDR estimator $\widehat{\ve \beta}_{\mbox{\tiny GMDR}}(\mathcal{I})$ is not invariant to a scale transformation of the predictors. 
    Therefore, we recommend standardizing each predictor
    before implementing GMDR, especially in high-throughput sequencing studies where different variables may have different scales.
However, $\widehat{\ve \beta}_{\mbox{\tiny GMDR}}(\mathcal{I})$ is invariant to a scale transformation of $\bf H$ and $\bf Q$. 
% To see this, consider ${\bf H}_2 = \rho {\bf H}_1$ and ${\bf Q}_2 = \tau {\bf Q}_1$. Denoting the GMD of ${\bf X}$ with respect to ${\bf H}_j$ and ${\bf Q}_j$ by ${\bf U}_j {\bf S}_j {\bf V}_j^\intercal$ for $j = 1, 2$, we get
% \(  
% {\bf U}_1^\intercal{\bf H}_1{\bf U}_1 = {\bf U}_2^\intercal{\bf H}_2{\bf U}_2 = {\bf I}_K \mbox{ and } {\bf V}_1^\intercal{\bf Q}_1{\bf V}_1 = {\bf V}_2^\intercal{\bf Q}_2{\bf V}_2 = {\bf I}_K.
% \)
% Some algebra yields that ${\bf U}_1 = \sqrt{\rho}{\bf U}_2$ and 
% ${\bf V}_1 = \sqrt{\tau}{\bf V}_2$.
% Also, since ${\bf U}_1 {\bf S}_1 {\bf V}_1^\intercal = {\bf U}_2 {\bf S}_2 {\bf V}_2^\intercal$, we get
% ${\bf S}_1 = (\rho\tau)^{-1/2}{\bf S}_1$, leading to \(
% {\bf Q}_1{\bf V}_1{\bf W}_{\mathcal{I}}{\bf S}_1^{-1}{\bf U}_1^\intercal{\bf H}_1{\bf y} = {\bf Q}_2{\bf V}_2{\bf W}_{\mathcal{I}}{\bf S}_2^{-1}{\bf U}_2^\intercal{\bf H}_2{\bf y}. 
% \)
\end{remark}
%where ${\bf W}_{\mathcal{I}} = \text{diag}(w_1, \ldots, w_K)$. 

%To illustrate the motivation behind $\widehat{\ve \beta}_{\mbox{\tiny GMDR}}(\mathcal{I})$, we first consider the low-dimensional case where $K = p$ and $\mathcal{I} = \mathcal{I}_p = \{1, \ldots, p\}$. In this case, it can be seen that 
% \begin{equation*}
%     \mathbb{E}\left(\widehat{\ve \beta}_{\mbox{\tiny GMDR}}(\mathcal{I}_p)\right) = {\bf QVS}^{-1}{\bf U}^\intercal{\bf HUSV}^\intercal\ve \beta^* = \ve \beta^*,
% \end{equation*}
% indicating that $\widehat{\ve \beta}_{\mbox{\tiny GMDR}}(\mathcal{I}_p)$ is an unbiased estimator of $\ve \beta^*$ in the low-dimensional setting. 

%Eq. (\ref{gmdr:bias}) implies that the bias of $\widehat{\ve \beta}_{\mbox{\tiny GMDR}}(\mathcal{I})$ comes from two sources: (i) the choice of ${\bf W}_{\mathcal{I}}$ and (ii) the fact that $K < p$. We hereafter refer to the bias coming from these two sources as the estimation bias and the projection bias, respectively.
%-------------------------------------------------------

%-------------------------------------------------------
% %These observations made above mean that theoretically, how $\bf Q$ informs the signal structure significantly impacts the accuracy of the proposed GMDR estimator. 
% %Unfortunately, in practice it is never checkable how $\ve\beta^*$ aligns with the top eigenvectors of $\bf Q$. Thus, how to select the ``optimal" $\mathcal{I}$ that leads to the best prediction accuracy becomes crucial. 

% It is critical to select the ``best" index set $\mathcal{I}$ that leads to the best prediction performance.

{The prediction performance of $\widehat{\ve \beta}_{\text{GMDR}}(\mathcal{I})$ depends on the choice of the index set $\mathcal{I}$, which can be seen as a tuning parameter.}
%the ``optimal" index set $\mathcal{I}$.
Note that, if ${\bf Q} = {\bf I}_p$ and ${\bf H} = {\bf I}_n$, then GMDR reduces to PCR. Thus, analogous to PCR, a natural way to select $\mathcal{I}$ is to consider GMD components that correspond to large GMD values, referred to as top GMD components hereafter. However, 
{since PCs are constructed without using the outcome, 
top PCs are not necessarily more predictive of the outcome than tail PCs \citep{cook2007}. Thus, we propose an alternative approach to find the most predictive $\mathcal{I}$ among all subsets of $\{1, \ldots, K\}$.}
Note that an exhaustive search over all $2^K$ subsets of $\{1, \ldots, K\}$ is computationally infeasible even for moderate $K$. 
To address this problem, we propose a procedure that weighs the importance of each GMD component by its contribution to the prediction of the outcome. Our idea is to decompose the total $R^2$ of the model into $K$ terms, each corresponding to a GMD component.
Specifically, we first regress ${\bf y}$ on all GMD components $\ve \Upsilon$ with respect to the $\bf H$-norm, and obtain 
\begin{align}\label{gammahat}
\widehat{\ve \gamma} = \text{argmin}_{\ve \gamma}\left\|{\bf y} - {\ve \Upsilon}\ve \gamma\right\|^2_{\bf H}. 
\end{align}
It can then be seen that the total $R^2$ for the model is given by
$R^2 = \left\| {\ve \Upsilon}\widehat{\ve\gamma}\right\|^2_{\bf H}/\left\|{\bf y}\right\|_{\bf H}^2$. 
%the total variance of ${\bf y}$ and the variance explained by ${\ve \Upsilon}$ are, respectively, given by, \textcolor{red}{AS: While I like this, I wonder whether we can use lasso to define $\widehat\gamma$ -- don't we just want a sparse $\widehat\gamma$? Also, what are our assumptions, are we basically assuming that both $\gamma$ and $L\beta$ are sparse? If yes, where is the explicit assumption about sparsit of $\gamma$?} 
%\[SS_{\text{tot}} = \left\|{\bf y}\right\|_{\bf H}^2 ~~~ \text{and} ~~~ SS_{\text{reg}} = \left\| {\ve \Upsilon}\widehat{\ve \gamma}\right\|^2_{\bf H}.\]
 %Thus, the total $R^2$ can be obtained as $R^2 = SS_{\text{reg}}/SS_{\text{tot}}.$  
 %Since any pair of GMD scores are $\bf H-$orthogonal, i.e, $\ve \eta_i^\intercal{\bf H}\ve \eta_j = 0$ for any $i \neq j$, we can decompose the total $R^2$ into $n$ terms, each of which corresponds to each GMD score. 
 Letting $\widehat{\ve \gamma} = \left(\widehat{\gamma}_1, \ldots, \widehat{\gamma}_K\right)^\intercal$, we can write %$SS_{reg} = \sum_{l=1}^n \left\|\ve \eta_l\widehat{\gamma}_l\right\|^2_{\bf H}$,
%which thus yields a decomposition of $R^2$, i.e,
$R^2 = \sum_{j=1}^K r_j^2$, with each $r_j$ represented explicitly in terms of $\ve \nu_j$, $\sigma_j^2$, and $\widehat{\gamma}_j$ as
\[r_j^2 = \frac{\|\ve \nu_j \widehat{\gamma}_j\|^2_{\bf H}}{\left\|{\bf y}\right\|_{\bf H}^2} = \frac{\sigma_j^2\widehat{\gamma}_j^2}{\left\|{\bf y}\right\|_{\bf H}^2}, ~\mbox{for}~j = 1, \ldots, K.\]
Here, we use the fact that $\ve \nu_i^\intercal{\bf H}\ve \nu_j = 0$ for any $i \neq j$. Since $r_1^2, \ldots, r_K^2$ share the same denominator, we define the {\it variable importance} (VI) score of the $j$-th GMD component as 
$\text{VI}_{j} =  \sigma_j^2\widehat{\gamma}_j^2$ for $j = 1,\ldots, K$, with a higher score being more predictive of the outcome.

Based on $\text{VI}_1, \ldots, \text{VI}_K$, we select the most predictive $\mathcal{I}$ in three steps: 
\begin{itemize}
\item[(i)] Sort $\{\text{VI}_j: j = 1, \ldots, K\}$ in nonincreasing order: $\text{VI}_{j_1} \geq \text{VI}_{j_2} \geq \cdots \geq \text{VI}_{j_K}.$ 
  \item[(ii)] For each $k = 1, \ldots, K$, consider $\mathcal{I}_k = \{j_1, \ldots, j_k\}$ and calculate the generalized cross-validation (GCV) statistic:
    \begin{equation}\label{est:9}
    \text{GCV}(k) = \frac{\left\|\left({\bf I}_n - {\bf G}(k)\right){\bf y}\right\|_{\bf H}^2}{\left(\text{tr}\left({\bf I}_n - {\bf G}(k)\right)\right)^2} = \frac{\left\|\left({\bf I}_n - {\bf G}(k)\right){\bf y}\right\|_{\bf H}^2}{\left(n - k\right)^2},
    \end{equation}
    where ${\bf G}(k) = {\ve \Upsilon}_{\mathcal{I}_k}\left({\ve \Upsilon}_{\mathcal{I}_k}^\intercal {\bf H}{\ve \Upsilon}_{\mathcal{I}_k}\right)^{-1}{\ve \Upsilon}_{\mathcal{I}_k}^\intercal{\bf H}.$
    \item[(iii)] Find $k_{opt} = \text{argmin}_{k}\text{GCV}(k)$, and obtain $\mathcal{I}_{k_{opt}} = \{j_1, \ldots, j_{k_{opt}}\}$.
\end{itemize}

%\begin{remark}\label{remark:selection}
%The proposed GCV procedure selects the GMD components that are most predictive of the outcome in a computationally feasible way.  Note that ${\widehat{\gamma}_l} = \sigma_l^{-1}{\bf u}_l^\intercal{\bf Hy}$, thus $\mbox{Var}(\widehat{\gamma}_l | {\bf X}) = \sigma_\epsilon^2 \sigma_l^{-2}$. This indicates that when the total $R^2$ is low ($\sigma_\epsilon^2$ is relatively large), for large $l$ (small $\sigma_l$), $\widehat{\gamma}_l$ may be unstable due to its large variance. 
%and VI$_l$ may be unreliable due to the small GMD value $\eta_l$.
%Thus, we recommend excluding the GMD components with extremely small GMD values before applying the GCV procedure.   
%\end{remark}

%\begin{remark}
%The idea of our GCV procedure is related to the supervised PCA (SPCA) \citep{bair2006prediction}, and it is worth highlighting the difference between them. SPCA performs the standard PCA on the reduced subset of variables, which are marginally associated with the outcome. Our GCV procedure selects the most predictive GMD components after performing the GMD using all variables. Although SPCA performs well for prediction, it cannot be used for inference because the estimator resulted from SPCA is dimension-reduced. 
%\end{remark}

%-----------------------------------------------------------------------
Having selected the most predictive GMD components, we now return to the estimation of regression coefficients. It can be seen from (\ref{gmdr:est}) that our GMDR estimator $\widehat{\ve \beta}_{\mbox{\tiny GMDR}}(\mathcal{I})$ belongs to the following class of estimators:
\begin{align}\label{b:gmd}
  \mathcal{B}_{\mbox{\tiny GMD}} = \{\ve \beta^w \in \mathbb{R}^p: \ve \beta^w = {\bf QVWS}^{-1}{\bf U}^\intercal{\bf Hy}\} 
\end{align}
for some weight matrix $ {\bf W} = \text{diag}(w_1, \ldots, w_K )$, where $ w_j \geq 0$ for $j = 1, \ldots, K.$ 
{In addition to letting ${\bf W}$ depend on the tuning index set $\mathcal{I}$, as done for GMDR, one can instead let ${\bf W}$ depend on a tuning parameter $\eta$.} 
For example, letting $w_j = w_j(\eta) = (\sigma_j^2 + \eta)^{-2}\sigma_j^2$ and ${\bf W}_\eta = \text{diag}(w_1(\eta), \ldots, w_K(\eta))$, one can obtain another estimator in $\mathcal{B}_{\mbox{\tiny GMD}}$ as $\ve \beta^w(\eta) = {\bf QVW_\eta S}^{-1}{\bf U}^\intercal{\bf Hy}$. 
 %If both $\bf H$ and $\bf Q$ are non-singular, 
It can be shown that (see Section 1 of the supplement \citep{wang2023gmdrsupp})
\begin{align}\label{KPR:0}
 \ve \beta^w(\eta) = \text{argmin}_{\ve \beta}\left\{ \left\|{\bf y } - {\bf X}\ve \beta \right\|^2_{\bf H} + \eta\left\|\ve \beta\right\|^2_{{\bf Q}^{-1}} \right\}:=
\widehat{\ve \beta}_{\mbox{\tiny KPR}}(\eta),
\end{align}
where $\widehat{\ve \beta}_{\mbox{\tiny KPR}}(\eta)$ is the estimator obtained from the kernel penalized regression (KPR, \citealp{randolph2018}).
 %of the kernel penalized regression (KPR, \citealp{randolph2018}). 
 %and $\eta > 0$ is a tuning parameter.
 %We provide the derivations of (\ref{KPR:0}) in the Appendix. 
 % KPR is a general penalization framework for two-way structured regression, and covers many existing approaches, including
 %generalized ridge regression, also known as the Tikhonov regularization. Please see, e.g., 
% \cite{hemmerle1975explicit} and \cite{randolph2012structured}. 
Although the motivations behind KPR and GMDR are quite different, (\ref{KPR:0}) implies that they share many features. First, both $\widehat{\ve \beta}_{\mbox{\tiny GMDR}}(\mathcal{I})$ and $\widehat{\ve \beta}_{\mbox{\tiny KPR}}(\eta)$ are in the column space of $\bf Q$, indicating that both estimators incorporate information from $\bf Q$ in similar ways. 
%a subspace of $\mathbb{R}^p$. 
Second, both estimators exert shrinkage effects on the GMD components through the weight matrix $\bf W$. The difference is that $\widehat{\ve \beta}_{\mbox{\tiny GMDR}}(\mathcal{I})$ exerts discrete shrinkage by truncation, nullifying the contribution of the GMD components that are not selected, while $\widehat{\ve \beta}_{\mbox{\tiny KPR}}(\eta)$ exerts a smooth shrinkage effect through the tuning parameter $\eta$ inherently involved in its construction. This connection between GMDR and KPR is similar to that between PCR and the ridge regression (see Section 3.4 in \cite{friedman2001elements} for more details).

%---------------------------------------------------------
\section{The GMD Inference}\label{sec:gmdi}
In this section, we propose a high-dimensional inferential framework for testing $H_0: \beta_l^* = 0$ for $l = 1, \ldots, p$, 
%the parameter of interest $\ve \beta^*$, 
called the GMD inference (GMDI). The proposed framework is based on any arbitrary estimator in the class $\mathcal{B}_{\mbox{\tiny GMD}}$, given in (\ref{b:gmd}). 
%{In particular, the GMDI can be implemented with the KPR estimator (\ref{KPR:0}), which addresses the lack of a valid inference procedure for KPR \citep{randolph2018}. 
%to assess associations between the covariates and outcome; 
%however, such an inferential procedure is lacking in \cite{randolph2018}.
%}
The GMDI procedure and its theoretical properties are presented in Section \ref{sec:gmdi:1}. In Section \ref{sec:gmdi:2}, we provide additional discussions on key assumptions made for GMDI. 
%made in Section \ref{sec:gmdi:1}. 
Section \ref{sec:krv:mirkat} introduces methods to assess the informativeness of the pre-specified $\bf H$ and $\bf Q$ to avoid violations of the assumptions that may impact type-I error and power.
Section \ref{sec:r-GMDI} proposes a robust GMDI procedure to incorporate partially informative structures for controlling type-I error rates and guaranteeing power.

%Recalling that $\mathbb{E}({\bf y}) = {\bf X}\ve \beta^*$, we denote $\ve \epsilon = {\bf y - X}\ve\beta^*$.
Recall from (\ref{main:model:1}) that ${\bf y} = {\bf X}\ve \beta^* + \ve \epsilon$, where $\mathbb{E}[\ve \epsilon \mid {\bf X}] = {\bf 0}_n$ and $\mbox{Cov}(\ve \epsilon \mid {\bf X}) = \ve \Psi$. Letting $\ve \Psi = {\bf L}_\psi^\intercal{\bf L}_\psi$ and
%model (\ref{gmdr:motiv}) as 
% $  {\bf y} = {\bf X}\ve \beta^* + \ve \epsilon $, 
$\ve \epsilon = {\bf {\bf L}_\psi}^\intercal \widetilde{\ve \epsilon}$ with $\widetilde{\ve \epsilon} = (\widetilde{\epsilon}_1, \ldots, \widetilde{\epsilon}_n)^\intercal$, %where ${\bf L}_\psi$ is an unknown matrix that characterizes the covariance of $\ve \epsilon$ amd
%$\widetilde{\ve \epsilon} = (\widetilde{\epsilon}_1, \ldots, \widetilde{\epsilon}_n)^\intercal$. 
we assume that $\widetilde{\epsilon}_1, \ldots, \widetilde{\epsilon}_n$ are $i.i.d.$ {sub-Gaussian} random variables with mean 0 and variance $1$; that is, there exists a constant $C > 0$ such that 
   \begin{align}\label{subgaussian}
   \mathbb{E}\left[ \exp(t\widetilde{\epsilon}_i)\right] \leq \exp \left(\frac{Ct^2}{2} \right) ~~~ \mbox{for all} ~ t \in \mathbb{R} ~ \mbox{and}~ i = 1, \ldots, n.
   \end{align}
%Assumption (\ref{subgaussian}) is less restrictive than the Gaussianity assumed in \cite{buhlmann2013} and \cite{zhangzhang2014}. 
This sub-Gaussianity assumption is only considered for ease of presentation; our results can be easily extended to other distributions with certain tail bounds, such as sub-exponential distributions (Chapter 2, \citealp{wainwright2019high}). 
% Note that $\mbox{Cov}({\bf y}) = {\bf L}_\psi^\intercal {\bf L}_\psi$, which is not necessarily equal to ${\bf I}_n$ (up to a constant).
% This 
% %indicates samples in two-way structured data can be c
% allows potential correlated and heteroscedastic samples. 
%and further highlight major distinctions of the proposed GMDI. 
%among the existing literature in high-dimensional inference.
\subsection{The GMDI Procedure}\label{sec:gmdi:1}
Let $\ve \beta^w = (\beta_1^w, \ldots, \beta_p^w)^\intercal$ be an arbitrary estimator from $\mathcal{B}_{\mbox{\tiny GMD}}$ in (\ref{b:gmd}) with a fixed weight matrix $\bf W$. %Our goal is to test the null hypothesis $H_{0,j}: \beta_j^* = 0$ for some $j = 1, \ldots, p$.
%where $\ve \beta^*$ is defined in (\ref{gmdr:motiv}).
We first note that $\beta_j^w$ can be a biased estimator of $\beta_j^*$. Letting $B_j$ denote the bias of $\beta_j^w$, one can see that
 \[ B_j =  \left({\bf QVW\bf V}^\intercal \ve \beta^*\right)_j - \beta_j^* =  \sum_{m \neq j}\xi_{jm}^w\beta_m^*  + (\xi_{jj}^w - 1)\beta_j^*,\]
 where $\xi_{jm}^w = ({\bf QVWV}^\intercal)_{(j,m)}$, for $j, m = 1, \ldots, p$.
Under $H_{0,j}$, it holds that for any $h_j \in \mathbb{R}$,
\(
B_j = B_j(h_j) := \sum_{m \neq j}\xi_{jm}^w\beta_m^*  + h_j(\xi_{jj}^w - 1)\beta_j^*.
\)
To construct a statistic for testing $H_{0,j}$ based on $\beta^w_j$, we correct the bias $B_j(h_j)$ using a consistent initial estimator of $\ve \beta^*$.  
Denoting by $\ve \beta^{init} = (\beta^{init}_1, \ldots, \beta^{init}_p)^\intercal$ such an initial estimator (to be discussed in detail later in this section), we can estimate $B_{j}(h_j)$ by 
\begin{equation}\label{bc:est}
\widehat{B}_{j}(h_j) = \sum_{m \neq j} \xi_{jm}^w \beta^{init}_m + h_j(\xi_{jj}^w - 1)\beta_j^{init}.
\end{equation}
%We make the following definition. 
Then, our bias-corrected estimator 
%is defined as follows. 
%This yields the following bias-corrected estimator:
%\begin{equation}\label{gmdr:corr}
%\begin{definition}
%Let $\widetilde{\bf y} = {\bf H}^{1/2}{\bf y}$ denote the transformed outcome. 
%Let ${\bf H} = {\bf L_HL_H}^\intercal$ denote the Cholesky decomposition of $\bf H$.
   % For $j = 1, \ldots, p$, the bias-corrected estimator 
    of $\beta_j^*$ is given by
    \begin{equation}\label{bc:def}
    \widehat{\beta}^w_j(h_j) = \beta^w_j - \widehat{B}_{j}(h_j), ~~j=1, \ldots, p.
    %= \sum_{i=1}^n a_{ji}\widetilde{y}_i - \widehat{B}_{j}(h_j),
    \end{equation}
   % where $\widetilde{\bf y} = {\bf H}^{1/2}{\bf y}$  and     $a_{ji} = \left( {\bf QVW\bf S}^{-1}{\bf U}^\intercal {\bf H}^{1/2}\right)_{(j,i)}$, for $i = 1, \ldots, n$.
%\end{equation}%
%\end{definition}

%As pointed out in \cite{buhlmann2013}, we only need to consider the bias under $H_{0,j}$:
%\[B_{j,H_{0,j}} = \sum_{m \neq j} \left({\bf QVW\bf V}^\intercal\right)_{(j,m)} \beta^*_m.\]
%\begin{remark}\label{projectionbias}
{Our bias-correction procedure is motivated by the ridge test proposed in \cite{buhlmann2013} and the grace test proposed in 
\cite{zhao2016}. Note that this is different from the widely used de-biased Lasso \citep{zhangzhang2014, vandegeer2014}, where the key step is to construct a projection direction that satisfies some ``orthogonality property''. However, in the high-dimensional setting, such a projection direction may not exist for highly correlated variables, which is common for two-way structured data. Our bias-correction procedure overcomes this issue since it only requires a consistent initial estimator of $\ve \beta^*$. This comes with the cost of not having an optimal test, which we discuss in detail in the remark below Theorem~\ref{power}.
%A potential shortcoming of our bias-correction procedure is that it would lead to a conservative test and hence the power of the GMDI may be compromised. 
}
% Let $\ve \theta^*$ be the projection of $\ve \beta^*$ onto the column space of $\bf QV$. 
% %It can then be seen that, 
% Similar to \cite{shao2012}, 
% when $p > n$,  $B_j$ can be decomposed as the sum of the estimation and projection biases, given by
% $ \left({\bf QVW\bf V}^\intercal\ve \beta^*\right)_j - \theta_j ~\mbox{and}~ \theta_j -\beta^*_j,$ respectively.
% Unlike \cite{buhlmann2013} that only corrects the projection bias, i.e., $\theta_j -\beta^*_j$, our bias-corrected estimator in (\ref{bc:def}) provides tighter bias correction by correcting both the estimation and projection biases.
%\end{remark}

\begin{remark}\label{remark:3}
The two most intuitive choices of $h_j$ are 0 and 1, which are, respectively, considered in \cite{buhlmann2013} and \cite{zhao2016}.
By considering $h_j = 0$, one only corrects the bias under the null hypothesis, while $h_j = 1$ corrects the general bias regardless of $\beta_j^*$. {While other choices of $h_j$ are mathematically valid, they are practically less meaningful. Thus, we shall limit the following discussion to consider $h_j = 0$ or $1$.}
%Our bias-corrected estimator (\ref{bc:def}) generalizes these two choices. 
%In the following asymptotic analyses, $h_j$ is allowed to be any arbitrary real number,but in our numerical studies, for the fair comparison between our method and existing methods, we only consider $h_j = 0$ and $1$. 
%As suggested in \cite{zhao2016}, one may directly correct $B_j$ instead of $B_{j,H_{0,j}}$. It can be seen that $B_j = B_{j,H_{0,j}} + \left({\bf QVW\bf V}^\intercal\right)_{(j,j)}\beta_j^* - \beta^*_j.$ Using the same initial estimator $\ve \beta^{init}$, the bias-corrected estimator resulted from the correction of $B_j$ is given by
%\begin{equation}\label{pvalue:rel}
%\widecheck{\beta}_j^w = \widehat{\beta}_j^w + \beta^{init}_j - \left({\bf QVW\bf V}^\intercal\right)_{(j,j)}\beta^{init}_j.
%\end{equation}
%One can see that $\widecheck{\beta}^w_j$ involves the initial estimate of $\beta_j^*$, whereas $\widehat{\beta}^w_j$ does not. Under the same set of conditions, we will show that both bias-correction methods yield asymptotically controlled type-I error. However, in finite sample case, there may be differences between $\widehat{\beta}^w_j$ and $\widecheck{\beta}^w_j$ in terms of type-I error rates and power.
%if $\beta^{init}_j$ does not accurately estimate $\beta^*_j$ under the null, then there may be an observed inflation of type-I error associated with $\widecheck{\beta}^w_j.$ 
%In following discussions, we primarily focus on $\widehat{\beta}_j^w$ and then discuss  $\widecheck{\beta}_j^w$ at the end of this section.
\end{remark}

%Letting $a_{ji} = \left( {\bf QVW\bf S}^{-1}{\bf U}^\intercal {\bf H}^{1/2}\right)_{(j,i)}$ for $j = 1, \ldots, p$ and $i = 1, \ldots, n$, 
{Recall that for model (\ref{main:model:1}), $\mbox{Cov}(\ve \epsilon \mid {\bf X}) = \ve \Psi = {\bf L}_\psi^\intercal {\bf L}_\psi$, $\ve \epsilon = {\bf {\bf L}_\psi}^\intercal \widetilde{\ve \epsilon}$, and $\widetilde{\ve \epsilon} = (\widetilde{\epsilon}_1, \ldots, \widetilde{\epsilon}_n)^\intercal$, where
$\widetilde{\epsilon}_1, \ldots, \widetilde{\epsilon}_n$ are $i.i.d.$ {sub-Gaussian} random variables with mean 0 and variance $1$.}
The following result characterizes the asymptotic distribution of $\widehat{\beta}^w_j(h_j)$ as $n \rightarrow \infty$.
\begin{proposition}\label{corr:test}
For $j = 1, \ldots, p$, consider the bias-corrected estimator $\widehat{\beta}^w_j(h_j)$ {with any fixed weight matrix $\bf W$}, given in (\ref{bc:def}). 
Letting  ${\bf A}= {\bf QVW\bf S}^{-1}{\bf U}^\intercal {\bf HL}_\psi^\intercal = \left( a_{ji}\right)_{j = 1, \ldots, p \text{ and } i = 1, \dots, n}$, 
if 
 \begin{align}\label{kappa}
 \lim_{n \rightarrow \infty} \frac{\max_{i = 1, \ldots, n}|a_{ji}|}{\sqrt{\sum_{i=1}^n a_{ji}^2}} = 0,
 \end{align}
{then for $h_j \in \{0, 1\}$}, 
 \begin{align}\label{prop1}
 & \widehat{\beta}^w_j(h_j) = \left((1 - h_j)\xi_{jj}^w + h_j\right)\beta_j^* + \sum_{m \neq j}  \xi_{jm}^w(\beta^*_m - \beta^{init}_m) + h_j(\xi_{jj}^w - 1)(\beta_j^* - \beta_j^{init}) + z_j^w.
 \end{align}
Here, $z_j^w = \sum_{i = 1}^n a_{ji} \widetilde{\epsilon}_i$ and 
\(
{{(\Omega_{jj}^w)^{-1/2}}}{z_j^w} \stackrel{d}{\rightarrow} N(0,1) ~\mbox{as}~ n \rightarrow \infty, 
\)
 %\sim N(0, \sigma_\epsilon^2\Omega_{jj}^w)
where
\( \Omega_{jj}^w = \left( {\bf AA}^\intercal \right)_{(j,j)}.
%\sum_{l = 1}^K \left\{w_l^2 \sigma_l^{-2}\left(\sum_{t = 1}^{p} q_{jt} v_{tl}  \right)^2 \right\}.
\)
\end{proposition}
 %As shown in (\ref{bc:def}), $a_{ji}$ is the weight for the $i$-th transformed outcome $\widetilde{y}_i$ in the definition of $\widehat{\beta}_j^w(h_j)$. Since $\mbox{Cov}(\widetilde{\bf y}) = \sigma_\epsilon^2 {\bf I}_n$, condition (\ref{kappa}) may thus imply that more information can be obtained as more samples are collected. The following corollary of Proposition \ref{corr:test} serves the basis for testing $H_{0,j}: \beta_j^* = 0$.
{
%If $\ve \beta^{init}$ is a consistent estimator of $\ve \beta^*$, i.e., $\|\ve \beta^* - \ve \beta^{init}\|_{1} = o_p(1)$, 
With a consistent initial estimator $\ve \beta^{init}$ that will be discussed later, 
Proposition \ref{corr:test} suggests using
%, and {${\bf L}_\psi$ is known a priori},
%can be consistently estimated, 
%with ${\bf L}_\psi$ replaced by its estimate, 
$\left|\widehat{\beta}^w_j(h_j)\right|$ as an asymptotically valid test statistic for testing $H_{0,j}$. However, its asymptotic variance $\Omega_{jj}^w$ involves the unknown quantity ${\bf L}_{\psi}$, which is not estimable in high-dimensional settings. }
The GMDI overcomes this difficulty by leveraging the relationship between the auxiliary information $\bf H$ and ${\bf L}_\psi$. More specifically, we assume  
%in terms of the spectral norm of ${\bf L}_\psi{\bf HL}_\psi^\intercal$, as stated in the following assumption.
%More specifically, the first assumption states that $\bf H$ informs ${\bf L}_\psi$ in terms of the spectral norm of ${\bf L}_\psi{\bf HL}_\psi^\intercal$.
\begin{itemize}
\item[(A1)] {As $n \rightarrow \infty$, there exists $\sigma^2 > 0$ such that $\|{\bf L}_\psi {\bf H} {\bf L}_\psi^\intercal - \sigma^2{\bf I}_n \|_2 = o(1)$.}
\end{itemize}
An alternative assumption is that  ${\ve  \Psi} = {\bf H}^{-1}$, which, however, is stringent in practice because it requires $\bf H$ to fully capture the unknown covariance $\ve \Psi$.
%In this case, one can check that ${\bf H} = \{\mbox{Cov}({\ve \epsilon})\}^{-1}$.
%indicating that $\bf H$ fully informs the variance component ${\bf L}_\psi$; i.e., $\bf H$ is a function of  ${\bf L}_\psi$. 
Our Assumption (A1) is thus more flexible 
because it only requires ${\bf H}^{-1}$ to be close to $\ve \Psi$ in terms of the spectral norm up to a scale transformation. 
%it allows the error term $\ve \epsilon$ and the row structure $\bf H$ to have different scales thanks to the parameter $\sigma^2$. 
%only requires that $\bf H - \sigma^2\ve \Psi^{-1}$ is close to 0 in terms of the spectral norm for some $\sigma^2 > 0$. 
%in the sense that the spectral norm of ${\bf L}_\psi{\bf HL}_\psi^\intercal$ is close to some unknown constant $\sigma^2$. 
Here, we assume $\bf H$ directly informs $\ve \Psi^{-1}$, not $\ve \Psi$; that is, $\bf H$ informs the {\it conditional} similarities between samples.
% i.e., similarities between samples after removing the effect of other samples.  
It is well-known that such conditional similarities can be characterized by partial correlations, which are closely related to the inverse covariance matrix.  
{In the following discussions, we will first develop the GMDI procedure by assuming $\sigma^2$ is known and then discuss procedures for estimating $\sigma^2$.}

The next proposition states that if Assumption (A1) holds, then $z_j^w$ (see Proposition \ref{corr:test}) converges in distribution to $N(0, R_{jj}^w)$ as $n \rightarrow \infty$, where $R_{jj}^w = \sigma^2\{{\bf QVW}^2{\bf S}^{-2}{\bf V}^\intercal{\bf Q}\}_{(j,j)}$. 
\begin{proposition}\label{var:test}
Consider the $z_j^w$ defined in Proposition \ref{corr:test}. 
Suppose that Assumption (A1) and condition (\ref{kappa}) hold. Then, we have 
\(
{(R_{jj}^w)^{-1/2}}{z_j^w} \stackrel{d}{\rightarrow} N(0,1) ~\mbox{as}~ n \rightarrow \infty.
\)
\end{proposition}

The proofs of Propositions \ref{corr:test} and \ref{var:test} are given in Section 2 of the supplement \citep{wang2023gmdrsupp}.
Next, we elaborate on how to obtain a consistent estimator $\ve \beta^{init}$.
Existing HDI tools that also perform bias correction use 
the {lasso} estimator \citep{tibshirani1996regression} as the initial estimator \citep{buhlmann2013, zhao2016}. Consistency of the {lasso} estimator requires that (i) the true regression coefficient vector is sparse, and (ii) the design matrix satisfies a restricted eigenvalue-type condition \citep{van2009conditions}. However, for two-way structured regression, due to potential strong correlations among variables, the true coefficients may not be sparse, 
%Also, since the variables/samples in two-way structured data may be highly correlated, 
and any restricted eigenvalue-type condition may fail; see \cite{van2009conditions} for more discussions.
%Unlike existing HDI procedures which assume the sparsity of $\ve \beta^*$, 

{As an alternative to those assumptions},
%is sparse (as done by existing HDI procedures), 
we assume that $\ve \beta^*$ is informed by the eigenvectors of $\bf Q$.
%To obtain a $\ve \beta^{init}$ that can yield negligible $\|\ve \beta^{init} - \ve \beta^*\|_1$, 
%We first introduce some additional notations. 
Roughly speaking, we assume that the majority of the signals in $\ve \beta^*$ can be captured by a few eigenvectors of $\bf Q$. More specifically, denoting by ${\bf Q = D \Delta D}^\intercal$ the eigen-decomposition of $\bf Q$ and $\widetilde{\ve \beta}^* = {\bf D}^\intercal\ve \beta^*$, we assume 
\begin{itemize}
\item[(A2)] For some $S_0 \subset \{1, \ldots, p\}$ with $s_0 = |S_0|$, $\left\| \widetilde{\ve \beta}^*_{S_0^c} \right\|_1 \leq \eta_1$, where $S_0^c$ is the complement of $S_0$,
    \(
    \eta_1 = O\left(\sqrt{ n^{-1} s_0 \log p}\right) \mbox{ and } s_0 = o\left\{\big(n/\log p\big)^{r}\right\} \mbox{ for some } r \in (0,1/2)
    \)
    as $n \rightarrow \infty$.
\end{itemize}
Under Assumption (A2), $\|\ve \beta^*\|_{{\bf Q}^{-1}}$, the penalty term of KPR in  (\ref{KPR:0}) is likely to be small. Thus,
Assumption (A2) is in fact aligned with the key idea of KPR.  Indeed, in Section \ref{sec:gmdi:2}, we will show that any estimator from the class $\mathcal{B}_{\mbox{\scriptsize GMD}}$ is less biased if $\ve \beta^*$ satisfies Assumption (A2).

 Our third assumption characterizes how $\bf H$ and $\bf Q$, respectively, inform the row and column structures of the design matrix $\bf X$. As mentioned earlier, any restricted eigenvalue-type condition may break down due to potentially strong correlations in $\bf X$. We assume that $\bf H$ and $\bf Q$ can help decorrelate the rows and columns of $\bf X$, respectively, so that the decorrelated design matrix satisfies a restricted eigenvalue-type condition. More specifically, we assume 
  \begin{itemize}
  \item[(A3)]  For some constants $0 < c_* < c^* < \infty$,   
      %The sparse Riesz condition (SRC) with rank $q^*$ for the transformed design matrix $\check{\bf X} = {\bf H}^{1/2} {\bf X} L_Q$; that is, 
    \[
    c_* \leq \frac{\|\check{\bf X}_A{\bf v}\|^2}{n\|{\bf v}\|^2} \leq c^* ~~~ \mbox{ for any } A \subset \{1, \ldots, p\} \mbox{ with } |A| = q^* \mbox{ and } {\bf v} \in \mathbb{R}^{*},
    \]
    where $\check{\bf X} = {\bf H}^{1/2} {\bf X}{\bf D}\ve \Delta^{1/2}$,
    $q^* \geq M_1^*s_0 + 1$ with $s_0$ given in Assumption (A2) and $M_1^*$ specified in Section 3 of the supplement \citep{wang2023gmdrsupp}. 
   \end{itemize}
   Letting $\check{\ve \Sigma}_A = n^{-1}\check{\bf X}_A^\intercal \check{\bf X}_A$, Assumption (A3) implies that all eigenvalues of $\check{\ve \Sigma}_A$ are inside the interval $[c_*, c^*]$ when the size of $A$ is no greater than $q^*$.
 This assumption is called the sparse Riesz condition \citep{zhang2008sparsity}.
  According to Proposition 1 in \cite{zhang2008sparsity}, if there exists some $q^*$ such that the maximum correlation between the variables in $\check{\bf X}$ is bounded by $\delta/(q^* - 1)$ for some $\delta < 1$, then this condition holds with rank $q^*$, $c_* = 1 - \delta$ and $c^* = 1 + \delta$. 

Under assumptions (A1)--(A3), we introduce the following three-step procedure to construct the bias-corrected estimator of $\beta_j^*$ based on an arbitrary estimator $\beta_j^w$ from $\mathcal{B}_{\text{GMD}}$ for $j = 1, \ldots, p$. 

{
{\bf GMDI bias-correction procedure: }
\begin{itemize}
\item[(B1)] For a fixed tuning parameter $\lambda$, find 
\begin{equation}\label{gmdr:init}
\widetilde{\ve \beta}(\lambda) = \mbox{argmin}_{\ve \beta} \left\{\frac{1}{2} \left\| {\bf y} - {\bf XD}\ve \beta \right\|^2_{\bf H} + \lambda \left\|{\bf \Delta}^{-1/2}\ve \beta \right\|_1 \right\}.
\end{equation}
\item[(B2)] Calculate $\ve \beta^{init} = {\bf D}\widetilde{\ve \beta}(\lambda)$. 
\item[(B3)] For a fixed $h_j \in \{0, 1\}$, let  $\widehat{\beta}_j^w(h_j) = \beta_j^w - \widehat{B}_j(h_j)$ with $\widehat{B}_j(h_j)$ defined in (\ref{bc:est}).
\end{itemize}
}
%Here, $\lambda > 0$ is a tuning parameter.  
We use a weighted $l_1  $ penalty in (\ref{gmdr:init}) with the weights equal to the inverse of the square root of the eigenvalues of $\bf Q$. We will explain the rationale behind this weight choice in Section \ref{sec:gmdi:2}. Also, consistency of $\ve \beta^{\text{init}}$ requires certain conditions on $\lambda$, which will be specified in Theorem \ref{pvalue}.

 Letting $\zeta_j(h_j) = \sum_{m \neq j}  \xi_{jm}^w(\beta^*_m - \beta^{init}_m) + h_j(\xi_{jj}^w - 1)(\beta_j^* - \beta_j^{init})$ for  $j = 1, \ldots, p$ and $\ve \Xi = \text{diag}(\xi_{11}^w, \ldots, \xi_{pp}^w)$, the following result serves as the basis for constructing an asymptotically valid test for $H_{0,j}$
% shows that the type-I error of testing $H_{0,j}$ can be asymptotically controlled using a test statistic 
using the bias-corrected estimator $\widehat{\beta}_j^w(h_j)$ given in (\ref{prop1}). 
  In the following theorems, without loss of generality, we assume that $\bf Q$ is appropriately scaled such that $\|{\bf Q}\|_2 = 1$. 
  \begin{theorem}\label{pvalue}
  Suppose  the
  columns of ${\bf X}$ are standardized such that $\|{\bf Xd}_j\|_{\bf H}^2 = n$, where ${\bf d}_j$ is the $j$-th column of $\bf D$, 
  %where ${\bf d}_j$ is the $j$-th column of $\bf D$, 
  for $j = 1, \ldots, p$. For $\widetilde{\ve \beta}(\lambda)$ in (\ref{gmdr:init}), 
 consider 
 \(
 \lambda = 2\sqrt{2c^* n\log p(1 + c_0) \|{\bf L}_\psi {\bf H L}_\psi^\intercal\|_2}
 \)
 with any $c_0 > 0$, where $c^*$ is given in Assumption (A3).
%  where $C$ is given in (\ref{subgaussian}). 
  For $h_j \in \{0, 1\}$, denote
  \begin{equation}\label{psij:hj}
   \Psi_j(h_j) =\left\|\left[\left({\bf QVW\bf V}^\intercal - (1 - h_j)\ve \Xi - h_j{\bf I}_p \right){\bf D}\right]_{(j,\cdot)}\right\|_\infty\left(\frac{\log p}{n}\right)^{1/2 - r},
   \end{equation}
where for any matrix $\bf M$, ${\bf M}_{(j,\cdot)}$ denotes the $j$-{th} row of $\bf M$.
%where $({\bf QV}W{\bf V}^\intercal{\bf L_Q})_{(j,\cdot)}$ is the $j^{th}$ row of ${\bf QV}W{\bf V}^\intercal{\bf L_Q}$. 
Then,
under condition (\ref{kappa}) and Assumptions (A1)--(A3),  we have
%\begin{equation}\label{test:7}
    %\left| \zeta_j  \right| \lesssim ^ { a s y } \cdot \Psi _ { j }  ~~~ \text{or equivalently} ~~~ 
    $\lim _ { n \rightarrow \infty } \operatorname { Pr } \left( \left| \zeta _ { j }(h_j)  \right| \leq \Psi _ { j }(h_j)  \right) = 1$.
%\end{equation}
 Furthermore, under $H_{0,j}$, for any $\alpha > 0$, 
\begin{equation}\label{bound:4}
    \limsup _{n \rightarrow \infty} \operatorname{Pr}\left(\left|\widehat{\beta}_{j}^{w}(h_j)\right|>\alpha \right) \leq \limsup _{n \rightarrow \infty} \operatorname{Pr}\left(\left|Z_{j}^{w}\right|+ \Psi_{j}(h_j)>\alpha \right),
\end{equation}
where $Z_j^w$ is given in Proposition \ref{corr:test}.
\end{theorem}

Combining Theorem \ref{pvalue} with Proposition \ref{var:test},  we can test $H_{0,j}$ using the asymptotically valid two-sided $p$-value
\begin{equation}\label{bound:5}
    P_{j}^{w}(h_j)=2\left\{1-\Phi\left({(R_{jj}^w)}^{-1/2} {\left\{\left|\widehat{\beta}_{j}^{w}(h_j)\right|-\Psi_{j}(h_j)\right\}_{+}}\right)\right\},
\end{equation}
where $\Phi(\cdot)$ is the cumulative distribution function of the standard normal distribution and $a_+ = \max(a,0)$. 
{
Calculating $P_{j}^{w}(h_j)$ requires obtaining a consistent estimator of $\sigma^2$. 
In this paper, we use the organic lasso \citep{yu2019estimating} to estimate $\sigma^2$ by regressing ${\bf H}^{1/2}{\bf y}$ against $\check{\bf X}$ with $\check{\bf X}$ defined in Assumption (A3), but other approaches, such as the scaled lasso \citep{sun2012scaled}, may also be used. 
%In the numerical applications, we choose $\lambda$ using 10-fold cross validation. 
}

Our next result guarantees the power of GMDI when the size of the true regression coefficient is sufficiently large.
%guarantees the detection of the alternatives using the $p$-value defined in (\ref{bound:5}).
\begin{theorem}\label{power}
Assume the conditions in Theorem \ref{pvalue} hold. For $h_j \in \{0, 1\}$,
if there exists some $0 < \alpha < 1 $ and $0 < \psi < 1$ such that
\begin{equation}\label{power:cond}
\left|\beta_{j}^{*}\right| \geq |(1 - h_j)\xi_{jj}^w + h_j|^{-1} \left(2 \Psi_{j}(h_j) + \left(q_{(1-\alpha / 2)} + q_{(1-\psi / 2)}\right)  \sqrt{R_{jj}^w} ~\right),
\end{equation}
where $\Phi(q_{t}) = t$ for any $t \in (0,1)$ and $\Psi_j(h_j)$ is defined in (\ref{psij:hj}), then
\(\lim_{n \rightarrow \infty} \operatorname{Pr}\left(P_j^w(h_j) \leq \alpha\right) \geq \psi.\)
\end{theorem}
%Theorem \ref{power} guarantees the power of the proposed test when the signal is sufficiently large. 
{It should be noted that condition (\ref{power:cond}) does not hold when $h_j = 0$ and $\xi_{jj}^w = 0$; however, this rarely happens and can be easily checked in advance. In cases where (\ref{power:cond}) is not true, 
%If somehow (\ref{power:cond}) fails to hold, 
$h_j = 1$ can be used. Proofs of Theorems \ref{pvalue} and \ref{power} are provided in Sections 3 and 4 of the supplement \citep{wang2023gmdrsupp}, respectively.}
%-----------------------------------------------------------------
\begin{remark}
{Similar to the ridge test and the Grace test, (\ref{bound:4}) implies that GMDI may be conservative.
%, and hence, the power of the GMDI may be compromised. 
%Nonetheless, we show in Section \ref{sec:simu} that the GMDI outperforms existing methods in terms of the type-I error rate and power in various simulation settings. 
Also, theoretical guarantees of GMDI require using a fixed weight matrix $\bf W$, but in practice, to achieve the optimal prediction performance, $\bf W$ is chosen via cross-validation (e.g., the proposed VI-based approach in Section \ref{sec:gmdr}).
When samples are {\it i.i.d}, one could address this issue by splitting the data into two parts, and then use one part to select $\bf W$ and the other part to perform inference. However, this data-splitting procedure becomes non-trivial, if not impossible, for two-way structured data. {An alternative way is to select top GMD components for GMDR and a fixed tuning parameter for KPR. In these cases, $\bf W$ becomes deterministic, but the prediction/estimation accuracy of GMDR/KPR may be compromised.}
Nonetheless, despite these two potential limitations, we show in Section \ref{sec:simu}, through extensive simulation studies, that the GMDI is more powerful than existing HDI methods with well-controlled type-I error rates.
% After realizing this methodological challenge, we decide to address this issue numerically: In Section \ref{sec:simu}, we first show that the GMDI procedure implemented with $\widehat{\ve \beta}_{\text{GMDR}}(\mathcal{I})$, where $\mathcal{I}$ is selected via the VI-based approach, can still lead to well-controlled type-I error rates and high power. Then, we show that when $\mathcal{I}$ is completely determined by the GMD values, and hence is deterministic, the GMDI also has decent performance despite the less optimal choice of $\mathcal{I}$.
}
\end{remark}

\subsection{On GMDI Assumptions}\label{sec:gmdi:2}
In this section, we discuss Assumptions (A2) and the weighted $l_1$ penalty used in (B1) from the perspective of the bias of any arbitrary estimator
%the class of estimators 
in $\mathcal{B}_{\mbox{\scriptsize GMD}}$.
Recall that the bias of $\ve \beta^w = {\bf QVWS}^{-1}{\bf U}^\intercal{\bf Hy}$ is given by
%\begin{equation}\label{gmdr:bias}
$
   \mbox{Bias}\left(\ve \beta^w \right) =  \mathbb{E}\left(\ve \beta^w \right) - \ve \beta^* = {\bf QVW}{\bf V}^\intercal\ve \beta^* - \ve \beta^*,
$
%\end{equation}
%To interpret $ \mbox{Bias}\left(\ve \beta^w| {\bf X}\right)$, we introduce some additional notations.
%Let ${\bf Q = DR D}^\intercal$ denote the eigen-decomposition of $\bf Q$, where ${\bf D} \in \mathbb{R}^{p \times q}$ and $q = \mbox{rank}(\bf Q)$. 
%It can be seen that ${\bf Q}^- = {\bf DR}^{-1}{\bf D}^\intercal$.
%Let $\mathcal{P}_{\bf Q}$ denote the projection operator from $\mathbb{R}^p$ to the column space of $\bf Q$. It can be seen that $\mathcal{P}_{\bf Q} = {\bf QQ}^{-}$.
%Letting $\mathcal{P}_{\bf Q}$ denotes the orthogonal projection onto the column space of $\bf Q$, 
%As $\ve \beta^w$ lies in the column space of $\bf Q$, 
%for any $\ve \beta^* \in \mathbb{R}^p$, we can write $\ve \beat^* = \mathcal{P}_{Q}\ve \beta^*$
%\(\ve \beta^* = \mathcal{P}_{\bf Q}\ve \beta^* + ({\bf I}_p - \mathcal{P}_{\bf Q})\ve \beta^*.\)
%where $\mathcal{P}_{\bf Q}$ and $({\bf I}_p - \mathcal{P}_{\bf Q})\ve \beta^*$ can be viewed as the portion of $\ve \beta^*$ that can and cannot be explained by $\bf Q$. 
which can be rewritten as
\begin{equation}\label{gmdr:bias2}
    \mbox{Bias}\left( \ve \beta^w \right) = {\bf Q}\left({\bf VW}{\bf V}^\intercal{\bf Q} - {\bf I}_p\right){\bf Q}^{-1}\ve \beta^*. 
    %+ \left({\bf QVW}{\bf V}^\intercal - {\bf I}_p\right)\left({\bf I}_p - \mathcal{P}_{\bf Q}\right)\ve \beta^*.
\end{equation}
Recalling $K = \mbox{rank}({\bf XQX}^\intercal {\bf H})$, we make the following observations from (\ref{gmdr:bias2}).
\begin{itemize}
    \item[(O1)] Suppose $K = p$. Let $\ve \beta^w$ be the GMDR estimator with all GMD components selected. 
    In this case, ${\bf W = I}_p$ and 
    it can be seen that ${\bf VW}{\bf V}^\intercal{\bf Q} = {\bf I}_p$. Thus, $\mbox{Bias}\left(\ve \beta^w \right) = 0$. This demonstrates that in the low-dimensional case ($K = p \leq n$), the GMDR estimator based on all GMD components is an unbiased estimator of $\ve \beta^*$ for any $\ve \beta^* \in \mathbb{R}^p$. 
    
    \item[(O2)] Suppose $K < p$, a common scenario in high-dimensional settings ($n < p$).
      In this case, it can be seen that ${\bf VW}{\bf V}^\intercal{\bf Q} \neq {\bf I}_p$ for any weight matrix $\bf W$. Then, using (\ref{gmdr:bias2}), we have  
   % \begin{equation*}\label{gmdr:bias2.5}
   $
    \left\|\mbox{Bias}\left( \ve \beta^w   \right) \right\|_2 \leq \left\|{\bf Q} \right\|_2 \left\|{\bf VW}{\bf V}^\intercal{\bf Q} - {\bf I}_p \right\|_2 \left\|{\bf Q}^{-1}\ve \beta^* \right\|_2,
    $
   % \end{equation*}
    indicating that $\ve \beta^w$ is less biased if $\left\|{\bf Q}^{-1}\ve \beta^*\right\|_2$ is small. Since ${\bf Q = D\Delta D}^\intercal$,     %Denoting ${\bf R} = \mbox{diag}(\lambda_1, \ldots, \lambda_q)$, where $q = \mbox{rank}({\bf Q})$,
    %let ${\bf Q} = \sum_{j = 1}^q \lambda_j {\bf d}_j {\bf d}_j^\intercal$ denote the eigen-decomposition of $\bf Q$, and 
    %let ${\bf R} = \mbox{diag}(\lambda_1, \ldots, \lambda_q)$ and ${\bf D} = ({\bf d}_1, \ldots, {\bf d}_q)$, and 
    it can be seen that
    \begin{equation}\label{gmdr:bias3}
    \left\|{\bf Q}^{-1}\ve \beta^* \right\|^2_2 =  \sum_{j = 1}^p \delta_j^{-2} \left({\bf d}_j^\intercal \ve \beta^*\right)^2, 
    \end{equation}
    where ${\bf d}_j$ is the $j$-th column of $\bf D$, i.e., the $j$-th eigenvector of $\bf Q$.
   Since $\delta_1 \geq \cdots \geq \delta_p > 0$, (\ref{gmdr:bias3}) implies that $\ve \beta^w$ is less biased if (a) only a few $|{\bf d}_j^\intercal\ve \beta^*|$ are non-zero, or (b) for large $j$ (small $\delta_j$), ${\bf d}_j^\intercal\ve \beta^* = 0$. Thus, Assumption (A2) aligns well with (a) because it indicates that the majority of the signals in $\ve \beta^*$ lie in the space spanned by a few eigenvectors of $\bf Q$. The weighted $l_1$ penalty in (B1) encourages ${\bf d}_j^\intercal \ve \beta^*$ to be 0 for large $j$ and thus aligns with (b). {Note that (b) also aligns with the heuristic of KPR, where $\ve \beta^*$ is assumed to be informed by the top eigenvectors of $\bf Q$.}
   %where ${\bf d}_j$ is the $j$-th column of $\bf D$.
   
  % $\ve \beta^*$ does not lie in the span of the tail eigenvectors of $\bf Q$, then .
   %An example is that $\ve \beta^* = a{\bf d}_1$ for some $a \in \mathbb{R}$, simply meaning that the direction of the first eigenvector of $\bf Q$ completely aligns with the signal direction. Hereafter, we refer to this particular case as that $\bf Q$ fully informs the signal structure.
    
 %   \item[(O3)] Consider the case where $\bf Q$ is singular. In this case, $K < p$ and $\mathcal{P}_{\bf Q} \neq {\bf I}_p$. Hence, besides the bias that is governed by $\|{\bf Q}^- \ve \beta^*\|_2$, as shown in (O2), additional bias arises if some portion of $\ve \beta^*$ lies in the orthogonal complement of the column space of $\bf Q$.
\end{itemize}
%-------------------------------------------------------------
\subsection{{Tests for informative $\bf H$ and $\bf Q$}}\label{sec:krv:mirkat}
{%The proposed GMDR and GMDI require $\bf H$ and $\bf Q$ to be informative for $\bf X$ and $\bf y$, 
%which can be checked in priori.  
{Informative $\bf H$ and $\bf Q$ required by the proposed GMDR and GMDI can be obtained from auxiliary data sources, which are common in omics studies. 
For example, the row and column structures used to construct Fig. \ref{AOAS:intro:fig1}C are estimated from the phylogenetic tree and the metagenomics data, respectively. 
}
However, in practice, one may get uninformative ${\bf H}$ and/or ${\bf Q}$, which
 may impact the type-I error and power.
 
To avoid uninformative external structures, 
we propose to use the kernel RV coefficient (KRV, \citealp{zhan2017fast}) to examine the informativeness of $\bf Q$ with respect to the column structure of $\bf X$. Specifically, 
we define ${\bf Q}_x = {\bf X}^\intercal {\bf X}$ to measure the Euclidean similarities between variables. Since $\bf Q$ is assumed to characterize conditional similarities, 
we test the association between ${\bf Q}^{-1}$ and ${\bf Q}_x$ using 
\[
%\mbox{KRV}_H = \mbox{tr}({\bf H_x H}^{-1}) \mbox{ and }
\mbox{KRV}({\bf Q}_x, {\bf Q}) = \frac{\mbox{tr}(\widetilde{\bf Q}_x \widetilde{\bf Q})}{\sqrt{\mbox{tr}(\widetilde{\bf Q}_x^2) \mbox{tr}(\widetilde{\bf Q}^2)}},
\]
where $\widetilde{\bf Q}_x = \mathbb{C}_p {\bf Q}_x \mathbb{C}_p$ and $\widetilde{\bf Q} = \mathbb{C}_p {\bf Q}^{-1} \mathbb{C}_p$ with 
\begin{equation}\label{gower:center}
\mathbb{C}_p = {\bf I}_p - p^{-1}{\bf 1}_p {\bf 1}_p^\intercal.
\end{equation}
%, and $\mbox{tr}({\bf M})$ denotes the trace of any square matrix ${\bf M}$. 
%Under appropriate normalization of $\bf X, H$, and $\bf Q$, larger values of $\mbox{HSIC}_H$ (or $\mbox{HSIC}_Q)$ indicate more informativeness between $\bf X$ and $\bf H$ (or $\bf Q$). 
A permutation test with a fast approximation of the permutation null distribution is used to test whether the true KRV is 0 \citep{zhan2017fast}. If the permutation $p$-value is less than a pre-selected significance level, say 0.05, then we consider $\bf Q$ an informative column structure.}

{
Similarly, defining ${\bf H}_x = {\bf XX}^\intercal$, one can calculate
$\mbox{KRV}({\bf H}_x, {\bf H})$ with a permutation-based $p$-value.
%its significance determined by permutation.  
As ${\bf H}^{-1}$ captures sample-wise similarities, we also examine the association between ${\bf H}^{-1}$ and the outcome $\bf y$ using the microbiome regression-based association tests (MiRKAT, \citealp{zhao2015testing}). MiRKAT is not performed for $\bf Q$ because the dimension of $\bf Q$ is incompatible with that of ${\bf y}$.
MiRKAT is built upon a mixed-effect model, where the microbiome abundances are modeled as random effects with the covariance matrix  $\tau {\bf H}^{-1}$ for some $\tau \geq 0$. 
%Thus, the statistical significance of the MiRKAT test (i.e., $\tau > 0$) indicates that the sample-wise covariance is equal to ${\bf H}^{-1}$ up to a constant, which shares a similar flavor to Assumption (A1).
Thus, the statistical significance of the MiRKAT test (i.e., $\tau$ > 0) rejects the hypothesis that the sample-wise covariance is substantially distinct from (a constant multiple of)  ${\bf H}^{-1}$. Hence, this test is in the spirit of our Assumption (A1).
If both the KRV and MiRKAT tests are statistically significant, we consider $\bf H$ an informative row structure. }

{
In Section \ref{sec:simu}, 
%we will show that informative row and column structures can lead to more powerful tests, whereas non-informative row or column structures can result in inflated type-I error rates and/or compromised power. 
we will also demonstrate the effectiveness of the KRV and MiRKAT tests in terms of excluding uninformative row and column structures. 
% For $\mbox{HSIC}_H$, we first permute the rows of $\bf X$ and then recompute the value of $\mbox{HSIC}_H$ based on the permuted data matrix. Repeating the permutation many times, we find the proportion of the permuted $\mbox{HSIC}_H$ values that are greater than or equal to the value calculated from the original data. The smaller the proportion is, the more informative $\bf H$ is. If the proportion is below some pre-specfied threshold (e.g., 0.05), one can proceed to use $\bf H$ for GMDR/GMDI.   
% The same permutation procedure can be applied to examine the ``significance" of $\mbox{HSIC}_Q$, except that one should permute the columns of $\bf X$ instead of the rows. 
} 

% This idea, which was also considered in \cite{zhao2016}, can be straightforwardly extended to multiple auxiliary structures, ${\bf Q}_1, \ldots, {\bf Q}_{L}$, for some $L \geq 2$. Let $\ve \pi = (\pi_1, \ldots, \pi_{L})^\intercal$ with $\pi_l \geq 0$ for $l = 1, \ldots, L$ and $\sum_{l=1}^L \pi_l \leq 1$, and one can consider ${\bf Q}(\ve \pi) = \sum_{l=1}^L\pi_l{\bf Q}_l + \left(1-\sum_{l=1}^L\pi_l \right){\bf I}$. {In practice, one can find the $\ve \pi$  that yields the best prediction accuracy.
% %Such a data-driven ${\bf Q}(\ve \pi)$ may be a better approximation to the underlying true column structure than every observed one. 
% We leave the investigation of this data adaptive procedure to future research.}
%------------------------------------------------------------------
\subsection{Robust GMDI with partially informative structures}\label{sec:r-GMDI}
While KRV and MiRKAT can help avoid uninformative row structures, they may identify partially informative structures that do not guarantee valid inference results. 
To address this issue, we propose a robust procedure to determine how much information from the external structures should be incorporated. 
The main idea is to
find a linear combination of a partially informative structure and the identity matrix through an optimal weighting scheme. 
    %to incorporate part of the information from $\bf H$ in Section 3.4 of the revised manuscript. 
    More specifically, consider model (\ref{main:model:1})
    with a partially informative structure $\bf H$ and a fully informative structure $\bf Q$. Without loss of generality, we assume $\|{\bf H}\|_2 = 1$. 
    In this case,  
    we define a weighted structure ${\bf H}(\tau) = \tau {\bf H} + (1 - \tau){\bf I}_n$ 
    %and ${\bf Q}(\tau_Q) = \tau_Q {\bf Q} + (1 - \tau_Q){\bf I}_p$
    with $\tau \in (0, 1)$. 
    %Note that when $\tau = 0$, ${\bf H}(\tau) = {\bf I}$, meaning that ${\bf H}$ is uninformative. 
    %On the other hand, when $\tau = 1$, ${\bf H}(\tau) = {\bf H}$, meaning that $\bf H$ is completely informative. 
    Motivated by the connection among $l_2$-penalized regression, dimension reduction-based regression, and linear mixed models (LMM) \citep{liu2007semiparametric, zhang2015principal, randolph2018}, 
    we find the optimal value of $\tau$ by considering the following LMM:
    \begin{equation}\label{LMM:1}
    {\bf y} = {\bf X}\ve \beta^* + \ve \epsilon, \mbox{ with } \ve \beta^* \sim N_p\left({\bf 0}, c_Q{\bf Q}\right), \ve \epsilon \sim N_n\left({\bf 0}, c_H{\bf H}(\tau)^{-1}\right)
    \end{equation}
    %${\bf y} = {\bf X}\ve \beta^* + \ve \epsilon$,where $\ve \beta^* \sim N_p({\bf 0}, c_Q{\bf Q})$ and $\ve \epsilon \sim N_n({\bf 0}, c_H{\bf H}(\tau)^{-1})$ 
    for some $c_H, c_Q > 0$. 
    Letting $\Omega(\tau) = c_Q{\bf X}{\bf Q}{\bf X}^\intercal + c_H{\bf H}(\tau)^{-1}$, one can see that ${\bf y} \sim N_n({\bf 0}, \Omega( \tau))$, leading to the following likelihood function: 
    \[
    l_n(c_H, c_Q, \tau) = \frac{\exp\left( -1/2 {\bf y}^\intercal \Omega(\tau)^{-1} {\bf y} \right)}{\sqrt{(2\pi)^n |\Omega(\tau)| }}.
    \]
    Since $c_H$ and $c_Q$ are identifiable only up to a scale transformation, we reparametrize the likelihood by defining $\lambda_{HQ} = c_H/c_Q$ and $l_n(\lambda_{HQ}, \tau) = l_n(c_H, c_Q, \tau)$. 
    Then, the maximum likelihood estimate (MLE) of  
    $\lambda_{HQ}$ and $\tau$ is 
    \begin{align*}
        \{\widehat{\lambda}_{HQ}, \widehat{\tau}\} = &~ \argminB \left\{
        {\bf y}^\intercal \Omega(\tau)^{-1} {\bf y} + \log |\Omega( \tau)| \right\}
         \mbox{ subject to } \lambda_{HQ} > 0, 1 > \tau > 0. 
    \end{align*}
    We use an augmented Lagrangian method to solve the optimization problem.
    %as implemented in the \texttt{KPR()} function in the R package \texttt{KPR}[cite]. 
    Having found $\widehat{\tau}$, one can implement GMDI with ${\bf H}(\widehat{\tau})$ and ${\bf Q}$, referred to as the robust GMDI procedure (r-GMDI) hereafter.
    We will demonstrate the effectiveness of  r-GMDI using simulations and real data applications.

%-------------------------------------------------------------------
\section{Simulation Studies}\label{sec:simu}
We conducted two simulation studies, each containing multiple settings, to compare the proposed GMDI with five existing high-dimensional inferential procedures: (i) the low-dimensional projection estimator (LDPE, \citealp{zhangzhang2014}); (ii) the ridge-based high-dimensional inference (Ridge, \citealp{buhlmann2013});
(iii) the de-correlated score test (dscore, \citealp{ning2017}); 
(iv) inference for the graph-constrained estimator (Grace, \citealp{zhao2016}) and (v) the non-sparse high-dimensional inference (ns-hdi, \citealp{zhubradic2018}). 
In the first study, we performed data-driven simulations based on a real microbiome data set.
In the second study, we simulated two-way structured data using a matrix variate normal distribution \citep{gupta2018matrix} with pre-specified row and column covariance matrices. 
We used a two-sided significance level $\alpha = 0.05$ for all tests.

As GMDI works for the entire family of estimators $\mathcal{B}_{\mbox{\tiny GMD}}$, we considered two specific estimators from $\mathcal{B}_{\mbox{\tiny GMD}}$: (i) the proposed GMDR estimator in (\ref{gmdr:est}) and (ii) the KPR estimator in (\ref{KPR:0}). We denote the resulting tests for the GMDR and KPR estimators by GMDI-d and GMDI-k, respectively, because GMDR exerts \textit{discrete} shrinkage effects on GMD components, whereas KPR exerts continuous shrinkage effects through a $\textit{kernel}$ function. For the selection of the index set $\mathcal{I}$ of the GMDR estimator $\widehat{\ve \beta}_{\mbox{\tiny GMDR}}(\mathcal{I})$, GMD components that explain less than 0.1\% of the total variance are excluded because the estimated coefficients corresponding to those components with low variances may be unstable. To see this, recall from (\ref{gammahat}) that $\widehat{\ve \gamma} = \text{argmin}_{\ve \gamma}\left\|{\bf y} - {\ve \Upsilon}\ve \gamma\right\|^2_{\bf H}$. Then, ${\widehat{\gamma}_l} = \sigma_l^{-1}{\bf u}_l^\intercal{\bf Hy}$ and $\mbox{Var}(\widehat{\gamma}_l) = \sigma_\epsilon^2 \sigma_l^{-2}$, for $l = 1, \ldots, K$. This indicates that when the total $R^2$ is low ($\sigma_\epsilon^2$ is relatively large), for large $l$ (small $\sigma_l$), $\widehat{\gamma}_l$ may be unstable due to its large variance. The index set $\mathcal{I}$ is then selected by the proposed GCV procedure based on the remaining GMD components. For the KPR estimator $\widehat{\ve \beta}_{\mbox{\tiny KPR}}(\eta)$, the tuning parameter $\eta$ is selected by 10-fold cross validation. For GMDI, the bias-correction parameter $h_j$ (see Proposition \ref{corr:test}) is set to be 1 for all $j$, as done for Grace; the tuning parameter $\lambda$ in (\ref{gmdr:init}) is set to be $2\sqrt{ 3n \log p}$, and the sparsity parameter $r$ is set to be $0.05$. 
For LDPE and Ridge, we used the implementation in the R package {\tt hdi}, and for the Grace test, we used the implementation in the R package {\tt Grace}. For LDPE, Ridge, and Grace,
the tuning parameters are selected using 10-fold cross-validation. 
%----------------------------------------------------------------

%--------------------------------------------------------------------------
% data-driven simulations
\subsection{Simulation 1}\label{sec:simu:1}
In this study, we performed \emph{data-driven simulations} using data collected as part of the ``Carbohydrates and Related Biomarkers" (CARB) study, conducted between June 2006 and July 2009 at the Fred Hutchinson Cancer Center. CARB was a randomized, controlled, crossover feeding study aimed at evaluating the effects of glycemic load on a variety of biomarkers, such as systemic inflammation, insulin resistance, and adipokines \citep{neuhouser2012low}. Participants were randomized based on body mass index and sex, and fed two controlled diets (randomly assigned order) for 28 days, with a 28-day washout period between diets. 
{The 16S rRNA genus abundance data used here are from 58 participants sampled at each of the three time points, resulting in 174 observations.
%16S rRNA taxon abundance data for 58 participants were obtained  at each of the three timepoints, resulting in 174 observations.  
To classify bacterial taxonomy, sequences were processed using QIIME \citep{caporaso2010qiime}. This processing produced a complete phylogenetic tree with 1054 leaves corresponding to level-7 taxa (species) defined by 97\% similarity and 151 genera (level 6 of the tree). Our simulation used 114 genera after filtering out those that did not appear in at least 30\% of the 174 samples. We correspondingly trimmed the tree back to the genus level with 114 leaves.}
%which will provides the structure defining Q in the analysis.
%and 114 genera were obtained after quality control. 
%Stool and blood samples were collected at baseline (entrance to the study) and after each dietary period. 
%to exploit extrinsic structures from a phylogenetic tree to achieve greater power of signal detection. 
%[add more info on this study] 

Let ${\bf X} \in \mathbb{R}^{174 \times 114}$ be the sample-by-taxon matrix with entries being taxon counts.
 Let $g({\bf z}) = \left( \prod_{k=1}^p z_k \right)^{1/p}$ denote the geometric mean of ${\bf z} = (z_1, \ldots, z_p)^\intercal$. The centered log-ratio (CLR) transformation of ${\bf z}$ is defined as
 \begin{equation}\label{clr}
 \mbox{clr}({\bf z}) = \left[ \log \frac{z_1}{g({\bf z})}, \ldots, \log \frac{z_p}{g({\bf z})} \right].
 \end{equation}
 Since the CLR transformation is not well defined when ${\bf z}$ contains zero entries, we added a pseudo count of 1 to all entries in ${\bf X}$ and then constructed the CLR transformed data matrix $\tilde{\bf X}$ by applying the CLR transformation (\ref{clr}) to each row of $\bf X$.
 %  Denoting by $\bf X$ the sample-by-taxon matrix whose rows are vectors $\{{\bf x}^i\}_{i=1}^n$, 
%   we constructed the centered log-ratio (CLR) transformed \citep{aitchison1982statistical} abundance matrix $\widetilde{\bf X} = [\tilde{\bf x}^1, \ldots, \tilde{\bf x}^n]^\intercal$ by
%  \begin{equation}\label{clr}
%  \tilde{\bf x}^i = \text{clr}({\bf x}^i + {\bf 1}_p) = \left[ \log \left\{ \frac{x_{i1} + 1}{g({\bf x}^i + {\bf 1}_p)} \right\}, \ldots, \log \left\{ \frac{x_{ip} + 1}{g({\bf x}^i + {\bf 1}_p)} \right\} \right],
%  \end{equation}
%  where the constant 
 %of $p = 114$ taxa from $n = 174$ observations. 
%Let $\mathcal{J} = {\bf I}_n - \frac{1}{n} {\bf 1}{\bf 1}^\intercal$, where $\bf 1$ is the $n \times 1$ vector of ones.
The auxiliary row structure was derived from the weighted UniFrac distance between observations  \citep{lozupone2005unifrac}. Specifically, letting $\ve \Delta_{\bf U} \in \mathbb{R}^{n \times n}$ be the squared weighted UniFrac distance matrix, we obtained ${\bf H} = \left( -\frac{1}{2}\mathbb{C}_n\ve \Delta_{\bf U} \mathbb{C}_n \right)^{-1}$, where the centering matrix $\mathbb{C}_n$ is defined in (\ref{gower:center}) in
Section \ref{sec:krv:mirkat}. 
%by ${\bf H} = -\frac{1}{2}\mathbb{C}_n\ve \Delta_{\bf H} \mathbb{C}_n$, where $\ve \Delta_{\bf U}$ is the squared weighted UniFrac distance matrix between observations  \citep{lozupone2005unifrac}.
%{This implies that the sample-wise covariance is informed by $-\frac{1}{2}\mathbb{C}_n\ve \Delta_{\bf U} \mathbb{C}_n$ from both the repeated measures and the UniFrac similarities among samples.} 
{
The column structure ${\bf Q} = \left( -\frac{1}{2}\mathbb{C}_p\ve \Delta_{\bf P} \mathbb{C}_p \right)^{-1}$, where
$\ve \Delta_{\bf P}$
is the squared patristic distance between taxa obtained from the phylogenetic tree. 
}
The KRV test yields a zero $p$-value for $\mbox{KRV}({\bf H}_x, {\bf H})$ and a $p$-value of $0.025$ for $\mbox{KRV}({\bf Q}_x, {\bf Q})$, indicating that ${\bf H}$ and ${\bf Q}$ are informative for the row and column structures of ${\bf X}$, respectively.

Letting ${\bf d}_j$ denote the $j$-th eigenvector of $\bf Q$, we set
{
\(
\ve \beta_0 = 5\sum_{j=1}^{10} \left\{2+3(j-1)\right\}^{-1/2} {\bf d}_j.
\)
%which resides in the space spanned by the $2$nd, $5$th, $8$th, $\ldots$, 29th eigenvectors of $\bf Q$. 
We then defined the true signal $\ve \beta^*$ as a thresholded version of $\ve \beta_0$:
\(
\ve \beta^* = s(\ve \beta_0, 0.1),
\)
where $s(x, \tau)$ is the hard-thresholding operator; i.e., $s(x, \tau) = x\mathbbm{1}(|x| > \tau)$, and the threshold $\tau = 0.1$  was selected so that 81 entries of $\ve \beta^*$ are non-zeros. 
}
The reason why we considered this thresholded parameter as our true parameter is two-fold. First, $\ve \beta^*$ has both zero and non-zero entries, allowing us to evaluate the type-I error rate from testing the zero coefficients and the power from testing the non-zero coefficients. In comparison, all entries of $\ve \beta_0$ are non-zero due to the structure of $\bf Q$.
%, and thus we cannot assess the type-I error rate if we use $\ve \beta_0$ as our true parameter. 
{
Second, 
%while $\ve \beta_0$ is completely informed by the top eigenvectors of $\bf Q$, 
the thresholded parameter $\ve \beta^*$ is no longer fully informed by the top eigenvectors of $\bf Q$, which is more realistic in practice.
%Also, $\ve \beta^*$ enjoys sparsity, which favors many existing HDI methods, such as the LDPE and the Ridge test. 
}
%As we discussed in Setting II of Simulation I, in practice, it is more common that $\bf Q$ partially informs the true parameter, which is the case for $\ve \beta^*$.
% Also, the derived theoretical results still hold when a small portion of the true parameter is not aligned with the top eigenvectors of $\bf Q$ (see Assumptions (A2)).

Let ${\bf H} = \sum_{j = 1}^n \lambda_{j,H} {\bf d}_{j,H}{\bf d}_{j,H}^\intercal$ denote the eigen-decomposition of $\bf H$, where $\lambda_{1, H} \geq \lambda_{2,H} \geq \cdots \geq \lambda_{n,H} > 0$ are the eigenvalues, and ${\bf d}_{1,H}, \ldots, {\bf d}_{n,H}$ are the corresponding eigenvectors. Defining
\(
\ve \Psi = \sum_{j=1}^n \left(\lambda_{j,H}^{-1} + \delta \lambda_{1,H}^{-1}\right) {\bf d}_{j,H} {\bf d}_{j, H}^\intercal,
%{\bf H}^{-1} + \delta \lambda_{1,H}^{-1} {\bf d}_{n,H}{\bf d}_{n,H}^\intercal,
%({\lambda}^{-1}_{n,H} + \delta){\bf d}_{n,H}{\bf d}_{n,H}^\intercal + \sum_{j=1}^{n-1} \lambda^{-1}_{j,H} {\bf d}_{j,H}{\bf d}_{j,H}^\intercal,
\)
we generated $\ve \epsilon$ from a multivariate normal distribution with mean $\bf 0$ and covariance $\ve \Psi$, and simulated the response $\bf y = \widetilde{\bf X}\ve \beta^* + \ve \epsilon$. In this case,
we can calculate
\(
\|{\bf L}_\psi {\bf H} {\bf L}_\psi^\intercal - {\bf I}_n \| = \delta,
\)
where $\ve \Psi = {\bf L}_\psi^\intercal  {\bf L}_\psi$. Thus, according to Assumption (A1), a smaller $\delta$ indicates that $\bf H$ better informs $\ve \Psi$; in particular, $\delta = 0$ means that $\bf H$ fully informs $\ve \Psi$.

\begin{figure}[t!]
    \centering
    \includegraphics[width = 0.7\textwidth, height = 0.3\textheight]{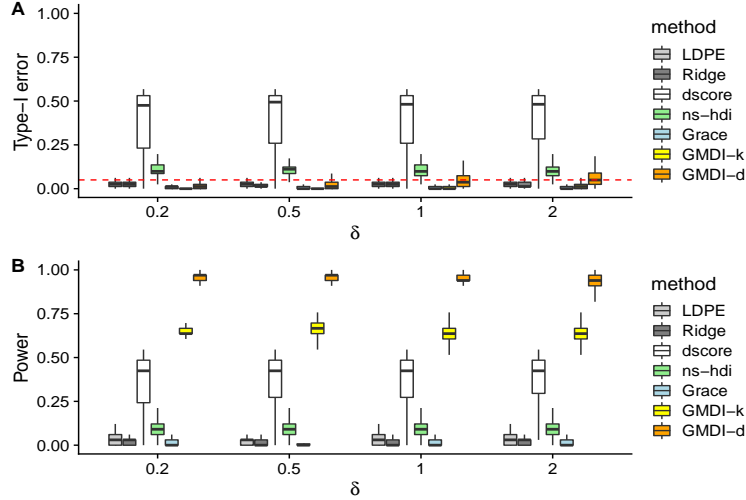}
    \caption{Boxplots of the type-I error (A) and power (B) over 500 replications for Simulation II with $\delta = 0.2, 0.5, 1$, and $2$: {Both GMDI-d and GMDI-k can roughly control the type-I error and have considerably higher power than the existing HDI methods.}
    }
    \label{bact}
\end{figure}

{
 We consider four values of $\delta$: 0.2, 0.5, 1, and 2.}
 %: $0.05, 0.2$ and $0.5$ 
% such that $\|{\bf L}_\psi {\bf H} {\bf L}_\psi^\intercal - {\bf I}_n \|$ equals $0.1, 1$ and $10$, respectively.}
The results are summarized in Fig. \ref{bact}.
{All existing HDI methods fail to differentiate between zero and non-zero entries. More specifically, LDPE, Ridge, and Grace have almost no power, while dscore and ns-hdi have highly inflated type-I error rates. This is because none of these methods can handle correlated samples. The proposed GMDI-k and GMDI-d show better performances. 
 %Inspecting the type-I error of the tests in Fig. \ref{bact}A shows that 
 {
 Both the GMDI-k and GMDI-d show decent power with roughly controlled type-I error rates. 
 %When $\delta = 2$, despite that both GMDI-k and GMDI-d have more inflated type-I error rates, they can still differentiate between zero and non-zero entries. 
}
% meaning that $\bf H$ becomes less informative, GMDI-d has an inflated type-I error rate. Inspecting Fig. 4B shows that both GMDI-k and GMDI-d have considerably higher power than existing methods. Similar to Simulation I, compared to GMDI-d, GMDI-k shows more stringent type-I error rate control and lower power. 
}

%---------------------------------------------------------------
\subsection{Simulation 2}
We considered four settings in this study. In Settings I and II, we considered data with column structures and examined how different choices of $\bf Q$ affect the performance of GMDI and the Grace test. In Setting III, we demonstrated the effectiveness of the KRV and MiRKAT in terms of detecting informative structures. 
In Setting IV, we demonstrated the effectiveness of the proposed robust GMDI in terms of handling partially informative structures. 

%We first introduce some notations that are used throughout the entire section. 
{\bf Setting I}:  We first simulated ${\bf X} \in \mathbb{R}^{200 \times 300}$ from a matrix variate normal distribution with mean $\bf 0$, row covariance ${\bf I}_{200}$ and column covariance $\ve \Sigma$, where
%Let $\bf Q^*$ denote a $300 \times 300$ block diagonal matrix, whose $(i,j)$ entry is given by 
\[ \left(\ve \Sigma^{-1}\right)_{(i,j)} = \begin{cases}
1,&  i = j \\
0.9^{|i - j|}, & i \neq j, i \leq 150, j \leq 150 \\
0.5^{|i - j|}, & i \neq j, i > 150, j > 150 \\
0, & ~~\mbox{otherwise}.
\end{cases}
\]
%\subsection{Setting I}
%  We first simulate ${\bf X} \in \mathbb{R}^{200 \times 300}$ from the matrix-variate model with mean $\bf 0$, row covariance ${\bf I}_{200}$ and column covariance $({\bf Q^*})^{-1}$.
% , which is implemented using the \texttt{R} package ``\texttt{LaplacesDemon}" \citep{LaplacesDemon2018}. 
%Letting ${\bf Q = \Sigma}^{-1}$, we  we defined \( \ve \beta^* = \sum_{j = 1}^{10} j^{-1/2} {\bf f}_j\), where ${\bf f}_j$ the $j$-th eigenvector of $\bf Q$.
Letting ${\bf Q} = {\bf \Sigma}^{-1}$ and denoting by ${\bf f}_j$ the $j$-th eigenvector of $\bf Q$, for $j = 1, \ldots, 300$, we defined \( \ve \beta^* = \sum_{j = 1}^{10} j^{-1/2} {\bf f}_j\), which aligns with the top 10 eigenvectors of $\bf Q$.
The response $\bf y$ was generated according to
%\begin{equation*}\label{simu}
\({\bf y} = {\bf X}\ve \beta^* + \ve \epsilon,\)
%\end{equation*}
where $\ve \epsilon $ was simulated from a multivariate normal distribution with mean $\bf 0$ and covariance $\ve \Psi = \sigma_\epsilon^{2} {\bf I}_{200}$ with  $\sigma_\epsilon^2$ selected to achieve an $R^2$ of $0.4, 0.6$ or $0.8$.  
Our GMDI was implemented using ${\bf H} = {\bf I}_{200}$ and ${\bf Q} = \ve \Sigma$, and $\sigma_\epsilon^2$ was estimated using the organic lasso \citep{yu2019estimating}. 
One can easily check that the pre-specified ${\bf H}$ and ${\bf Q}$ satisfy Assumptions (A1)-(A3). 
%{In this case, one can check that Assumption (A1) holds with ${\bf H} = \ve \Psi^{-1}$, and Assumption (A2) holds with $S_0 = \{1, \ldots, 10\}$.
%Also, with the selected $\bf H$ and $\bf Q$, the transformed data matrix $\check{\bf X}$ (see Assumption (A3) for its definition) follows a matrix variate normal distribution with diagonal row and column covariance matrices, indicating that Assumption (A3) holds with high probability.
%}
% (iii) Assumption (A3) holds with high probability (see pro)
%Assumptions (A1)--(A3) should hold perfectly (Assumption (A3) holds with high probability). Thus, we call this choice of $\bf H$ and $\bf Q$ fully informative.
%Here, $\ve \beta^*$ is generated as
%\[ \ve \beta^* = \sum_{j = 1}^{10} j^{-1/2} {\bf d}_j,\]
%indicating that $\bf Q^*$ informs $\ve \beta^*$ through its top 10 eigenvectors.
%Hence, in this case, it is easy to see that $\bf Q$ fully informs the direction of the signal; that is, $\ve \beta^*$ is the first eigenvector of $\bf Q^*$. 
%We consider two scenarios of $\ve \beta^*$: (S1): $\ve \beta^* = {\bf d}_1$ and (S2): $\ve \beta^* = {\bf d}_2$.
%The $\ve \beta^*_a = a{\bf d}_1$, indicating that the first eigenvector of $\bf Q$ informs the entire $\ve \beta^*_a$. Different $a$'s are considered to yield $R^2$'s of $0.2, 0.4, 0.6$ and $0.8$.  
%is simulated in two steps: we first simulate \[\widetilde{\ve \beta}^*_a = (\underbrace{a, \ldots, a}_{5}, \underbrace{0, \ldots, 0}_{295})^\intercal,\]
%and then set $\ve \beta^*_a = {\bf L_Q}\widetilde{\ve \beta}^*_a$.
By the block diagonal design of $\bf Q$, we know that the first 150 coefficients of ${\bf \ve \beta}^*$ 
are non-zero, while the rest are zero. 
%This indicates that $\ve \beta^*$ is not truly sparse, and thus the LASSO, Ridge, dscore and Grace are lacking theoretical support. 
%In this case, we assume that the observed $\bf Q$ and $\bf H$ are exactly equal to the truth, i.e., $\bf Q = Q^*$ and ${\bf H = I}_{200}$.  
%Our GMDI is implemented with respect to ${\bf H} = {\bf I}_{200}$ and ${\bf Q} = {\bf Q}^*$, and Grace is implemented using ${\bf L} = ({{\bf Q}^*})^{-1}$ (see \citealp{zhao2016} for details).
This enables us to evaluate the power from testing the non-zero coefficients and the type-I error rate from testing the zero coefficients.
\begin{figure}[t!]
    \centering
    \includegraphics[width = 0.7\textwidth]{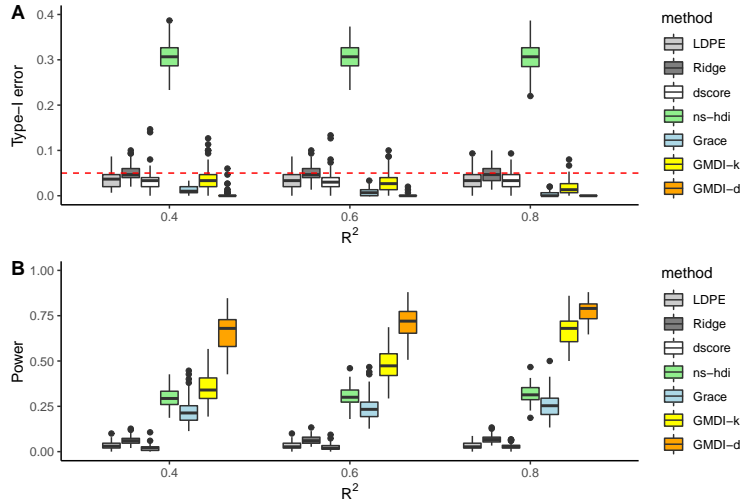}
    \caption{Boxplots of the type-I error (A) and power (B) over 500 replications for Setting I with $R^2 = 0.4, 0.6$ and $0.8$: Both GMDI-d and GMDI-k can control the type-I error, and have considerably higher power than other methods.}
    \label{simu1}
\end{figure}
The results are summarized in Fig. \ref{simu1}.
Figure \ref{simu1}A shows that all methods except ns-hdi can control the type-I error rate. This is likely because in this setting, the precision matrix of the variables, $\ve \Sigma^{-1}$, does not satisfy the row sparsity condition required by ns-hdi. 
%when assumption A1 and A2 hold, in that the median of the observed type-I error rates are not greater than $\alpha = 0.05$. %The GMDR0 yields very conservative control of the type-I error rate. 
The power comparison in Fig. \ref{simu1}B shows that
both GMDI-k and GMDI-d have considerably higher power than the existing methods. 
More specifically, LDPE, Ridge, and dscore have very low power since they completely ignore the column structure of $\bf X$ and $\ve \beta^*$ is not sparse. Because the Grace estimator can incorporate the column structure (Grace is implemented using ${\bf L} = {\ve \Sigma}$; see \citealp{zhao2016} for details), the Grace test gains more power than LDPE and Ridge. However, since the Grace test still requires the sparsity of $\ve \beta^*$, which is not satisfied in this setting, it is not as powerful as GMDI-d or GMDI-k. These results clearly demonstrate the importance of incorporating informative column structures for gaining more power.  As $R^2$ increases, GMDI-k and GMDI-d both yield more stringent control of the type-I error and more power at the same time. GMDI-d has higher power than GMDI-k, especially for low $R^2$ values; this is accompanied by the observation that GMDI-k yields more conservative control of the type-I error rate than GMDI-d. This difference between GMDI-d and GMDI-k may be attributed to 
the fact that GMDI-k shrinks all components, whereas GMDI-d only selects a subset of components without adding any shrinkage effect. 
%{Both GMDI-d and GMDI-k lead to some outliers with relatively low power. This is possibly due to sub-optimal selection of GMD components and tuning parameter for GMDI-d and GMDI-k, respectively.}
%the proposed selection procedure of the GMD components that ensures high prediction accuracy of the GMDR estimator, upon which GMDI-d is based.

{We also evaluated the prediction performance of GMDR by considering two methods for selecting the GMD components: the proposed VI-based procedure and the classical procedure that selects top GMD components, referred to as VI and TOP, respectively. 
Specifically, for each $i = 1, \ldots, 200$, 
we obtained a prediction of $y_i$ based on the leave-one-out cross-validation (LOOCV), denoted by $\widehat{y}_i$.  Letting $\widehat{\bf y} = (\widehat{y}_1, \ldots, \widehat{y}_{200})^\intercal$, we calculated the relative mean squared error (RMSE) according to 
\(
\mbox{RMSE} = {\left\|{\bf y} - \widehat{\bf y}\right\|^2}/{\left\| {\bf y} \right\|^2}.
\)
Table \ref{gmdr:pred} shows the mean and standard deviation (sd) of the  RMSEs over 500 replications. As $R^2$ increases, both methods show better prediction performance. For all values of $R^2$, the VI method shows lower average prediction errors than the TOP method with similar standard deviations, demonstrating the effectiveness of the proposed VI method. 
\begin{table}[!t]
    \caption{The mean (sd) of the RMSEs for the methods of VI and TOP over 500 replications.}
    \centering
    \begin{tabular}{|c|c|c|c|}
        \hline
        \diagbox[width=8em]{Method}{$R^2$} & 0.4 & 0.6 & 0.8 \\ \hline
        VI  &  0.946 &  0.895 &  0.832 \\
            &  (0.082) & (0.088) & (0.091) \\
         \hline
        TOP &  0.967 &  0.934 & 0.859  \\
            & (0.065) & (0.093) & (0.102) \\
        \hline
    \end{tabular}
    \label{gmdr:pred}
\end{table}
}

{\bf Setting II}: In the previous setting, our GMDI was implemented using correctly specified $\bf H$ and $\bf Q$. In practice, the auxiliary structures may be mis-specified.
%one may not have correctly specified structural information.
%it is more common that the auxiliary information is partially informative, meaning that Assumptions (A1)--(A3) hold approximately. 
In this simulation, 
we examined how different choices of $\bf Q$ affect the performance of GMDI and the Grace test. The simulation setting is mostly the same as in Setting I,  
%$\bf y$ is generated according to (\ref{simu}), where $\ve \beta^* = {\bf d}_1$ and $\sigma_\epsilon^2$ is selected to achieve an $R^2$ of 0.4. The main distinction between this setting and Setting I is that the true precision matrix $\bf Q^*$ is no longer observed. 
except that instead of using $\ve \Sigma^{-1}$ as $\bf Q$, we considered two perturbed matrices: ${\bf Q}^{(1)}$ and ${\bf Q}^{(2)}$. Here, ${\bf Q}^{(1)}$ is defined similar to ${\ve \Sigma}^{-1}$, except that ${\bf Q}^{(1)}_{(i,j)} = 0.1^{|i-j|}$ for all $(i,j) \in \{(a,b):  (a - 150)(b-150) < 0 \}$, 
%whose $(i,j)$ entry are, respectively, given by
%\[
%{\bf Q}^{(1)}_{(i,j)} = \begin{cases}
%1,&  i = j \\
%0.9^{|i - j|}, & i \neq j, i \leq 150, j \leq 150 \\
%0.5^{|i - j|}, & i \neq j, i > 150, j > 150 \\
%0.1^{|i-j|}, & ~~\mbox{otherwise}
%\end{cases}
%\]
and { ${\bf Q}^{(2)} = 0.9 \times {\bf I}_{300} + 0.1 \times {\bf 1}_{300} {\bf 1}_{300}^\intercal$.} 
%${\bf Q}^{(2)}_{(i,j)} = 0.6^{|i-j|}$ for all $i, j = 1, \ldots, 300$. 
{Under the significance level 0.05, 
492 out of 500 independent realizations of $\bf X$ lead to statistically significant results for testing $\mbox{KRV}({\bf Q}_x, {\bf Q}^{(1)})$, whereas only five are statistically significant for testing $\mbox{KRV}({\bf Q}_x, {\bf Q}^{(2)})$. This indicates that ${\bf Q}^{(1)}$ is still informative in spite of small perturbations, but ${\bf Q}^{(2)}$ is completely mis-specified. 
}
%One can check that Assumption (A2) roughly holds for ${\bf Q}^{(1)}$, but is completely violated for ${\bf Q}^{(2)}$. Thus, ${\bf Q}^{(1)}$ and ${\bf Q}^{(2)}$ are, respectively, an example of partially informative and non-informative auxiliary column structure.

%While ${\bf Q}^{(2)}$ completely misspecifies the true structure, ${\bf Q}^{(1)}$ roughly captures the structure of ${\bf Q}^*$ with some off-diagonal errors, which is a more common scenario in practice. 
% ${\bf Q}^{(1)}$ exhibit some similarities to ${\bf Q}^*$. 
%It can be seen that ${\bf Q}^{(1)}$ correctly separates two major clusters, while misspecifies the amount of correlation within each cluster. On the contrary, ${\bf Q}^{(2)}$ correctly captures the amount of correlation within each cluster, but the separation between two major clusters becomes less clear due to the error in the off-diagonal blocks. 
\begin{figure}[t!]
    \centering
    \includegraphics[width = 0.55\textwidth, height = 0.25\textheight]{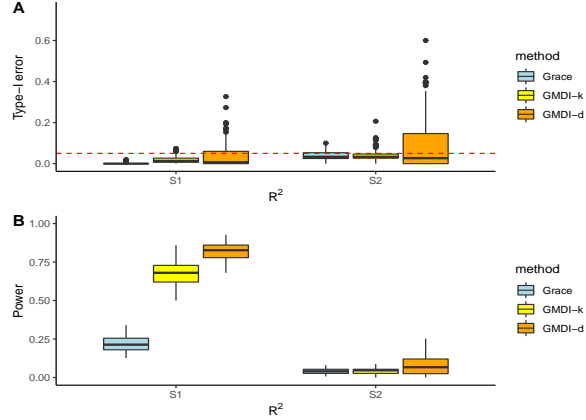}
    \caption{Boxplots of the type-I error (A) and the power (B) over 500 replications for Setting II with $R^2=0.8$.
    The S1 and S2 on the x-axis represent ${\bf Q}^{(1)}$ and ${\bf Q}^{(2)}$ respectively: Both GMDI-d and GMDI-k work well under small perturbations of $\bf Q$. With a completely mis-specified $\bf Q$, GMDI-d and GMDI-k have limited power. This mis-specified choice of $\bf Q$ can be avoided by the KRV test.
  }
    \label{simu2}
\end{figure}
\begin{figure}[t!]
    \centering
    \includegraphics[width = 0.7\textwidth]{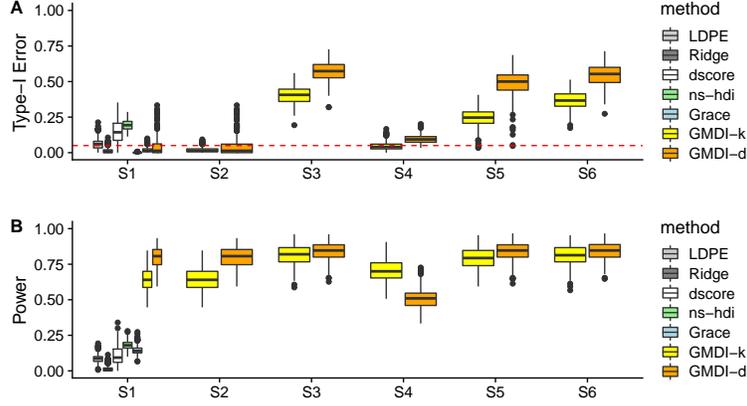}
    \caption{Boxplots of the type-I error (A) and power (B) over 500 replications for Setting III (S1): ${\bf H}^{(1)}$; (S2): ${\bf H}^{(2)}$; (S3): ${\bf H}^{(3)}$; (S4): ${\bf H}^{(4)}$; (S5): ${\bf H}^{(5)}$; (S6): ${\bf H}^{(6)}$. None of the existing HDI methods can differentiate between zero and non-zero entries. 
    GMDI-k and GMDI-d have highly inflated type-I error rates for ${\bf H}^{(3)}, {\bf H}^{(5)}$, and ${\bf H}^{(6)}$, which are uninformative structures according to the KRV and MiRKAT tests in Table \ref{krv}. 
    }
    \label{simu3}
\end{figure}

{The results of Grace, GMDI-d and GMDI-k for $R^2 = 0.8$ are summarized in Fig. \ref{simu2}.}
%, since other methods are not affected by the choice of $\bf Q$.
%Comparing Fig. \ref{simu2} with Fig. \ref{simu1}, 
It can be seen that with small perturbations, i.e., ${\bf Q}^{(1)}$, all three methods can still control the type-I error, and GMDI has higher power than Grace. 
When $\bf Q$ is uninformative, i.e., ${\bf Q}^{(2)}$, none of the three methods can differentiate between zero and non-zero entries. 
%More specifically, GMDI-d suffers from highly inflated type-I error rates, while Grace and GMDI-k have very limited power. 
This simulation also indicates the importance and effectiveness of using the KRV test to examine the informativeness of the column structures before implementing the GMDI.  
{\bf Setting III}: 
Next, we assessed the effectiveness of KRV and MiRKAT in terms of identifying informative sample (row) structures. We simulated $\bf X$ from the matrix variate normal distribution with mean $\bf 0$, row covariance ${\bf R}$ and column covariance $\ve \Sigma$, where $\ve \Sigma$ is defined in Setting I, and
%In this setting, we consider that $\bf X$ has both row and column structures, given by $\bf H^*$ and $\bf Q^*$ respectively.
%We incorporate additional structures among samples, i.e, $\bf H^*$, into Setting I. 
%whose $(i,j)$ entry is given as follows:
\[ \left({\bf R}^{-1}\right)_{(i,j)} = \begin{cases}
1,&  i = j \\
0.9^{|i - j|}, & i \neq j, i \leq 100, j \leq 100 \\
0.5^{|i - j|}, & i \neq j, i > 100, j > 100 \\
0, & ~~\mbox{otherwise}.
\end{cases}
\]
 %the Cholesky decomposition of $\bf H^*$ is given by ${\bf H^* = \bf L_H^*L_H^*}^\intercal$.
%Let ${\bf X} = {\bf H}^{1/2}^{-1}\check{\bf X}{\bf \Delta}^{-1/2}{\bf D}^{T}$, and 
Finally, 
we simulated 
\({\bf y} = 5{\bf X}{\ve \beta}^* + \ve \epsilon,\)
where {$\ve \beta^*$ is the same as defined in Setting I, and $\ve \epsilon$ follows a multivariate normal distribution with mean $\bf 0$ and covariance ${\bf R}$. Here, we multiplied $\ve \beta^*$ by $5$ such that the model $R^2$ is approximately 0.5. 
%$\|{\bf X}\widetilde{\ve \beta}^*\|^2_{\bf R}/\|{\bf y}\|_{\bf R}^2 \approx 60\%$.
}
%$\ve \epsilon \sim N_{200}(0, {\bf R}^{-1})$. 
%and $\sigma_\epsilon^2$ is chosen to achieve an $R^2$ of $0.6$. 
%We also want to see how perturbations of the row structure would affect the performance of the proposed GMDI procedure. 
%As analogous to Setting II, we allow our observed $\bf H$ to be perturbed from ${\bf H}^*$.
%We assume that both $\bf Q^*$ $\bf H^*$ may suffer from misspecification. 
We considered 
{six} choices of $\bf H$: 
{
${\bf H}^{(1)} = {\bf R}^{-1}$, the true row structure; 
%${\bf R}, {\bf H}^{(1)}$ and ${\bf I}_{200}$.
%Here, 
%(S1) ${\bf H} = {\bf H}^*$; (S2) ${\bf H} = {\bf H}^{(1)}$ and (S3) ${\bf H} = {\bf I}_{200}$, where 
${\bf H}^{(2)}$ has slightly mis-specified off-diagonal entries, defined similar to ${\bf H}^{(1)}$ except that $ {\bf H}^{(2)}_{(i,j)} = 0.1^{|i - j|}$ for all $(i,j) \in \{(a,b): (a - 100)(b-100) < 0\}$; 
${\bf H}^{(3)}$ captures the block diagonal structure of the true row correlation but has mis-specified entries:
\[ \left({\bf H}^{(3)}\right)_{(i,j)} = \begin{cases}
1,&  i = j \\
(-0.4)^{|i - j|}, & i \neq j, i \leq 100, j \leq 100 \\
(-0.8)^{|i - j|}, & i \neq j, i > 100, j > 100 \\
0, & ~~\mbox{otherwise}; 
\end{cases}
\]
${\bf H}^{(4)}$ correctly specifies the correlation structure among the first 100 individuals but has a mis-specified structure for the other individuals: \[ \left({\bf H}^{(4)}\right)_{(i,j)} = \begin{cases}
1,&  i = j \\
0.9^{|i - j|}, & i \neq j, i \leq 100, j \leq 100 \\
(-1)^{|i - j|}\times 0.002, & ~~\mbox{otherwise}; 
\end{cases}
\]
${\bf H}^{(5)}$ correctly specifies the correlation structure among the first 20 individuals but has a mis-specified structure for the other individuals: \[ \left({\bf H}^{(5)}\right)_{(i,j)} = \begin{cases}
1,&  i = j \\
0.9^{|i - j|}, & i \neq j, i \leq 20, j \leq 20 \\
(-1)^{|i - j|}\times 0.005, & ~~\mbox{otherwise}; 
\end{cases}
\]
${\bf H}^{(6)}$ has completely mis-specified structures with $\left({\bf H}^{(6)}\right)_{ij} = (-0.5)^{|i-j|}$ for $i, j = 1, \ldots, 200$. 
Here, the coefficients $0.002$ and $0.005$ were selected such that the smallest eigenvalues of ${\bf H}^{(4)}$ and ${\bf H}^{(5)}$ are both around $0.05$. 
%Among 500 independent realizations of $\bf X$, the number of  significant KRV test results for ${\bf H}^{(1)}, {\bf H}^{(2)},$ and ${\bf H}^{(3)}$ are $500, 487$, and $2$, respectively. This is further supported by the MiRKAT results: 
To test whether the six choices of $\bf H$ are informative, we applied the KRV and MiRKAT tests using the \texttt{R} functions \texttt{KRV()} and \texttt{MiRKAT()}, respectively \citep{zhao2015testing}. 
Table \ref{krv} summarizes the proportion of the statistically significant tests based on 500 simulated data sets under the significance level $0.01$.  
As expected, both ${\bf H}^{(1)}$ and ${\bf H}^{(2)}$
are informative, because they are the same as or very close to the true row structure.
%Similarly, ${\bf H}^{(2)}$ is informative, because it resembles the true row correlation matrix despite some mis-specified off-diagonal entries. 
Notably, ${\bf H}^{(4)}$ is also deemed informative in spite of only capturing the true correlations among half of the total individuals. 
 Since ${\bf H}^{(6)}$ is completely mis-specified, its lack of informativeness can be foreseen. However, ${\bf H}^{(3)}$ is also deemed uninformative in spite of correctly capturing the block-diagonal structure of the true correlation matrix. As we will see in Fig. \ref{simu3}, ${\bf H}^{(1)}$, ${\bf H}^{(2)}$, and ${\bf H}^{(4)}$ can lead to well-controlled type-I error rates and decent powers for GMDI-k, whereas ${\bf H}^{(3)}$ and ${\bf H}^{(6)}$ can yield highly inflated type-I error rates. 
 Among the six choices, ${\bf H}^{(5)}$ is the most special because all of the KRV tests are statistically significant but only 21\% of the MiRKAT tests are statistically significant. This indicates that ${\bf H}^{(5)}$ is informative of the row structure of $\bf X$ but not predictive of the outcome $\bf y$. As discussed in Section \ref{sec:krv:mirkat}, such a structure is not regarded as informative and should not be used in practice. Indeed, as we will see in Fig. \ref{simu3}, ${\bf H}^{(5)}$ can lead to inflated type-I error rates. 
}
\begin{table}
\caption{The proportion of statistically significant KRV and MiRKAT tests based on 500 independent data sets under the significance level 0.01 (\%).}
%for ${\bf H}^{(1)}, {\bf H}^{(2)}$, and ${\bf H}^{(3)}$.}
\begin{tabular}{ccccccc}
\hline
~ & ${\bf H}^{(1)}$ & ${\bf H}^{(2)}$ & ${\bf H}^{(3)}$ & ${\bf H}^{(4)}$ & ${\bf H}^{(5)}$ & ${\bf H}^{(6)}$\\
KRV    & 100    & 100  & 0 & 100 & 100 & 0\\
MiRKAT & 100   & 100  & 0 & 100 & 21 & 0\\
\hline
\end{tabular}
\label{krv}
\end{table}

%the $(i,j)$ entry of ${\bf H}^{(1)}$ is given by
%\[ {\bf H}^{(1)}_{(i,j)} = \begin{cases}
%1,&  i = j \\
%0.9^{|i - j|}, & i \neq j, i \leq 100, j \leq 100 \\
%0.5^{|i - j|}, & i \neq j, i > 100, j > 100 \\
%0.1^{|i - j|}, & ~~\mbox{otherwise}.
%\end{cases}
%\]
% One can check that Assumption (A1) holds when ${\bf H} = {\bf R}$ and holds approximately when ${\bf H} = {\bf H}^{(1)}$, but is completely violated when ${\bf H} = {\bf I}_{200}$.
%except that we select $a = 0.2, 0.3,0.4, 0.5, 0.7$ and $0.9$, which yield an $R^2$ of $0.2, 0.4, 0.6, 0.7, 0.8$ and $0.9$ respectively. The implementation of all methods is the same as that in Setting I. 
%The results are summarized in Fig. \ref{simu3}. 
{
We implemented the proposed GMDI-k and GMDI-d with respect to ${\bf Q} = \ve \Sigma^{-1}$ and
all six choices of ${\bf H}$.  We only reported the performance of existing HDI methods under ${\bf H}^{(1)}$, because these methods are not affected by the selection of $\bf H$.
}
All the existing methods fail to differentiate non-zero coefficients from zero ones because they assume $i.i.d$ samples, which is violated in this setting.
In particular, the dscore test can control the type-I error in Setting I, but it fails in this setting where  samples are correlated. 
%Also, the Grace test performs slightly better  thanks to incorporating the column structure, but compared to Fig. \ref{simu1}B, its power is comprised due to ignoring the correlated samples. 
%The left panel of Fig. \ref{simu3}A shows that when $\bf H$ is correctly specified, all methods but dscore and ns-hdi can (asymptotically) control the type-I error. Since Fig. \ref{simu1}A shows that the dscore test can control the type-I error when samples are independent, Fig. 4A thus indicates that the dscore test may not be appropriate for correlated samples. 
{When the selected ${\bf H}$ is correctly specified (e.g., ${\bf H}^{(1)}$) or has small perturbations (e.g., ${\bf H}^{(2)}$), both GMDI-k and GMDI-d show well-controlled type-I error rates, and GMDI-d shows the highest power; this is consistent with Fig. \ref{simu1}. When the selected ${\bf H}$ is partially informative (e.g., ${\bf H}^{(4)}$), GMDI-k shows better controlled type-I error rates and higher power, compared to GMDI-d. 
%This phenomenon is different from that in Fig. \ref{bact}, where GMDI-d outperforms GMDI-k. 
This may indicate GMDI-k is more robust regarding partially informative structures.
When the selected $\bf H$ is uninformative (e.g., ${\bf H}^{(3)}$, ${\bf H}^{(5)}$, and ${\bf H}^{(6)}$), both GMDI-d and GMDI-k suffer from a large inflation of the type-I error rate.
}
%These results demonstrate the importance of  informative row structures.
This simulation demonstrates the effectiveness of using the KRV and MiRKAT tests to avoid uninformative row structures before implementing the GMDI.  
%-------------------------------------------

{\bf Setting IV:} We examine the robust GMDI procedure in Section \ref{sec:r-GMDI} using a simulation study with partially informative row structures. 
Similar to Setting III, we simulated $\bf X$ from the matrix variate normal distribution with mean $\bf 0$, row covariance $\bf R$, and column covariance $\ve \Sigma$, where $\ve \Sigma$ and $\bf R$ are, respectively, defined in Setting I and III. 
We then generated the response ${\bf y} = 10{\bf X}\ve \beta^* + \ve \epsilon$, where $\ve \beta^*$ is defined in Setting I, and $\ve \epsilon \sim N_n({\bf 0}, {\bf R})$. 
By design, the model $R^2$ is approximately 0.85.
According to Assumptions (A1)-(A3), ${\bf R}^{-1}$ and ${\ve \Sigma}^{-1}$ are fully informative row and column structures, respectively. 
We next constructed partially informative row structures by thresholding the tail eigenvalues of $\bf R$. 
Specifically, letting ${\bf R} = \sum_{i=1}^n d_{r,i}{\bf v}_{r,i}{\bf v}_{r,i}^\intercal$ denote the eigen-decomposition of $\bf R$, we defined  %let $k(\theta)$ denote the smallest integer such that $\sum_{i=1}^{k(\theta)} d_{r,i} / \sum_{i=1}^{n} d_{r,i} \geq \theta$, where $d_{r,i}$ is the $i$-th eigenvalue of $\bf R$. 
\(
{\bf H}(\theta) = \sum_{i=1}^{k(\theta)} d_{r,i}^{-1} {\bf v}_{r,i}{\bf v}_{r,i}^\intercal,
\)
where $k(\theta)$ is the smallest integer such that $\sum_{i=1}^{k(\theta)} d_{r,i} / \sum_{i=1}^{n} d_{r,i} \geq \theta$ 
for any given threshold $\theta \in (0, 1]$.
Note that ${\bf H}(1) = {\bf R}^{-1}$, which is a fully informative row structure. 
When $\theta < 1$, ${\bf H}(\theta)$ is partially informative with larger values of $\theta$ leading to a more informative structure. 

\begin{figure}[ht!]
    \centering
    \includegraphics[width = 0.5\textwidth, height = 0.25\textheight]{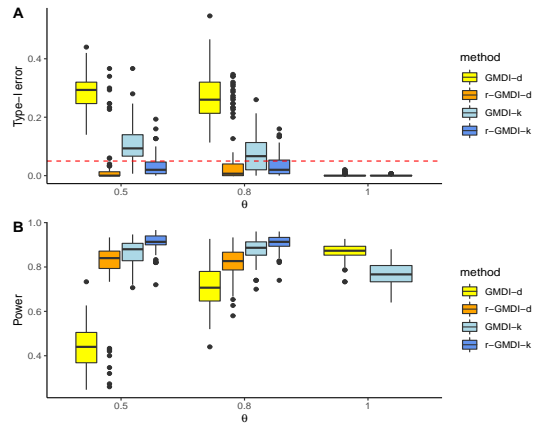}
    \caption{Boxplots of the type-I error (A) and power (B) over 500 replications for Setting IV with $\theta = 0.5, 0.8$, and $1$. The proposed robust GMDI procedure can control the type-I error rate and enhance power when the auxiliary row structure is partially informative. 
    }
    \label{simu4}
\end{figure}

We implemented the GMDI-k and GMDI-d with respect to ${\bf Q} = {\ve \Sigma}^{-1}$ and ${\bf H} = {\bf H}(\theta)$
for $\theta = 0.5, 0.8$, and 1. 
For $\theta = 0.5$ and $0.8$, we also implemented the proposed robust GMDI procedure with ${\bf Q} = {\ve \Sigma}^{-1}$ and  
${\bf H} = \tau{\bf H}(\theta) + (1 - \tau){\bf I}_n$, as described in Section \ref{sec:r-GMDI}.
We denote the robust procedures for GMDI-k and GMDI-d by r-GMDI-k and r-GMDI-d, respectively.  
Figure \ref{simu4} shows the type-I error rates and powers for all the scenarios over 500 independent replications. When $\bf H$ is partially informative, i.e., $\theta = 0.5$ and $0.8$, both GMDI-k and GMDI-d have inflated type-I error rates, and GMDI-d has compromised power. 
GMDI-k shows more robustness to partially informative row structures than GMDI-d, which is consistent with S4 in Setting III. 
The robust GMDI procedures have significantly better performance in terms of better-controlled type-I error rates and enhanced powers.  
In particular, the robust GMDI-k procedure even has higher power than the GMDI with a fully informative row structure. 
This may be due to the fact that GMDI yields conservative $p$-values (the type-I error rates are mostly 0 when $\theta = 1$), which could be alleviated by the robust GMDI procedure. 

%--------------------------------

%-------------------------------------------------------  
\section{Analysis of Gut Microbiome Data}\label{sec:realdata}
In this section, we illustrate the proposed GMDR and GMDI by analyzing a gut microbiome data set from \cite{Yatsunenko2012}, which was described briefly in the Introduction. We kept $p = 149$ bacterial genera that were present in at least 25\% of the  $n = 100$ samples.  
To make the measurements comparable between subjects, we applied the CLR transformation to obtain a $100\times 149$ data matrix $\bf X$, as done in Section \ref{sec:simu:1}. For the column structure, we used the inverse of the $p\times p$ matrix of patristic similarities between the tips of the phylogenetic tree, as in Section \ref{sec:simu:1}. 
%The row structure was derived from counts of EC numbers which specify enzyme-catalyzed reactions based on the metagenomic content. Because evolutionary diversity in bacteria is strongly correlated with metabolic diversity, this provides an informative auxiliary view of sample similarities based on bacterial function. 
The row structure is derived from sample similarities based on Enzyme Commission (EC) numbers which provide insights into the microbial function: counts of EC numbers specify enzyme-catalyzed reactions based on bacterial genomic content. This gives a reasonable auxiliary view of microbial community similarity since evolutionary diversity in bacteria is correlated with metabolic diversity. 
Specifically, these EC data represent counts of $432$ classes of enzymes observed in the bacteria from the same $n=100$ individuals. We applied the CLR transformation to rows of the EC data and centered its columns to have a mean of zero. The resulting $100 \times 432$ matrix is denoted by $\bf Z$. The row similarity structure is then estimated by the inverse Euclidean kernel ${\bf H}=n({\bf ZZ}^\intercal)^{-1}$. For clarity, in this example we denote the row and column structure respectively by ${\bf H}^\text{M}$ and ${\bf Q}^\text{M}$. The KRV test yields zero $p$-values for both ${\bf H}^\text{M}$ and ${\bf Q}^\text{M}$, indicating the informativeness of ${\bf H}^\text{M}$ and ${\bf Q}^\text{M}$. 

%Let ${\bf y} \in \mathbb{R}^n$ denote the vector of outcome  logarithm of age. 

{ We aim to identify bacterial taxa associated with age. The human microbiome is a complex ecosystem and plays a crucial role in the host's development, nutrition, and immunity \citep{belkaid2014role,bana2019microbiome}. The human microbiome has been found to be associated with many age-related diseases, including cancer and neurodegenerative disorders \citep{sepich2021microbiome,fang2020microbiome}. 
 Therefore, identifying age-associated taxa is important for uncovering the mechanistic link between the microbiome and aging. 
%the association between bacterial genus and age. 
In this dataset, the individuals' ages range from 6 months to 53 years. 
As the distribution of age is highly skewed (around 70\% of the samples are below 3 years of age), we use the logarithm of age as our response variable, denoted by $\bf y$. 
{MiRKAT yields a zero $p$-value when testing the association between ${\bf H}^{\mbox{\tiny M} }$ and $\bf y$, indicating the row structure ${\bf H}^{\mbox{\tiny M} }$ also informs the outcome $\bf y$. }
}

Besides the marginal analysis result shown in Fig. \ref{AOAS:intro:fig1}B, it is more interesting to examine the conditional association between each bacterial genus and age, as bacteria do not live independently.
% {
% We first constructed the GMDR estimator based on ${\bf H}^{\mbox{\tiny M} }$ and ${\bf Q}^{\mbox{\tiny M} }$. Specifically, 33 out of the 100 GMD components were excluded for having less than $0.1\%$ of the total variance.  
% %and evaluated the prediction performance of these estimators. 
% {
% %The corresponding GCV statistic of the GMDR estimator is 2.35.  
% For comparison, we also constructed the KPR estimator, which
% had a larger GCV statistic of $6.77$, indicating that the GMDR estimator has better prediction performance. This may be because the GMDR tends to select the most predictive GMD components. 
% }
% } 
{
We implemented r-GMDI-k and r-GMDI-d
to detect conditional associations between bacterial genera and age;  
the estimated robust row structure was ${\bf H}^{\text{R}} = 0.996 {\bf H}^{\text{M}} + 0.004 {\bf I}_n$. This again indicates the strong informativeness of ${\bf H}^{\text{M}}$. 
The GMDI bias-correction procedure yielded a sparse estimator $\widetilde{\ve \beta}(\lambda) \in \mathbb{R}^{149}$ with 13 non-zero entries scattered over the index space $\{1, \ldots, 149\}$ (see (\ref{gmdr:init}) for the definition of $\widetilde{\ve \beta}(\lambda)$). 
This indicates that the initial estimate $\ve \beta^{\text{init}}$ aligns with the space spanned by 13 eigenvectors of ${\bf Q}^{\text{M}}$.
For r-GMDI-d, only 2 out of the 100 GMD components were excluded for having less than $0.1\%$ of the total variance, and 31 GMD components were selected by the proposed VI-based procedure. 
We found that the organic lasso procedure for estimating $\sigma^2$ (see the definition of $\sigma^2$ in Assumption (A1)) is numerically unstable, which may yield slightly different GMDI results for different runs.  
Thus, we fitted the organic lasso 100 times and obtained the average estimate of $\sigma^2$, based on which
% motivated by the idea of stability selection \citep{meinshausen2010stability}, 
we implemented the robust GMDI-d and GMDI-k with ${\bf H}^{\text{R}}$ and ${\bf Q}^{\text{M}}$. 
%and reported microbes that were significant in at least 90\% of those repetitions. 
%By carefully examining the indices of the non-zero entries, 
As a reference, we also implemented the Grace test \citep{zhao2016}, Ridge test \citep{buhlmann2013}, and LDPE \citep{zhangzhang2014}.} 
The dscore and ns-hdi tests were not implemented because they failed to control the type-I error rates in Fig. \ref{bact}.
The Grace test was implemented using ${\bf L} = ({\bf Q}^{\text{M}})^{-1}$. 
% for the KPR1, Ridge and {lasso} estimator, respectively.
%assessed the  significance of individual taxon using the Ridge test \citep{buhlmann2013}, LDPE \citep{zhangzhang2014} and the Grace test \citep{zhao2016}. 
%We display the $p$-values of all 9 methods in Fig. 4B, and the $p$-values corresponding to the marginal associations is included as a reference. 
%It can be seen that by incorporating information from the phylogenetic structure and the EC data, GMDI allows the identification of more genera than methods that do not account for external structures, such as LDPE, and methods that fail to account for these structures in an efficient way, such as the Grace test.
%Scatter plots of the $p$-values for all pairwise comparisons among the four methods are displayed in Fig 5. 
We considered a two-sided significance level $\alpha = 0.05$ for all the tests.

Genera found statistically significantly associated with age after controlling for FDR at 0.1 are reported in Table \ref{annotation}. While the Ridge test results in no statistically significant genera, {the Grace test and LDPE are able to detect 10 and 3 statistically significant microbes, respectively.} 
%GMDI-d2 and GMDI-k2 identify 3 genera as well, but 
  By incorporating the auxiliary information, r-GMDI-d can detect more genera, whereas r-GMDI-k appears conservative. This is consistent with the results in Fig. \ref{simu1}.
  In addition, all the microbes detected by LDPE and r-GMDI-k are also detected by r-GMDI-d; five out of the ten microbes detected by Grace are also detected by r-GMDI-d.
  %which is more concordant with Fig. \ref{AOAS:intro:fig1}B. 
  {However, compared to the vast majority of taxa that are marginally associated with age shown in Fig. \ref{AOAS:intro:fig1}B, the number of statistically significant conditional associations is relatively small. This may indicate that only a limited number of microbes are near the end of the causal pathways linking the microbiome and age. However, without adjusting for potential confounders, we have to be cautious about making any causal interpretations, such as, which microbes are drivers or followers of the detected age-microbiome associations.}

 {The bacterial genus {\it Staphylococcus}, detected by LDPE, Grace, and r-GMDI-d, is known as a dominant microbe in newborns delivered by Cesarean section \citep{dominguez2010delivery}. 
 \textit{Bifidobacterium}, identified by Grace and r-GMDI-d, was highlighted in \cite{Yatsunenko2012} as one of the four dominant baby gut microbes. This may indicate the  informativeness of ${\bf Q}^{\mbox{\tiny M}}$ for identifying age-associated bacterial genera. 
 %Other bacterial genera listed in Table  are known to be associated with various human diseases. For example, 
 {\it Dialister}, detected by Grace and GMDI, has been 
 shown to play a role in age-related diseases, such as obesity and diabetes \citep{xu2020function, gurung2020role}. 
 %The baterial family {\it Enterobacteriaceae}, which the detected genus {\it Enterobacter} belongs to, is known associated with Crohn's Disease \citep{baldelli2021role}. 
 {\it Veillonella}, identified only by the two GMDI methods, is a signature of infant (4-month-old) microbiome and breastfeeding \citep{backhed2015dynamics}. 
 %are  associated with various soft tissue infections \citep{bhatti2000veillonella}. 
 One particular genus only detected by r-GMDI-d, {\it Catenibacterium}, has been shown to be associated with decreased lifetime cardiovascular disease risk \citep{kelly2016gut}.
 %, which may be of interest for future scientific investigation. 
}

%It should be noted, however, that most of our findings are not reported in \cite{Yatsunenko2012}.  Since the results of \cite{Yatsunenko2012} are primarily marginal correlations, our findings based on multivariate models and external structures may be complimentary to theirs. 

%--------------------------------------------------
\begin{table}
%\caption{Genera found to be associated with age when controlling for FDR}
%\footnotesize
%\caption{Genera found to be associated with age when controlling for FDR at 0.1 using LDPE, the Ridge test, the Grace test, GMDI-d1, GMDI-d2, GMDI-d3, GMDI-k1, GMDI-k2 and GMDI-k3.}
\caption[caption]{Genera found to be associated with age after controlling for FDR at 0.1 using the Ridge test,  LDPE,  the Grace test, r-GMDI-d, and r-GMDI-k. The microbes are arranged alphabetically according to their names.
}
\label{annotation}
   %{\begin{varwidth}[t]{\linewidth}Genera found to be associated with age when controlling for FDR at 0.1 using  LDPE,\\ the Ridge test, the Grace test, GMDI-d1, GMDI-d2, GMDI-d3, GMDI-k1, GMDI-k2 and GMDI-k3.\end{varwidth}}
\vspace{5 mm}
\begin{tabular}{ccc}
\hline
      & Genus  & Total \\ \hline
Ridge & {\it (none)}          & 0 \\ \hline
%LDPE & \makecell[c]{\textit{Campylobacter,}  \textit{Atopobium}, \textit{Peptoniphilus}} & 3 \\ \hline
LDPE & \makecell[c]{\textit{Desulfovibrio, Methanobrevibacter, Staphylococcus}
}
& 3 \\ \hline
Grace & \makecell[c]{\textit{Abiotrophia, Bifidobacterium, Desulfovibrio, Dialister, Holdemania},\\
{\it Lachnobacterium, Methanobrevibacter, Roseburia, Rothia, Staphylococcus} \\
}                 & 10 \\ \hline

r-GMDI-d & \makecell[c]{\textit{Adlercreutzia, Anaerococcus, Anaerotruncus, Atopobium, Bifidobacterium,} \\   
\textit{Catenibacterium, Desulfovibrio, Dialister, Diaphorobacter, Erwinia}, \\
\textit{Kocuria, Limnohabitans, Methanobrevibacter, Mitsuokella, Plesiomonas}, \\
{\it Proteus, Pseudobutyrivibrio, Staphylococcus, Streptococcus, Veillonella}\\
%\textit{Tetragenococcus, Treponema, Veillonella} \\
}      & 20 \\ \hline

r-GMDI-k & \makecell[c]{\textit{Atopobium, Dialister, Erwinia, Veillonella}  \\   
%\textit{Tetragenococcus, Treponema, Veillonella} \\
}      & 4 \\ \hline

\end{tabular}
\end{table}

\section{Discussion}\label{sec:discussion}
This paper proposes estimation and inference procedures for high-dimensional linear regression with two-way structured data. For estimation, we develop GMDR which accounts for arbitrary pre-specified two-way structures. 
For inference of individual regression coefficients, we propose GMDI, a general high-dimensional inferential framework for a large family of estimators that include the GMDR estimator.
%define a large family of estimators that include the GMDR estimator and further propose a general high-dimensional inferential framework for any arbitrary estimator in this family, called GMDI. 
Compared to existing high-dimensional inferential tools, GMDI does not require the true regression coefficients to be sparse, it allows dependent and heteroscedastic samples, and it provides flexibility for users to specify relevant auxiliary row and column structures.
%our GMDI can gain more power by allowing non-sparse regression coefficients and efficiently incorporating the information from the pre-specified two-way structures.   

{We have also proposed a robust GMDI procedure for incorporating a partially informative row structure. 
%assessing the informativeness of the pre-specified structures 
In practice, one may have multiple partially informative row structures obtained from different data sources.
We can extend the weighting scheme in Section \ref{sec:r-GMDI} to this scenario.
%incorporate a weighted average of these structures into the GMDR/GMDI framework. 
%Take the column structure $\bf Q$ as an example. 
%This idea, which was also considered in \cite{zhao2016}, can be straightforwardly extended to multiple auxiliary structures, 
Suppose we observe $N-1$ informative structures
${\bf H}_1, \ldots, {\bf H}_{N-1}$, for some $N \geq 2$. Let $\ve \pi = (\pi_1, \ldots, \pi_{N-1})^\intercal$ with $\pi_l \geq 0$ for $l = 1, \ldots, N-1$ and $\sum_{l=1}^{N-1} \pi_l \leq 1$, and one can consider ${\bf H}(\ve \pi) = \sum_{l=1}^{N-1}\pi_l{\bf H}_l + \left(1-\sum_{l=1}^{N-1}\pi_l \right){\bf I}_n$. {
One can find the $\ve \pi$  that yields the best prediction accuracy using a constrained optimization method.
The proposed robust GMDI procedure may be extended to handle a partially informative column structure $\bf Q$. 
However, simply taking a linear combination ${\bf Q}(\tau) = \tau {\bf Q} + (1 - \tau){\bf I}_p$ may not be effective because ${\bf Q}(\tau)$ has the same set of eigenvectors as $\bf Q$ for any $\tau \in (0, 1]$. 
As a result, ${\bf Q}(\tau)$ would not satisfy Assumption (A2) better than ${\bf Q}$. 
%Such a data-driven ${\bf Q}(\ve \pi)$ may be a better approximation to the underlying true column structure than every observed one. 
We leave these extensions as future investigations.}
}

{
The proposed GMDR and GMDI also provide a framework for supervised integrative analysis of multi-view data, i.e., data collected from multiple sources on the same subjects, which are becoming increasingly common in biology, neuroscience, and engineering \citep{li2018survey, zhang2019advancing, mars2021common}
%\citep{NingXing2010, guo2013convex, wang2013multi, li2019integrative}. 
As demonstrated in Section \ref{sec:realdata}, an informative row structure can be obtained from another data view that collects different features on the same set of samples. 
}
Analogously, when there are additional studies addressing the same scientific question, in other words, measuring the same set of variables, one can obtain the column structure from these studies in a similar way.

While the proposed method is motivated and illustrated using microbiome data, our method is generally applicable to arbitrary two-way structured data, such as gene expression data and neuroimaging data. 
%As illustrated in our numerical studies, the proposed GMDI can (asymptotically) control the type-I error rate and have higher power than existing methods even when small perturbations are added to the observed structures, and GMDR can lead to higher prediction accuracy when informative structures are incorporated. 
It is often possible to obtain informative auxiliary row and/or column structures for these data. 
%For example, as illustrated in Section 5,  a phylogenetic tree is often used to characterize the evolutionary relationship among taxa in microbiome studies. 
For the analysis of gene expression data, {one can obtain the gene pathway information from, for example, Kyoto Encyclopedia
of Genes and Genomes (KEGG, \citealp{kanehisa2000post}) or NCI Pathway Interaction Database \citep{schaefer2009pid} and define $\bf Q$ as the graph Laplacian of the gene pathway.
}
For the analysis of neuroimaging data, these structures are often defined as smoothing matrices relevant to the spatial/temporal structure of the images. 
{
Specifically,
for functional MRI (fMRI) studies that measure images of the brain over time, one can take $\bf Q$ to be the graph Laplacian of the graph connecting voxels in the brain \citep{karas2019brain}, and ${\bf H} = \left( h_{ij} \right)$ to be an exponentially smoothing matrix with $h_{ij} = \exp\{-(t_i - t_j)^2/\kappa \}$, where $t_i$ and $t_j$ are the $i$-th and $j$-th time points, respectively, and $\kappa > 0$ is a tuning parameter
\citep{Allen2014}.}  

{
It would be useful to extend GMDR and GMDI to account for confounders. 
Letting ${\bf Z} = \left( {\bf z}_1, \ldots, {\bf z}_n \right)^\intercal$ denote the low-dimensional matrix of confounders, 
we consider the following semi-parametric model 
\begin{equation}\label{confounder:1}
{\bf y} = {g}\left({\bf Z}\right) + {\bf X}\ve \beta^* + \ve \epsilon,
\end{equation}
where ${g}\left({\bf Z}\right) = \left( g({\bf z}_1), \ldots, g({\bf z}_n)  \right)^\intercal$ with $g(\cdot)$ being an unknown smooth function, 
$\mathbb{E}[\ve \epsilon \mid {\bf Z, X}] = {\bf 0}$, and $\mbox{Cov}[\ve \epsilon \mid {\bf Z, X}] = {\ve \Psi}$. 
To extend GMDR and GMDI to model (\ref{confounder:1}), we leverage the connection between model (\ref{confounder:1}) and the following linear mixed model \citep{liu2007semiparametric}:
\begin{align}\label{confounder:2}
    {\bf y} = {\bf g} + {\bf X}\ve \beta^* + \ve \epsilon;
\end{align}
here, ${g}$ is an $n \times 1$ vector of random effects with mean ${\bf 0}$ and covariance $\sigma_z^2 {\bf K_Z}$, where ${\bf K_Z} = \left( K({\bf z}_i, {\bf z}_j) \right)_{i,j = 1, \ldots, n}$ for some pre-specified kernel $K(\cdot, \cdot)$. Popular choices of $K(\cdot, \cdot)$ include the Gaussian kernel $K({\bf z}_i, {\bf z}_j) = \exp\left\{-\|{\bf z}_i - {\bf z}_j\|^2/\rho \right\}$ and the $d$-th polynomial kernel $K({\bf z}_i, {\bf z}_j) = \left( {\bf z}_i^\intercal {\bf z}_j + \rho \right)^d$, where $\rho$ and $d$ are tuning parameters. 
Letting $\ve \delta = {\bf g} + \ve \epsilon$, we obtain the marginal representation of model (\ref{confounder:2}):
\(
%\begin{align}\label{confounder:3}
    {\bf y} = {\bf X}\ve \beta^* + \ve \delta,
%\end{align}
\)
where $\mathbb{E}[\ve \delta \mid {\bf X, Z}] = {\bf 0}$ and $\mbox{Cov}[\ve \delta \mid {\bf X, Z}] = \sigma_z^2 {\bf K_Z} + \ve \Psi$. 
Since 
\(
\left( \sigma_z^2 {\bf K_Z} + \ve \Psi \right)^{-1} = \ve \Psi^{-1} \left( \sigma_z^2 {\bf K_Z} \ve \Psi^{-1} + {\bf I}_n \right)^{-1}, 
\)
one can then implement GMDR and GMDI with the row structure
$\sigma^{-2}{\bf H} \left( \sigma_z^2 {\bf K_Z} \sigma^{-2} {\bf H} + {\bf I}_n \right)^{-1}$ and the column structure $\bf Q$
for some ${\bf H}$ and ${\bf Q}$ satisfying Assumptions (A1)-(A3), where $\sigma^2$ is introduced in Assumption (A1). 
Assuming the normality of $\bf g$ and $\ve \epsilon$, the variance components $\sigma^2$ and $\sigma_z^2$ may be obtained by using penalized maximum likelihood estimation, which we leave for future investigation.
}

{
Finally, it would be interesting to extend GMDR and GMDI to analyze two-way structured categorical predictors.  However, since the GMD incorporates ${\bf H}$ and ${\bf Q}$ through the ${\bf H, Q}$-norm in (\ref{est:1}), which is not suitable for categorical data, the current GMDR and GMDI framework are not directly applicable to categorical data. 
To address this issue, 
an extension of GMD that replaces the ${\bf H, Q}$-norm with some appropriate norm for categorical variables is essential, which could be a fruitful future research direction.  
}

%\section*{Acknowledgement} We thank the editor, A.E., and reviewers for many helpful comments. 
%The authors of this work were supported by NIH Grants R01GM133848, R01GM145772, R01GM129512, R01HL155417, and P50CA228944. 
\begin{acks}[Acknowledgments]
The authors would like to thank the anonymous referees, an Associate
Editor and the Editor for their constructive comments that improved the
quality of this paper.
\end{acks}

\begin{supplement}
\stitle{Proofs of our main theoretical results.} 
\sdescription{This supplementary document provides proofs for eq. (\ref{KPR:0}) and all propositions and theorems in the main paper.}
\end{supplement}

\bibliographystyle{imsart-nameyear} % Style BST file
\bibliography{reference}       % Bibliography file (usually '*.bib')

\begin{thebibliography}{56}
% BibTex style file: imsart-nameyear.bst, 2017-11-03
% Default style options (sort=1,type=nameyear).
% Used options (sort=1,type=nameyear).

\bibitem[\protect\citeauthoryear{Allen, Grosenick and Taylor}{2014}]{Allen2014}
\begin{barticle}[author]
\bauthor{\bsnm{Allen},~\bfnm{Genevera~I.}\binits{G.~I.}},
  \bauthor{\bsnm{Grosenick},~\bfnm{Logan}\binits{L.}} \AND
  \bauthor{\bsnm{Taylor},~\bfnm{Jonathan}\binits{J.}}
(\byear{2014}).
\btitle{A Generalized Least-Square Matrix Decomposition}.
\bjournal{Journal of the American Statistical Association}
\bvolume{109}
\bpages{145-159}.
\bdoi{10.1080/01621459.2013.852978}
\end{barticle}
\endbibitem

\bibitem[\protect\citeauthoryear{B{\"a}ckhed
  et~al.}{2015}]{backhed2015dynamics}
\begin{barticle}[author]
\bauthor{\bsnm{B{\"a}ckhed},~\bfnm{Fredrik}\binits{F.}},
  \bauthor{\bsnm{Roswall},~\bfnm{Josefine}\binits{J.}},
  \bauthor{\bsnm{Peng},~\bfnm{Yangqing}\binits{Y.}},
  \bauthor{\bsnm{Feng},~\bfnm{Qiang}\binits{Q.}},
  \bauthor{\bsnm{Jia},~\bfnm{Huijue}\binits{H.}},
  \bauthor{\bsnm{Kovatcheva-Datchary},~\bfnm{Petia}\binits{P.}},
  \bauthor{\bsnm{Li},~\bfnm{Yin}\binits{Y.}},
  \bauthor{\bsnm{Xia},~\bfnm{Yan}\binits{Y.}},
  \bauthor{\bsnm{Xie},~\bfnm{Hailiang}\binits{H.}},
  \bauthor{\bsnm{Zhong},~\bfnm{Huanzi}\binits{H.}} \betal{et~al.}
(\byear{2015}).
\btitle{Dynamics and stabilization of the human gut microbiome during the first
  year of life}.
\bjournal{Cell host \& microbe}
\bvolume{17}
\bpages{690--703}.
\end{barticle}
\endbibitem

\bibitem[\protect\citeauthoryear{Bana and Cabreiro}{2019}]{bana2019microbiome}
\begin{barticle}[author]
\bauthor{\bsnm{Bana},~\bfnm{Bianca}\binits{B.}} \AND
  \bauthor{\bsnm{Cabreiro},~\bfnm{Filipe}\binits{F.}}
(\byear{2019}).
\btitle{The microbiome and aging}.
\bjournal{Annual Review of Genetics}
\bvolume{53}
\bpages{239--261}.
\end{barticle}
\endbibitem

\bibitem[\protect\citeauthoryear{Belkaid and Hand}{2014}]{belkaid2014role}
\begin{barticle}[author]
\bauthor{\bsnm{Belkaid},~\bfnm{Yasmine}\binits{Y.}} \AND
  \bauthor{\bsnm{Hand},~\bfnm{Timothy~W}\binits{T.~W.}}
(\byear{2014}).
\btitle{Role of the microbiota in immunity and inflammation}.
\bjournal{Cell}
\bvolume{157}
\bpages{121--141}.
\end{barticle}
\endbibitem

\bibitem[\protect\citeauthoryear{Belloni, Chernozhukov and
  Kato}{2015}]{belloni2015uniform}
\begin{barticle}[author]
\bauthor{\bsnm{Belloni},~\bfnm{Alexandre}\binits{A.}},
  \bauthor{\bsnm{Chernozhukov},~\bfnm{Victor}\binits{V.}} \AND
  \bauthor{\bsnm{Kato},~\bfnm{Kengo}\binits{K.}}
(\byear{2015}).
\btitle{Uniform post-selection inference for least absolute deviation
  regression and other Z-estimation problems}.
\bjournal{Biometrika}
\bvolume{102}
\bpages{77--94}.
\end{barticle}
\endbibitem

\bibitem[\protect\citeauthoryear{Benjamini and
  Yekutieli}{2001}]{benjamini2001control}
\begin{barticle}[author]
\bauthor{\bsnm{Benjamini},~\bfnm{Yoav}\binits{Y.}} \AND
  \bauthor{\bsnm{Yekutieli},~\bfnm{Daniel}\binits{D.}}
(\byear{2001}).
\btitle{The control of the false discovery rate in multiple testing under
  dependency}.
\bjournal{The Annals of Statistics}
\bvolume{29}
\bpages{1165--1188}.
\end{barticle}
\endbibitem

\bibitem[\protect\citeauthoryear{B{\"u}hlmann}{2013}]{buhlmann2013}
\begin{barticle}[author]
\bauthor{\bsnm{B{\"u}hlmann},~\bfnm{Peter}\binits{P.}}
(\byear{2013}).
\btitle{Statistical significance in high-dimensional linear models}.
\bjournal{Bernoulli}
\bvolume{19}
\bpages{1212--1242}.
\bdoi{10.3150/12-BEJSP11}
\end{barticle}
\endbibitem

\bibitem[\protect\citeauthoryear{Caporaso et~al.}{2010}]{caporaso2010qiime}
\begin{barticle}[author]
\bauthor{\bsnm{Caporaso},~\bfnm{J~Gregory}\binits{J.~G.}},
  \bauthor{\bsnm{Kuczynski},~\bfnm{Justin}\binits{J.}},
  \bauthor{\bsnm{Stombaugh},~\bfnm{Jesse}\binits{J.}},
  \bauthor{\bsnm{Bittinger},~\bfnm{Kyle}\binits{K.}},
  \bauthor{\bsnm{Bushman},~\bfnm{Frederic~D}\binits{F.~D.}},
  \bauthor{\bsnm{Costello},~\bfnm{Elizabeth~K}\binits{E.~K.}},
  \bauthor{\bsnm{Fierer},~\bfnm{Noah}\binits{N.}},
  \bauthor{\bsnm{Pena},~\bfnm{Antonio~Gonzalez}\binits{A.~G.}},
  \bauthor{\bsnm{Goodrich},~\bfnm{Julia~K}\binits{J.~K.}},
  \bauthor{\bsnm{Gordon},~\bfnm{Jeffrey~I}\binits{J.~I.}} \betal{et~al.}
(\byear{2010}).
\btitle{QIIME allows analysis of high-throughput community sequencing data}.
\bjournal{Nature methods}
\bvolume{7}
\bpages{335--336}.
\end{barticle}
\endbibitem

\bibitem[\protect\citeauthoryear{Cook}{2007}]{cook2007}
\begin{barticle}[author]
\bauthor{\bsnm{Cook},~\bfnm{R.~Dennis}\binits{R.~D.}}
(\byear{2007}).
\btitle{Fisher Lecture: Dimension Reduction in Regression}.
\bjournal{Statistical Science}
\bvolume{22}
\bpages{1--26}.
\bdoi{10.1214/088342306000000682}
\end{barticle}
\endbibitem

\bibitem[\protect\citeauthoryear{Cuesta et~al.}{2015}]{Cuesta2015}
\begin{barticle}[author]
\bauthor{\bsnm{Cuesta},~\bfnm{Sergio~Mart{\'\i}nez}\binits{S.~M.}},
  \bauthor{\bsnm{Rahman},~\bfnm{Syed~Asad}\binits{S.~A.}},
  \bauthor{\bsnm{Furnham},~\bfnm{Nicholas}\binits{N.}} \AND
  \bauthor{\bsnm{Thornton},~\bfnm{Janet~M}\binits{J.~M.}}
(\byear{2015}).
\btitle{The classification and evolution of enzyme function}.
\bjournal{Biophysical journal}
\bvolume{109}
\bpages{1082--1086}.
\end{barticle}
\endbibitem

\bibitem[\protect\citeauthoryear{Dominguez-Bello
  et~al.}{2010}]{dominguez2010delivery}
\begin{barticle}[author]
\bauthor{\bsnm{Dominguez-Bello},~\bfnm{Maria~G}\binits{M.~G.}},
  \bauthor{\bsnm{Costello},~\bfnm{Elizabeth~K}\binits{E.~K.}},
  \bauthor{\bsnm{Contreras},~\bfnm{Monica}\binits{M.}},
  \bauthor{\bsnm{Magris},~\bfnm{Magda}\binits{M.}},
  \bauthor{\bsnm{Hidalgo},~\bfnm{Glida}\binits{G.}},
  \bauthor{\bsnm{Fierer},~\bfnm{Noah}\binits{N.}} \AND
  \bauthor{\bsnm{Knight},~\bfnm{Rob}\binits{R.}}
(\byear{2010}).
\btitle{Delivery mode shapes the acquisition and structure of the initial
  microbiota across multiple body habitats in newborns}.
\bjournal{Proceedings of the National Academy of Sciences}
\bvolume{107}
\bpages{11971--11975}.
\end{barticle}
\endbibitem

\bibitem[\protect\citeauthoryear{Escoufier}{1987}]{escoufier1987duality}
\begin{bincollection}[author]
\bauthor{\bsnm{Escoufier},~\bfnm{Y}\binits{Y.}}
(\byear{1987}).
\btitle{The duality diagram: a means for better practical applications}.
In \bbooktitle{Develoments in Numerical Ecology}
\bpages{139--156}.
\bpublisher{Springer}.
\end{bincollection}
\endbibitem

\bibitem[\protect\citeauthoryear{Escoufier}{2006}]{Escoufier2006}
\begin{binproceedings}[author]
\bauthor{\bsnm{Escoufier},~\bfnm{Yves}\binits{Y.}}
(\byear{2006}).
\btitle{Operator related to a data matrix: a survey}.
In \bbooktitle{Compstat 2006 - Proceedings in Computational Statistics}
(\beditor{\bfnm{Alfredo}\binits{A.}~\bsnm{Rizzi}} \AND
  \beditor{\bfnm{Maurizio}\binits{M.}~\bsnm{Vichi}}, eds.)
\bpages{285--297}.
\bpublisher{Physica-Verlag HD}, \baddress{Heidelberg}.
\end{binproceedings}
\endbibitem

\bibitem[\protect\citeauthoryear{Fang et~al.}{2020}]{fang2020microbiome}
\begin{barticle}[author]
\bauthor{\bsnm{Fang},~\bfnm{P}\binits{P.}},
  \bauthor{\bsnm{Kazmi},~\bfnm{SA}\binits{S.}},
  \bauthor{\bsnm{Jameson},~\bfnm{KG}\binits{K.}} \AND
  \bauthor{\bsnm{Hsiao},~\bfnm{EY}\binits{E.}}
(\byear{2020}).
\btitle{The microbiome as a modifier of neurodegenerative disease risk}.
\bjournal{Cell Host \& Microbe}
\bvolume{28}
\bpages{201--222}.
\end{barticle}
\endbibitem

\bibitem[\protect\citeauthoryear{Friedman et~al.}{2001}]{friedman2001elements}
\begin{bbook}[author]
\bauthor{\bsnm{Friedman},~\bfnm{Jerome}\binits{J.}},
  \bauthor{\bsnm{Hastie},~\bfnm{Trevor}\binits{T.}},
  \bauthor{\bsnm{Tibshirani},~\bfnm{Robert}\binits{R.}} \betal{et~al.}
(\byear{2001}).
\btitle{The elements of statistical learning}
\bvolume{1}.
\bpublisher{Springer series in statistics New York}.
\end{bbook}
\endbibitem

\bibitem[\protect\citeauthoryear{Golub and Van~Loan}{2013}]{golub2013matrix}
\begin{bbook}[author]
\bauthor{\bsnm{Golub},~\bfnm{Gene~H}\binits{G.~H.}} \AND
  \bauthor{\bsnm{Van~Loan},~\bfnm{Charles~F}\binits{C.~F.}}
(\byear{2013}).
\btitle{Matrix computations}.
\bpublisher{JHU press}.
\end{bbook}
\endbibitem

\bibitem[\protect\citeauthoryear{Gupta and Nagar}{2018}]{gupta2018matrix}
\begin{bbook}[author]
\bauthor{\bsnm{Gupta},~\bfnm{Arjun~K}\binits{A.~K.}} \AND
  \bauthor{\bsnm{Nagar},~\bfnm{Daya~K}\binits{D.~K.}}
(\byear{2018}).
\btitle{Matrix variate distributions}.
\bpublisher{Chapman and Hall/CRC}.
\end{bbook}
\endbibitem

\bibitem[\protect\citeauthoryear{Gurung et~al.}{2020}]{gurung2020role}
\begin{barticle}[author]
\bauthor{\bsnm{Gurung},~\bfnm{Manoj}\binits{M.}},
  \bauthor{\bsnm{Li},~\bfnm{Zhipeng}\binits{Z.}},
  \bauthor{\bsnm{You},~\bfnm{Hannah}\binits{H.}},
  \bauthor{\bsnm{Rodrigues},~\bfnm{Richard}\binits{R.}},
  \bauthor{\bsnm{Jump},~\bfnm{Donald~B}\binits{D.~B.}},
  \bauthor{\bsnm{Morgun},~\bfnm{Andrey}\binits{A.}} \AND
  \bauthor{\bsnm{Shulzhenko},~\bfnm{Natalia}\binits{N.}}
(\byear{2020}).
\btitle{Role of gut microbiota in type 2 diabetes pathophysiology}.
\bjournal{EBioMedicine}
\bvolume{51}
\bpages{102590}.
\end{barticle}
\endbibitem

\bibitem[\protect\citeauthoryear{Hullar et~al.}{2021}]{hullar2021associations}
\begin{barticle}[author]
\bauthor{\bsnm{Hullar},~\bfnm{Meredith~AJ}\binits{M.~A.}},
  \bauthor{\bsnm{Jenkins},~\bfnm{Isaac~C}\binits{I.~C.}},
  \bauthor{\bsnm{Randolph},~\bfnm{Timothy~W}\binits{T.~W.}},
  \bauthor{\bsnm{Curtis},~\bfnm{Keith~R}\binits{K.~R.}},
  \bauthor{\bsnm{Monroe},~\bfnm{Kristine~R}\binits{K.~R.}},
  \bauthor{\bsnm{Ernst},~\bfnm{Thomas}\binits{T.}},
  \bauthor{\bsnm{Shepherd},~\bfnm{John~A}\binits{J.~A.}},
  \bauthor{\bsnm{Stram},~\bfnm{Daniel~O}\binits{D.~O.}},
  \bauthor{\bsnm{Cheng},~\bfnm{Iona}\binits{I.}},
  \bauthor{\bsnm{Kristal},~\bfnm{Bruce~S}\binits{B.~S.}} \betal{et~al.}
(\byear{2021}).
\btitle{Associations of the gut microbiome with hepatic adiposity in the
  Multiethnic Cohort Adiposity Phenotype Study}.
\bjournal{Gut microbes}
\bvolume{13}
\bpages{1965463}.
\end{barticle}
\endbibitem

\bibitem[\protect\citeauthoryear{Javanmard and Montanari}{2014a}]{javanmard14a}
\begin{barticle}[author]
\bauthor{\bsnm{Javanmard},~\bfnm{Adel}\binits{A.}} \AND
  \bauthor{\bsnm{Montanari},~\bfnm{Andrea}\binits{A.}}
(\byear{2014}a).
\btitle{Confidence Intervals and Hypothesis Testing for High-Dimensional
  Regression}.
\bjournal{Journal of Machine Learning Research}
\bvolume{15}
\bpages{2869-2909}.
\end{barticle}
\endbibitem

\bibitem[\protect\citeauthoryear{Javanmard and
  Montanari}{2014b}]{javanmard2014hypothesis}
\begin{barticle}[author]
\bauthor{\bsnm{Javanmard},~\bfnm{Adel}\binits{A.}} \AND
  \bauthor{\bsnm{Montanari},~\bfnm{Andrea}\binits{A.}}
(\byear{2014}b).
\btitle{Hypothesis testing in high-dimensional regression under the gaussian
  random design model: Asymptotic theory}.
\bjournal{IEEE Transactions on Information Theory}
\bvolume{60}
\bpages{6522--6554}.
\end{barticle}
\endbibitem

\bibitem[\protect\citeauthoryear{Kanehisa}{2000}]{kanehisa2000post}
\begin{bbook}[author]
\bauthor{\bsnm{Kanehisa},~\bfnm{Minoru}\binits{M.}}
(\byear{2000}).
\btitle{Post-genome informatics}.
\bpublisher{OUP Oxford}.
\end{bbook}
\endbibitem

\bibitem[\protect\citeauthoryear{Karas et~al.}{2019}]{karas2019brain}
\begin{barticle}[author]
\bauthor{\bsnm{Karas},~\bfnm{Marta}\binits{M.}},
  \bauthor{\bsnm{Brzyski},~\bfnm{Damian}\binits{D.}},
  \bauthor{\bsnm{Dzemidzic},~\bfnm{Mario}\binits{M.}},
  \bauthor{\bsnm{Go{\~n}i},~\bfnm{Joaqu{\'\i}n}\binits{J.}},
  \bauthor{\bsnm{Kareken},~\bfnm{David~A}\binits{D.~A.}},
  \bauthor{\bsnm{Randolph},~\bfnm{Timothy~W}\binits{T.~W.}} \AND
  \bauthor{\bsnm{Harezlak},~\bfnm{Jaroslaw}\binits{J.}}
(\byear{2019}).
\btitle{Brain connectivity-informed regularization methods for regression}.
\bjournal{Statistics in Biosciences}
\bvolume{11}
\bpages{47--90}.
\end{barticle}
\endbibitem

\bibitem[\protect\citeauthoryear{Kelly et~al.}{2016}]{kelly2016gut}
\begin{barticle}[author]
\bauthor{\bsnm{Kelly},~\bfnm{Tanika~N}\binits{T.~N.}},
  \bauthor{\bsnm{Bazzano},~\bfnm{Lydia~A}\binits{L.~A.}},
  \bauthor{\bsnm{Ajami},~\bfnm{Nadim~J}\binits{N.~J.}},
  \bauthor{\bsnm{He},~\bfnm{Hua}\binits{H.}},
  \bauthor{\bsnm{Zhao},~\bfnm{Jinying}\binits{J.}},
  \bauthor{\bsnm{Petrosino},~\bfnm{Joseph~F}\binits{J.~F.}},
  \bauthor{\bsnm{Correa},~\bfnm{Adolfo}\binits{A.}} \AND
  \bauthor{\bsnm{He},~\bfnm{Jiang}\binits{J.}}
(\byear{2016}).
\btitle{Gut microbiome associates with lifetime cardiovascular disease risk
  profile among bogalusa heart study participants}.
\bjournal{Circulation research}
\bvolume{119}
\bpages{956--964}.
\end{barticle}
\endbibitem

\bibitem[\protect\citeauthoryear{Li, Cai and Li}{2021}]{li2021inference}
\begin{barticle}[author]
\bauthor{\bsnm{Li},~\bfnm{Sai}\binits{S.}},
  \bauthor{\bsnm{Cai},~\bfnm{T~Tony}\binits{T.~T.}} \AND
  \bauthor{\bsnm{Li},~\bfnm{Hongzhe}\binits{H.}}
(\byear{2021}).
\btitle{Inference for high-dimensional linear mixed-effects models: A
  quasi-likelihood approach}.
\bjournal{Journal of the American Statistical Association}
\bpages{1--12}.
\end{barticle}
\endbibitem

\bibitem[\protect\citeauthoryear{Li, Yang and Zhang}{2018}]{li2018survey}
\begin{barticle}[author]
\bauthor{\bsnm{Li},~\bfnm{Yingming}\binits{Y.}},
  \bauthor{\bsnm{Yang},~\bfnm{Ming}\binits{M.}} \AND
  \bauthor{\bsnm{Zhang},~\bfnm{Zhongfei}\binits{Z.}}
(\byear{2018}).
\btitle{A survey of multi-view representation learning}.
\bjournal{IEEE transactions on knowledge and data engineering}
\bvolume{31}
\bpages{1863--1883}.
\end{barticle}
\endbibitem

\bibitem[\protect\citeauthoryear{Liu, Lin and
  Ghosh}{2007}]{liu2007semiparametric}
\begin{barticle}[author]
\bauthor{\bsnm{Liu},~\bfnm{Dawei}\binits{D.}},
  \bauthor{\bsnm{Lin},~\bfnm{Xihong}\binits{X.}} \AND
  \bauthor{\bsnm{Ghosh},~\bfnm{Debashis}\binits{D.}}
(\byear{2007}).
\btitle{Semiparametric regression of multidimensional genetic pathway data:
  Least-squares kernel machines and linear mixed models}.
\bjournal{Biometrics}
\bvolume{63}
\bpages{1079--1088}.
\end{barticle}
\endbibitem

\bibitem[\protect\citeauthoryear{Lozupone and
  Knight}{2005}]{lozupone2005unifrac}
\begin{barticle}[author]
\bauthor{\bsnm{Lozupone},~\bfnm{Catherine}\binits{C.}} \AND
  \bauthor{\bsnm{Knight},~\bfnm{Rob}\binits{R.}}
(\byear{2005}).
\btitle{UniFrac: a new phylogenetic method for comparing microbial
  communities}.
\bjournal{Applied and environmental microbiology}
\bvolume{71}
\bpages{8228--8235}.
\end{barticle}
\endbibitem

\bibitem[\protect\citeauthoryear{Mars, Jbabdi and
  Rushworth}{2021}]{mars2021common}
\begin{barticle}[author]
\bauthor{\bsnm{Mars},~\bfnm{Rogier~B}\binits{R.~B.}},
  \bauthor{\bsnm{Jbabdi},~\bfnm{Saad}\binits{S.}} \AND
  \bauthor{\bsnm{Rushworth},~\bfnm{Matthew~FS}\binits{M.~F.}}
(\byear{2021}).
\btitle{A common space approach to comparative neuroscience}.
\bjournal{Annual Review of Neuroscience}
\bvolume{44}.
\end{barticle}
\endbibitem

\bibitem[\protect\citeauthoryear{Mitra and Zhang}{2016}]{mitra2016benefit}
\begin{barticle}[author]
\bauthor{\bsnm{Mitra},~\bfnm{Ritwik}\binits{R.}} \AND
  \bauthor{\bsnm{Zhang},~\bfnm{Cun-Hui}\binits{C.-H.}}
(\byear{2016}).
\btitle{The benefit of group sparsity in group inference with de-biased scaled
  group Lasso}.
\bjournal{Electronic Journal of Statistics}
\bvolume{10}
\bpages{1829--1873}.
\end{barticle}
\endbibitem

\bibitem[\protect\citeauthoryear{Neuhouser et~al.}{2012}]{neuhouser2012low}
\begin{barticle}[author]
\bauthor{\bsnm{Neuhouser},~\bfnm{Marian~L}\binits{M.~L.}},
  \bauthor{\bsnm{Schwarz},~\bfnm{Yvonne}\binits{Y.}},
  \bauthor{\bsnm{Wang},~\bfnm{Chiachi}\binits{C.}},
  \bauthor{\bsnm{Breymeyer},~\bfnm{Kara}\binits{K.}},
  \bauthor{\bsnm{Coronado},~\bfnm{Gloria}\binits{G.}},
  \bauthor{\bsnm{Wang},~\bfnm{Chin-Yun}\binits{C.-Y.}},
  \bauthor{\bsnm{Noar},~\bfnm{Karen}\binits{K.}},
  \bauthor{\bsnm{Song},~\bfnm{Xiaoling}\binits{X.}} \AND
  \bauthor{\bsnm{Lampe},~\bfnm{Johanna~W}\binits{J.~W.}}
(\byear{2012}).
\btitle{A low-glycemic load diet reduces serum C-reactive protein and modestly
  increases adiponectin in overweight and obese adults}.
\bjournal{The Journal of nutrition}
\bvolume{142}
\bpages{369--374}.
\end{barticle}
\endbibitem

\bibitem[\protect\citeauthoryear{Ning and Liu}{2017}]{ning2017}
\begin{barticle}[author]
\bauthor{\bsnm{Ning},~\bfnm{Yang}\binits{Y.}} \AND
  \bauthor{\bsnm{Liu},~\bfnm{Han}\binits{H.}}
(\byear{2017}).
\btitle{A general theory of hypothesis tests and confidence regions for sparse
  high dimensional models}.
\bjournal{The Annals of Statistics}
\bvolume{45}
\bpages{158--195}.
\bdoi{10.1214/16-AOS1448}
\end{barticle}
\endbibitem

\bibitem[\protect\citeauthoryear{Randolph et~al.}{2018}]{randolph2018}
\begin{barticle}[author]
\bauthor{\bsnm{Randolph},~\bfnm{T.~W.}\binits{T.~W.}},
  \bauthor{\bsnm{Zhao},~\bfnm{Sen}\binits{S.}},
  \bauthor{\bsnm{Copeland},~\bfnm{Wade}\binits{W.}},
  \bauthor{\bsnm{Hullar},~\bfnm{Meredith}\binits{M.}} \AND
  \bauthor{\bsnm{Shojaie},~\bfnm{Ali}\binits{A.}}
(\byear{2018}).
\btitle{Kernel-penalized regression for analysis of microbiome data}.
\bjournal{The Annals of Applied Statistics}
\bvolume{12}
\bpages{540--566}.
\bdoi{10.1214/17-AOAS1102}
\end{barticle}
\endbibitem

\bibitem[\protect\citeauthoryear{Schaefer et~al.}{2009}]{schaefer2009pid}
\begin{barticle}[author]
\bauthor{\bsnm{Schaefer},~\bfnm{Carl~F}\binits{C.~F.}},
  \bauthor{\bsnm{Anthony},~\bfnm{Kira}\binits{K.}},
  \bauthor{\bsnm{Krupa},~\bfnm{Shiva}\binits{S.}},
  \bauthor{\bsnm{Buchoff},~\bfnm{Jeffrey}\binits{J.}},
  \bauthor{\bsnm{Day},~\bfnm{Matthew}\binits{M.}},
  \bauthor{\bsnm{Hannay},~\bfnm{Timo}\binits{T.}} \AND
  \bauthor{\bsnm{Buetow},~\bfnm{Kenneth~H}\binits{K.~H.}}
(\byear{2009}).
\btitle{PID: the pathway interaction database}.
\bjournal{Nucleic acids research}
\bvolume{37}
\bpages{D674--D679}.
\end{barticle}
\endbibitem

\bibitem[\protect\citeauthoryear{Sepich-Poore
  et~al.}{2021}]{sepich2021microbiome}
\begin{barticle}[author]
\bauthor{\bsnm{Sepich-Poore},~\bfnm{Gregory~D}\binits{G.~D.}},
  \bauthor{\bsnm{Zitvogel},~\bfnm{Laurence}\binits{L.}},
  \bauthor{\bsnm{Straussman},~\bfnm{Ravid}\binits{R.}},
  \bauthor{\bsnm{Hasty},~\bfnm{Jeff}\binits{J.}},
  \bauthor{\bsnm{Wargo},~\bfnm{Jennifer~A}\binits{J.~A.}} \AND
  \bauthor{\bsnm{Knight},~\bfnm{Rob}\binits{R.}}
(\byear{2021}).
\btitle{The microbiome and human cancer}.
\bjournal{Science}
\bvolume{371}
\bpages{eabc4552}.
\end{barticle}
\endbibitem

\bibitem[\protect\citeauthoryear{Sharifi and Ye}{2017}]{Sharifi2017}
\begin{bincollection}[author]
\bauthor{\bsnm{Sharifi},~\bfnm{F}\binits{F.}} \AND
  \bauthor{\bsnm{Ye},~\bfnm{Y}\binits{Y.}}
(\byear{2017}).
\btitle{From gene annotation to function prediction for metagenomics}.
In \bbooktitle{Protein Function Prediction}
\bpages{27--34}.
\bpublisher{Springer}.
\end{bincollection}
\endbibitem

\bibitem[\protect\citeauthoryear{Sun and Zhang}{2012}]{sun2012scaled}
\begin{barticle}[author]
\bauthor{\bsnm{Sun},~\bfnm{Tingni}\binits{T.}} \AND
  \bauthor{\bsnm{Zhang},~\bfnm{Cun-Hui}\binits{C.-H.}}
(\byear{2012}).
\btitle{Scaled sparse linear regression}.
\bjournal{Biometrika}
\bvolume{99}
\bpages{879--898}.
\end{barticle}
\endbibitem

\bibitem[\protect\citeauthoryear{Tibshirani}{1996}]{tibshirani1996regression}
\begin{barticle}[author]
\bauthor{\bsnm{Tibshirani},~\bfnm{Robert}\binits{R.}}
(\byear{1996}).
\btitle{Regression shrinkage and selection via the lasso}.
\bjournal{Journal of the Royal Statistical Society: Series B (Methodological)}
\bvolume{58}
\bpages{267--288}.
\end{barticle}
\endbibitem

\bibitem[\protect\citeauthoryear{van~de Geer et~al.}{2009}]{van2009conditions}
\begin{barticle}[author]
\bauthor{\bparticle{van~de} \bsnm{Geer},~\bfnm{Sara}\binits{S.}},
  \bauthor{\bsnm{B{\"u}hlmann},~\bfnm{Peter}\binits{P.}} \betal{et~al.}
(\byear{2009}).
\btitle{On the conditions used to prove oracle results for the Lasso}.
\bjournal{Electronic Journal of Statistics}
\bvolume{3}
\bpages{1360--1392}.
\end{barticle}
\endbibitem

\bibitem[\protect\citeauthoryear{van~de Geer et~al.}{2014}]{vandegeer2014}
\begin{barticle}[author]
\bauthor{\bparticle{van~de} \bsnm{Geer},~\bfnm{Sara}\binits{S.}},
  \bauthor{\bsnm{Bühlmann},~\bfnm{Peter}\binits{P.}},
  \bauthor{\bsnm{Ritov},~\bfnm{Ya’acov}\binits{Y.}} \AND
  \bauthor{\bsnm{Dezeure},~\bfnm{Ruben}\binits{R.}}
(\byear{2014}).
\btitle{On asymptotically optimal confidence regions and tests for
  high-dimensional models}.
\bjournal{The Annals of Statistics}
\bvolume{42}
\bpages{1166--1202}.
\bdoi{10.1214/14-AOS1221}
\end{barticle}
\endbibitem

\bibitem[\protect\citeauthoryear{Wainwright}{2019}]{wainwright2019high}
\begin{bbook}[author]
\bauthor{\bsnm{Wainwright},~\bfnm{Martin~J}\binits{M.~J.}}
(\byear{2019}).
\btitle{High-dimensional statistics: A non-asymptotic viewpoint}
\bvolume{48}.
\bpublisher{Cambridge University Press}.
\end{bbook}
\endbibitem

\bibitem[\protect\citeauthoryear{Wang et~al.}{2019}]{Wange00504-19}
\begin{barticle}[author]
\bauthor{\bsnm{Wang},~\bfnm{Yue}\binits{Y.}},
  \bauthor{\bsnm{Randolph},~\bfnm{Timothy~W.}\binits{T.~W.}},
  \bauthor{\bsnm{Shojaie},~\bfnm{Ali}\binits{A.}} \AND
  \bauthor{\bsnm{Ma},~\bfnm{Jing}\binits{J.}}
(\byear{2019}).
\btitle{The Generalized Matrix Decomposition Biplot and Its Application to
  Microbiome Data}.
\bjournal{mSystems}
\bvolume{4}.
\bdoi{10.1128/mSystems.00504-19}
\end{barticle}
\endbibitem

\bibitem[\protect\citeauthoryear{Wang et~al.}{2023}]{wang2023gmdrsupp}
\begin{barticle}[author]
\bauthor{\bsnm{Wang},~\bfnm{Yue}\binits{Y.}},
  \bauthor{\bsnm{Shojaie},~\bfnm{Ali}\binits{A.}},
  \bauthor{\bsnm{Randolph},~\bfnm{Timothy}\binits{T.}},
  \bauthor{\bsnm{Knight},~\bfnm{Parker}\binits{P.}} \AND
  \bauthor{\bsnm{Ma},~\bfnm{Jing}\binits{J.}}
(\byear{2023}).
\btitle{Supplement to “Generalized matrix decomposition regression:
  estimation and inference for two-way structured data.”}.
\end{barticle}
\endbibitem

\bibitem[\protect\citeauthoryear{Washburne et~al.}{2018}]{Washburne2018}
\begin{barticle}[author]
\bauthor{\bsnm{Washburne},~\bfnm{Alex~D}\binits{A.~D.}},
  \bauthor{\bsnm{Morton},~\bfnm{James~T}\binits{J.~T.}},
  \bauthor{\bsnm{Sanders},~\bfnm{Jon}\binits{J.}},
  \bauthor{\bsnm{McDonald},~\bfnm{Daniel}\binits{D.}},
  \bauthor{\bsnm{Zhu},~\bfnm{Qiyun}\binits{Q.}},
  \bauthor{\bsnm{Oliverio},~\bfnm{Angela~M}\binits{A.~M.}} \AND
  \bauthor{\bsnm{Knight},~\bfnm{Rob}\binits{R.}}
(\byear{2018}).
\btitle{Methods for phylogenetic analysis of microbiome data}.
\bjournal{Nature microbiology}
\bvolume{3}
\bpages{652--661}.
\end{barticle}
\endbibitem

\bibitem[\protect\citeauthoryear{Xu et~al.}{2020}]{xu2020function}
\begin{barticle}[author]
\bauthor{\bsnm{Xu},~\bfnm{Yu}\binits{Y.}},
  \bauthor{\bsnm{Wang},~\bfnm{Ning}\binits{N.}},
  \bauthor{\bsnm{Tan},~\bfnm{Hor-Yue}\binits{H.-Y.}},
  \bauthor{\bsnm{Li},~\bfnm{Sha}\binits{S.}},
  \bauthor{\bsnm{Zhang},~\bfnm{Cheng}\binits{C.}} \AND
  \bauthor{\bsnm{Feng},~\bfnm{Yibin}\binits{Y.}}
(\byear{2020}).
\btitle{Function of Akkermansia muciniphila in obesity: interactions with lipid
  metabolism, immune response and gut systems}.
\bjournal{Frontiers in microbiology}
\bvolume{11}
\bpages{219}.
\end{barticle}
\endbibitem

\bibitem[\protect\citeauthoryear{Yatsunenko et~al.}{2012}]{Yatsunenko2012}
\begin{barticle}[author]
\bauthor{\bsnm{Yatsunenko},~\bfnm{T.}\binits{T.}},
  \bauthor{\bsnm{Rey},~\bfnm{F.~E.}\binits{F.~E.}},
  \bauthor{\bsnm{Manary},~\bfnm{M.~J.}\binits{M.~J.}},
  \bauthor{\bsnm{Trehan},~\bfnm{I.}\binits{I.}},
  \bauthor{\bsnm{Dominguez-Bello},~\bfnm{M.~G.}\binits{M.~G.}},
  \bauthor{\bsnm{Contreras},~\bfnm{M.}\binits{M.}},
  \bauthor{\bsnm{Magris},~\bfnm{M.}\binits{M.}},
  \bauthor{\bsnm{Hidalgo},~\bfnm{G.}\binits{G.}},
  \bauthor{\bsnm{Baldassano},~\bfnm{R.~N.}\binits{R.~N.}},
  \bauthor{\bsnm{Anokhin},~\bfnm{A.~P.}\binits{A.~P.}},
  \bauthor{\bsnm{Heath},~\bfnm{A.~C.}\binits{A.~C.}},
  \bauthor{\bsnm{Warner},~\bfnm{B.}\binits{B.}},
  \bauthor{\bsnm{Reeder},~\bfnm{J.}\binits{J.}},
  \bauthor{\bsnm{Kuczynski},~\bfnm{J.}\binits{J.}},
  \bauthor{\bsnm{Caporaso},~\bfnm{J.~G.}\binits{J.~G.}},
  \bauthor{\bsnm{Lozupone},~\bfnm{C.~A.}\binits{C.~A.}},
  \bauthor{\bsnm{Lauber},~\bfnm{C.}\binits{C.}},
  \bauthor{\bsnm{Clemente},~\bfnm{J.~C.}\binits{J.~C.}},
  \bauthor{\bsnm{Knights},~\bfnm{D.}\binits{D.}},
  \bauthor{\bsnm{Knight},~\bfnm{R.}\binits{R.}} \AND
  \bauthor{\bsnm{Gordon},~\bfnm{J.~I.}\binits{J.~I.}}
(\byear{2012}).
\btitle{{{H}uman gut microbiome viewed across age and geography}}.
\bjournal{Nature}
\bvolume{486}
\bpages{222--227}.
\end{barticle}
\endbibitem

\bibitem[\protect\citeauthoryear{Yu and Bien}{2019}]{yu2019estimating}
\begin{barticle}[author]
\bauthor{\bsnm{Yu},~\bfnm{Guo}\binits{G.}} \AND
  \bauthor{\bsnm{Bien},~\bfnm{Jacob}\binits{J.}}
(\byear{2019}).
\btitle{Estimating the error variance in a high-dimensional linear model}.
\bjournal{Biometrika}
\bvolume{106}
\bpages{533--546}.
\end{barticle}
\endbibitem

\bibitem[\protect\citeauthoryear{Zeevi et~al.}{2019}]{zeevi2019structural}
\begin{barticle}[author]
\bauthor{\bsnm{Zeevi},~\bfnm{David}\binits{D.}},
  \bauthor{\bsnm{Korem},~\bfnm{Tal}\binits{T.}},
  \bauthor{\bsnm{Godneva},~\bfnm{Anastasia}\binits{A.}},
  \bauthor{\bsnm{Bar},~\bfnm{Noam}\binits{N.}},
  \bauthor{\bsnm{Kurilshikov},~\bfnm{Alexander}\binits{A.}},
  \bauthor{\bsnm{Lotan-Pompan},~\bfnm{Maya}\binits{M.}},
  \bauthor{\bsnm{Weinberger},~\bfnm{Adina}\binits{A.}},
  \bauthor{\bsnm{Fu},~\bfnm{Jingyuan}\binits{J.}},
  \bauthor{\bsnm{Wijmenga},~\bfnm{Cisca}\binits{C.}},
  \bauthor{\bsnm{Zhernakova},~\bfnm{Alexandra}\binits{A.}} \betal{et~al.}
(\byear{2019}).
\btitle{Structural variation in the gut microbiome associates with host
  health}.
\bjournal{Nature}
\bvolume{568}
\bpages{43--48}.
\end{barticle}
\endbibitem

\bibitem[\protect\citeauthoryear{Zhan et~al.}{2017}]{zhan2017fast}
\begin{barticle}[author]
\bauthor{\bsnm{Zhan},~\bfnm{Xiang}\binits{X.}},
  \bauthor{\bsnm{Plantinga},~\bfnm{Anna}\binits{A.}},
  \bauthor{\bsnm{Zhao},~\bfnm{Ni}\binits{N.}} \AND
  \bauthor{\bsnm{Wu},~\bfnm{Michael~C}\binits{M.~C.}}
(\byear{2017}).
\btitle{A fast small-sample kernel independence test for microbiome
  community-level association analysis}.
\bjournal{Biometrics}
\bvolume{73}
\bpages{1453--1463}.
\end{barticle}
\endbibitem

\bibitem[\protect\citeauthoryear{Zhang et~al.}{2008}]{zhang2008sparsity}
\begin{barticle}[author]
\bauthor{\bsnm{Zhang},~\bfnm{Cun-Hui}\binits{C.-H.}},
  \bauthor{\bsnm{Huang},~\bfnm{Jian}\binits{J.}} \betal{et~al.}
(\byear{2008}).
\btitle{The sparsity and bias of the lasso selection in high-dimensional linear
  regression}.
\bjournal{The Annals of Statistics}
\bvolume{36}
\bpages{1567--1594}.
\end{barticle}
\endbibitem

\bibitem[\protect\citeauthoryear{Zhang and Pan}{2015}]{zhang2015principal}
\begin{barticle}[author]
\bauthor{\bsnm{Zhang},~\bfnm{Yiwei}\binits{Y.}} \AND
  \bauthor{\bsnm{Pan},~\bfnm{Wei}\binits{W.}}
(\byear{2015}).
\btitle{Principal component regression and linear mixed model in association
  analysis of structured samples: competitors or complements?}
\bjournal{Genetic epidemiology}
\bvolume{39}
\bpages{149--155}.
\end{barticle}
\endbibitem

\bibitem[\protect\citeauthoryear{Zhang and Zhang}{2014}]{zhangzhang2014}
\begin{barticle}[author]
\bauthor{\bsnm{Zhang},~\bfnm{Cun~Hui}\binits{C.~H.}} \AND
  \bauthor{\bsnm{Zhang},~\bfnm{Stephanie~S.}\binits{S.~S.}}
(\byear{2014}).
\btitle{Confidence intervals for low dimensional parameters in high dimensional
  linear models}.
\bjournal{Journal of the Royal Statistical Society: Series B (Statistical
  Methodology)}
\bvolume{76}
\bpages{217-242}.
\end{barticle}
\endbibitem

\bibitem[\protect\citeauthoryear{Zhang et~al.}{2019}]{zhang2019advancing}
\begin{barticle}[author]
\bauthor{\bsnm{Zhang},~\bfnm{Xu}\binits{X.}},
  \bauthor{\bsnm{Li},~\bfnm{Leyuan}\binits{L.}},
  \bauthor{\bsnm{Butcher},~\bfnm{James}\binits{J.}},
  \bauthor{\bsnm{Stintzi},~\bfnm{Alain}\binits{A.}} \AND
  \bauthor{\bsnm{Figeys},~\bfnm{Daniel}\binits{D.}}
(\byear{2019}).
\btitle{Advancing functional and translational microbiome research using
  meta-omics approaches}.
\bjournal{Microbiome}
\bvolume{7}
\bpages{1--12}.
\end{barticle}
\endbibitem

\bibitem[\protect\citeauthoryear{Zhao and Shojaie}{2016}]{zhao2016}
\begin{barticle}[author]
\bauthor{\bsnm{Zhao},~\bfnm{S.}\binits{S.}} \AND
  \bauthor{\bsnm{Shojaie},~\bfnm{A.}\binits{A.}}
(\byear{2016}).
\btitle{{{A} significance test for graph-constrained estimation}}.
\bjournal{Biometrics}
\bvolume{72}
\bpages{484--493}.
\end{barticle}
\endbibitem

\bibitem[\protect\citeauthoryear{Zhao et~al.}{2015}]{zhao2015testing}
\begin{barticle}[author]
\bauthor{\bsnm{Zhao},~\bfnm{Ni}\binits{N.}},
  \bauthor{\bsnm{Chen},~\bfnm{Jun}\binits{J.}},
  \bauthor{\bsnm{Carroll},~\bfnm{Ian~M}\binits{I.~M.}},
  \bauthor{\bsnm{Ringel-Kulka},~\bfnm{Tamar}\binits{T.}},
  \bauthor{\bsnm{Epstein},~\bfnm{Michael~P}\binits{M.~P.}},
  \bauthor{\bsnm{Zhou},~\bfnm{Hua}\binits{H.}},
  \bauthor{\bsnm{Zhou},~\bfnm{Jin~J}\binits{J.~J.}},
  \bauthor{\bsnm{Ringel},~\bfnm{Yehuda}\binits{Y.}},
  \bauthor{\bsnm{Li},~\bfnm{Hongzhe}\binits{H.}} \AND
  \bauthor{\bsnm{Wu},~\bfnm{Michael~C}\binits{M.~C.}}
(\byear{2015}).
\btitle{Testing in microbiome-profiling studies with MiRKAT, the microbiome
  regression-based kernel association test}.
\bjournal{The American Journal of Human Genetics}
\bvolume{96}
\bpages{797--807}.
\end{barticle}
\endbibitem

\bibitem[\protect\citeauthoryear{Zhu and Bradic}{2018}]{zhubradic2018}
\begin{barticle}[author]
\bauthor{\bsnm{Zhu},~\bfnm{Yinchu}\binits{Y.}} \AND
  \bauthor{\bsnm{Bradic},~\bfnm{Jelena}\binits{J.}}
(\byear{2018}).
\btitle{Linear Hypothesis Testing in Dense High-Dimensional Linear Models}.
\bjournal{Journal of the American Statistical Association}
\bvolume{113}
\bpages{1583-1600}.
\bdoi{10.1080/01621459.2017.1356319}
\end{barticle}
\endbibitem

\end{thebibliography}

%% or include bibliography directly:
% \begin{thebibliography}{4}
% %%
% \bibitem[\protect\citeauthoryear{Billingsley}{1999}]{r1}
% \textsc{Billingsley, P.} (1999). \textit{Convergence of
% Probability Measures}, 2nd ed.
% Wiley, New York.

% \bibitem[\protect\citeauthoryear{Bourbaki}{1966}]{r2}
% \textsc{Bourbaki, N.}  (1966). \textit{General Topology}  \textbf{1}.
% Addison--Wesley, Reading, MA.

% \bibitem[\protect\citeauthoryear{Ethier and Kurtz}{1985}]{r3}
% \textsc{Ethier, S. N.} and \textsc{Kurtz, T. G.} (1985).
% \textit{Markov Processes: Characterization and Convergence}.
% Wiley, New York.

% \bibitem[\protect\citeauthoryear{Prokhorov}{1956}]{r4}
% \textsc{Prokhorov, Yu.} (1956).
% Convergence of random processes and limit theorems in probability
% theory. \textit{Theory  Probab.  Appl.}
% \textbf{1} 157--214.
% \end{thebibliography}

\end{document}